
%
%
%
%
\def\unredoffs{\hoffset-.14truein\voffset-.2truein} 
 
%
%
\newbox\leftpage \newdimen\fullhsize \newdimen\hstitle \newdimen\hsbody
\tolerance=1000\hfuzz=2pt
\catcode`\@=11 
%
\magnification=1095\unredoffs\baselineskip=16pt plus 2pt minus 1pt
\hsbody=\hsize \hstitle=\hsize 
%
%
%
\newcount\yearltd\yearltd=\year\advance\yearltd by -1900

%
%

\def\draftmode{\message{ DRAFTMODE }\def\draftdate{{\rm preliminary draft:
\number\month/\number\day/\number\yearltd\ \ \hourmin}}%
\headline={\hfil\draftdate}\writelabels\baselineskip=16pt plus 2pt minus 2pt
 {\count255=\time\divide\count255 by 60 \xdef\hourmin{\number\count255}
  \multiply\count255 by-60\advance\count255 by\time
  \xdef\hourmin{\hourmin:\ifnum\count255<10 0\fi\the\count255}}}
\def\nolabels{\def\wrlabeL##1{}\def\eqlabeL##1{}\def\reflabeL##1{}}
\def\writelabels{\def\wrlabeL##1{\leavevmode\vadjust{\rlap{\smash%
{\line{{\escapechar=` \hfill\rlap{\sevenrm\hskip.03in\string##1}}}}}}}%
\def\eqlabeL##1{{\escapechar-1\rlap{\sevenrm\hskip.05in\string##1}}}%
\def\reflabeL##1{\noexpand\llap{\noexpand\sevenrm\string\string\string##1}}}
\nolabels
%
\global\newcount\secno \global\secno=0
\global\newcount\meqno \global\meqno=1
\def\newsec#1{\global\advance\secno by1\message{(\the\secno. #1)}
\global\subsecno=0\eqnres@t\noindent{\bf\the\secno. #1}
\writetoca{{\secsym} {#1}}\par\nobreak\medskip\nobreak}
\def\eqnres@t{\xdef\secsym{\the\secno.}\global\meqno=1\bigbreak\bigskip}
\def\sequentialequations{\def\eqnres@t{\bigbreak}}\xdef\secsym{}
\global\newcount\subsecno \global\subsecno=0
\def\subsec#1{\global\advance\subsecno by1\message{(\secsym\the\subsecno. #1)}
\ifnum\lastpenalty>9000\else\bigbreak\fi
\noindent{\bf\secsym\the\subsecno. #1}\writetoca{\string\quad 
{\secsym\the\subsecno.} {#1}}\par\nobreak\medskip\nobreak}
\def\appendix#1#2{\global\meqno=1\global\subsecno=0\xdef\secsym{\hbox{#1.}}
\bigbreak\bigskip\noindent{\bf Appendix #1. #2}\message{(#1. #2)}
\writetoca{Appendix {#1.} {#2}}\par\nobreak\medskip\nobreak}
%
%
\def\eqnn#1{\xdef #1{(\secsym\the\meqno)}\writedef{#1\leftbracket#1}%
\global\advance\meqno by1\wrlabeL#1}
\def\eqna#1{\xdef #1##1{\hbox{$(\secsym\the\meqno##1)$}}
\writedef{#1\numbersign1\leftbracket#1{\numbersign1}}%
\global\advance\meqno by1\wrlabeL{#1$\{\}$}}
\def\eqn#1#2{\xdef #1{(\secsym\the\meqno)}\writedef{#1\leftbracket#1}%
\global\advance\meqno by1$$#2\eqno#1\eqlabeL#1$$}
%
\newskip\footskip\footskip14pt plus 1pt minus 1pt 
\def\footnotefont{\ninepoint}\def\f@t#1{\footnotefont #1\@foot}
\def\f@@t{\baselineskip\footskip\bgroup\footnotefont\aftergroup\@foot\let\next}
\setbox\strutbox=\hbox{\vrule height9.5pt depth4.5pt width0pt}
\global\newcount\ftno \global\ftno=0
\def\foot{\global\advance\ftno by1\footnote{$^{\the\ftno}$}}
%
\newwrite\ftfile   
\def\footend{\def\foot{\global\advance\ftno by1\chardef\wfile=\ftfile
$^{\the\ftno}$\ifnum\ftno=1\immediate\openout\ftfile=foots.tmp\fi%
\immediate\write\ftfile{\noexpand\smallskip%
\noexpand\item{f\the\ftno:\ }\pctsign}\findarg}%
\def\footatend{\vfill\eject\immediate\closeout\ftfile{\parindent=20pt
\centerline{\bf Footnotes}\nobreak\bigskip\input foots.tmp }}}
\def\footatend{}
%
%
\global\newcount\refno \global\refno=1
\newwrite\rfile
\def\ref{[\the\refno]\nref}
\def\nref#1{\xdef#1{[\the\refno]}\writedef{#1\leftbracket#1}%
\ifnum\refno=1\immediate\openout\rfile=refs.tmp\fi
\global\advance\refno by1\chardef\wfile=\rfile\immediate
\write\rfile{\noexpand\item{#1\ }\reflabeL{#1\hskip.31in}\pctsign}\findarg}
\def\findarg#1#{\begingroup\obeylines\newlinechar=`\^^M\pass@rg}
{\obeylines\gdef\pass@rg#1{\writ@line\relax #1^^M\hbox{}^^M}%
\gdef\writ@line#1^^M{\expandafter\toks0\expandafter{\striprel@x #1}%
\edef\next{\the\toks0}\ifx\next\em@rk\let\next=\endgroup\else\ifx\next\empty%
\else\immediate\write\wfile{\the\toks0}\fi\let\next=\writ@line\fi\next\relax}}
\def\striprel@x#1{} \def\em@rk{\hbox{}} 
\def\lref{\begingroup\obeylines\lr@f}
\def\lr@f#1#2{\gdef#1{\ref#1{#2}}\endgroup\unskip}
\def\semi{;\hfil\break}
\def\addref#1{\immediate\write\rfile{\noexpand\item{}#1}} 
\def\startrefs#1{\immediate\openout\rfile=refs.tmp\refno=#1}
\def\xref{\expandafter\xr@f}\def\xr@f[#1]{#1}
\def\refs#1{\count255=1[\r@fs #1{\hbox{}}]}
\def\r@fs#1{\ifx\und@fined#1\message{reflabel \string#1 is undefined.}%
\nref#1{need to supply reference \string#1.}\fi%
\vphantom{\hphantom{#1}}\edef\next{#1}\ifx\next\em@rk\def\next{}%
\else\ifx\next#1\ifodd\count255\relax\xref#1\count255=0\fi%
\else#1\count255=1\fi\let\next=\r@fs\fi\next}
%

%
\newwrite\ffile\global\newcount\figno \global\figno=1
\def\fig{fig.~\the\figno\nfig}
\def\nfig#1{\xdef#1{fig.~\the\figno}%
\writedef{#1\leftbracket fig.\noexpand~\the\figno}%
\ifnum\figno=1\immediate\openout\ffile=figs.tmp\fi\chardef\wfile=\ffile%
\immediate\write\ffile{\noexpand\medskip\noexpand\item{Fig.\ \the\figno. }
\reflabeL{#1\hskip.55in}\pctsign}\global\advance\figno by1\findarg}
\def\xfig{\expandafter\xf@g}\def\xf@g fig.\penalty\@M\ {}
\def\figs#1{figs.~\f@gs #1{\hbox{}}}
\def\f@gs#1{\edef\next{#1}\ifx\next\em@rk\def\next{}\else
\ifx\next#1\xfig #1\else#1\fi\let\next=\f@gs\fi\next}
\newwrite\lfile
{\escapechar-1\xdef\pctsign{\string\%}\xdef\leftbracket{\string\{}
\xdef\rightbracket{\string\}}\xdef\numbersign{\string\#}}

\def\writestop{\def\writestoppt{\immediate\write\lfile{\string\pageno%
\the\pageno\string\startrefs\leftbracket\the\refno\rightbracket%
\string\def\string\secsym\leftbracket\secsym\rightbracket%
\string\secno\the\secno\string\meqno\the\meqno}\immediate\closeout\lfile}}
\def\writestoppt{}\def\writedef#1{}
\def\seclab#1{\xdef #1{\the\secno}\writedef{#1\leftbracket#1}\wrlabeL{#1=#1}}
\def\subseclab#1{\xdef #1{\secsym\the\subsecno}%
\writedef{#1\leftbracket#1}\wrlabeL{#1=#1}}
\newwrite\tfile \def\writetoca#1{}
\def\leaderfill{\leaders\hbox to 1em{\hss.\hss}\hfill}
\def\writetoc{\immediate\openout\tfile=toc.tmp 
   \def\writetoca##1{{\edef\next{\write\tfile{\noindent ##1 
   \string\leaderfill {\noexpand\number\pageno} \par}}\next}}}

%
\catcode`\@=12 
%
\edef\tfontsize{\ifx\answ\bigans scaled\magstep3\else scaled\magstep4\fi}
 \tfontsize  \tfontsize
 \tfontsize \font\titlei=cmmi10 \tfontsize
\font\titleis=cmmi7 \tfontsize \font\titleiss=cmmi5 \tfontsize
\font\titlesy=cmsy10 \tfontsize \font\titlesys=cmsy7 \tfontsize
\font\titlesyss=cmsy5 \tfontsize  \tfontsize
\skewchar\titlei='177 \skewchar\titleis='177 \skewchar\titleiss='177
\skewchar\titlesy='60 \skewchar\titlesys='60 \skewchar\titlesyss='60
 \ifx\answ\bigans\else scaled\magstep1\fi
\ifx\answ\bigans\else

 \font\absi=cmmi10 scaled\magstep1
\font\absis=cmmi7 scaled\magstep1 \font\absiss=cmmi5 scaled\magstep1
\font\abssy=cmsy10 scaled\magstep1 \font\abssys=cmsy7 scaled\magstep1
\font\abssyss=cmsy5 scaled\magstep1 
\skewchar\absi='177 \skewchar\absis='177 \skewchar\absiss='177
\skewchar\abssy='60 \skewchar\abssys='60 \skewchar\abssyss='60
\fi
\font\ninerm=cmr9 \font\sixrm=cmr6 \font\ninei=cmmi9 \font\sixi=cmmi6 
\font\ninesy=cmsy9 \font\sixsy=cmsy6 \font\ninebf=cmbx9 
\font\nineit=cmti9 \font\ninesl=cmsl9 \skewchar\ninei='177
\skewchar\sixi='177 \skewchar\ninesy='60 \skewchar\sixsy='60 
\def\ninepoint{\def\rm{\fam0\ninerm}
\textfont0=\ninerm \scriptfont0=\sixrm \scriptscriptfont0=\fiverm
\textfont1=\ninei \scriptfont1=\sixi \scriptscriptfont1=\fivei
\textfont2=\ninesy \scriptfont2=\sixsy \scriptscriptfont2=\fivesy
\textfont\itfam=\ninei \def\it{\fam\itfam\nineit}\def\sl{\fam\slfam\ninesl}%
\textfont\bffam=\ninebf \def\bf{\fam\bffam\ninebf}\rm} 
%
%
\def\noblackbox{\overfullrule=0pt}
\hyphenation{anom-aly anom-alies coun-ter-term coun-ter-terms}
\def\inv{^{\raise.15ex\hbox{${\scriptscriptstyle -}$}\kern-.05em 1}}

\def\Dsl{\,\raise.15ex\hbox{/}\mkern-13.5mu D} 
\def\dsl{\raise.15ex\hbox{/}\kern-.57em\partial}

\def\lspace{\ifx\answ\bigans{}\else\qquad\fi}
\def\lbspace{\ifx\answ\bigans{}\else\hskip-.2in\fi} 
\def\boxeqn#1{\vcenter{\vbox{\hrule\hbox{\vrule\kern3pt\vbox{\kern3pt
	\hbox{${\displaystyle #1}$}\kern3pt}\kern3pt\vrule}\hrule}}}
\def\mbox#1#2{\vcenter{\hrule \hbox{\vrule height#2in
		\kern#1in \vrule} \hrule}}  
%

\def\darr#1{\raise1.5ex\hbox{$\leftrightarrow$}\mkern-16.5mu #1}

\def\half{{\textstyle{1\over2}}} 
\def\roughly#1{\raise.3ex\hbox{$#1$\kern-.75em\lower1ex\hbox{$\sim$}}}

\def\p2inf{\mathrel{\mathop{\sim}\limits_{\scriptscriptstyle
{p^2 \rightarrow \infty }}}}
\def\kap2inf{\mathrel{\mathop{\sim}\limits_{\scriptscriptstyle
{\kappa \rightarrow \infty }}}}
\def\x2inf{\mathrel{\mathop{\sim}\limits_{\scriptscriptstyle
{x \rightarrow \infty }}}}
\def\Lam2inf{\mathrel{\mathop{\sim}\limits_{\scriptscriptstyle
{\Lambda \rightarrow \infty }}}}
\def\frac#1#2{{{#1}\over {#2}}}
\def\half{\hbox{${1\over 2}$}}

\def\Gev{{\rm GeV}}

\def\Real{\Re e}

\def\lsim{\mathrel{mathpalette\@v1000ersim<}}
\def\gsim{\mathrel{mathpalette\@versim>}}

\catcode`@=11 
\def\slash#1{\mathord{\mathpalette\c@ncel#1}}
 \def\c@ncel#1#2{\ooalign{$\hfil#1\mkern1mu/\hfil$\crcr$#1#2$}}
\def\lsim{\mathrel{\mathpalette\@versim<}}
\def\gsim{\mathrel{\mathpalette\@versim>}}
 \def\@versim#1#2{\lower0.2ex\vbox{\baselineskip\z@skip\lineskip\z@skip
       \lineskiplimit\z@\ialign{$\m@th#1\hfil##$\crcr#2\crcr\sim\crcr}}}
\catcode`@=12 

\def\EPJ{{\it Eur. Phys. J.~}}
\def\PR{{\it Phys.~Rev.~}}

\def\NP{{\it Nucl.~Phys.~}}
\def\PL{{\it Phys.~Lett.~}}
\def\PRep{{\it Phys.~Rep.~}}

\def\SJNP{{\it Sov.~Jour.~Nucl.~Phys.~}}
\def\ZP{{\it Zeit.~Phys.~}}

\def\vol#1{{\bf #1}}
\def\vyp#1#2#3{\vol{#1} (#2) #3}

\def\Asl{\raise.15ex\hbox{/}\mkern-11.5mu A}
\def\psl{\lower.12ex\hbox{/}\mkern-9.5mu p}
\def\qsl{\lower.12ex\hbox{/}\mkern-9.5mu q}
\def\rsl{\lower.03ex\hbox{/}\mkern-9.5mu r}
\def\ksl{\raise.06ex\hbox{/}\mkern-9.5mu k}

\noblackbox

\pageno=0\nopagenumbers\tolerance=10000\hfuzz=5pt
\line{\hfill Cavendish-HEP-01/03}
\vskip 36pt
\centerline{\bf The Running Coupling BFKL Anomalous Dimensions}
\centerline{\bf and Splitting Functions}
\vskip 36pt
\centerline{Robert~S.~Thorne\foot{Royal Society University Research Fellow}}
\vskip 12pt
\centerline{\it Cavendish Laboratory, University of Cambridge}
\centerline{\it Madingley Road, Cambridge, CB3 0HE, U.K.}
\vskip 0.9in
{\narrower\baselineskip 10pt
\centerline{\bf Abstract}
\medskip
I explicitly calculate the anomalous dimensions and splitting
functions governing the $Q^2$ evolution 
of the parton densities and structure functions which result from the 
running coupling
BFKL equation at LO, i.e. I perform a resummation in powers of $\ln(1/x)$
and in powers of $\beta_0$ simultaneously.
This is extended as far as possible to NLO.  These are 
expressed in an exact, perturbatively calculable analytic form, 
up to small power-suppressed contributions which may also be modelled
to very good accuracy by analytic expressions. Infrared renormalons,
while in principle present in a solution in terms of powers in 
$\alpha_s(Q^2)$, are ultimately avoided.
The few higher twist contributions which are directly calculable 
are extremely small. The
splitting functions are very different from those obtained from the fixed
coupling equation, with weaker power-like growth $\sim x^{-0.25}$,
which does
not set in until extremely small $x$ indeed. The NLO BFKL corrections to
the splitting functions are moderate, both for the form of the
asymptotic power-like behaviour and more importantly for the
range of $x$ relevant for collider physics. 
Hence, a stable perturbative expansion and predictive power at small $x$ 
are obtained.}
   
\vskip 0.7in
\line{\hfill}
\line{March 2001\hfill}
\vfill\eject
\footline={\hss\tenrm\folio\hss}



\newsec{Introduction.}

Small $x$ physics has been a particularly  active area of particle physics
research in the past few years, driven largely by the first data for 
$x<0.005$ being obtained by the HERA experiments \ref\hone{H1 
collaboration: S. Aid {\it et al.},\NP  \vyp{B470}{1996}{3}\semi
H1 collaboration: C. Adloff {\it et al.}, \NP
\vyp{B497}{1997}{3}.}
\ref\zeus{ZEUS collaboration: M. Derrick {\it et al},
\ZP  \vyp{C69}{1996}{607}\semi ZEUS collaboration: M. Derrick {\it et al},
\ZP  \vyp{C72}{1996}{399}.}. However, as well as the need to describe this 
HERA data correctly, it will also be extremely important to  
understand the correct way of calculating physics at small $x$ in order to
interpret the results coming from the LHC in a truly quantitative manner.
For example, for the production of a particle of mass $\sim 100\Gev$ the 
typical value of $x$ probed (at central rapidity) is $0.005$, but values up 
to two orders 
of magnitude in either direction will also have an almost equally large 
influence.\foot{For an illustration of the $x$ and $Q^2$ of parton 
distributions sampled at the 
LHC see fig. 1 of \ref\MRSTa{A.D. Martin, R.G. Roberts, W.J. Stirling and 
R.S. Thorne, \EPJ \vyp{C14}{2000}{133}.}.}
       
The potential complication at small $x$ is that the splitting functions and
coefficient functions governing the evolution of parton distributions and 
their conversion to physical quantities have terms in their perturbative 
expansions which behave like $\alpha_s^n \ln^m(1/x)$, where $m$ can reach 
up to $n-1$. Therefore, as the power of the coupling increases the powers of
$\xi=\ln(1/x)$ also increase, and rapid perturbative convergence is not really
guaranteed if $\xi \gsim 1/\alpha_s$ i.e. $\sim 5$. This problem is not 
really diminished 
at the LHC, where the coupling is likely to be smaller than at HERA, since
the parton distributions to be used will be those measured at HERA at much
lower scales and evolved up to LHC scales. 
This question of large $\ln(1/x)$ terms is in principle addressed by the 
BFKL equation \ref\BFKL{L.N. Lipatov, 
\SJNP \vyp{23}{1976}{338}\semi
E.A. Kuraev, L.N. Lipatov and V.S. Fadin, {\it Sov.~Jour. JETP} 
\vyp{45}{1977}{199}\semi
Ya. Balitskii and L.N. Lipatov, \SJNP \vyp{28}{1978}{6}.}, 
which is an integral equation for the unintegrated 4-point  
gluon Green's function in the high energy limit. This sums the leading 
high-energy, or in the DIS case, small-$x$ behaviour, which is dominated by
the gluon, and thus allows the extraction of leading $\ln(1/x)$ terms for
relevant quantities, such as splitting functions. 

Hence, a major point of debate of the past decade has been
whether the standard DGLAP approach based on renormalization group
equations and conventionally ordered simply in powers of
$\alpha_s(Q^2)$, or the BFKL equation, 
which sums leading logarithms
in $(1/x)$, is most effective way of dealing with small $x$ physics
(most particularly structure functions), and/or whether the two approaches 
need to be combined in some way, and if so, how? 
While the conventional DGLAP approach has been
relatively successful, it does have some significant problems
(which are often overlooked): 
a valence-like, or even negative input gluon leading to a strange 
low-$Q^2$ $F_L(x,Q^2)$; 
undershooting of the data systematically for $x \sim 0.01$ at the highest 
$Q^2$ when a global fit is performed; and apparent instability at small $x$ 
order-by-order in $\alpha_s$ up to NLLO
\ref\mrstb{A.D. Martin, R.G. Roberts, W.J. Stirling and R.S. Thorne, 
\EPJ {\bf C16} (2000) 117.}.\foot{Of course the full 
NNLO splitting functions are not known, but good estimates are available
\ref\NNLOest{W.L. van Neerven and A. Vogt, \NP \vyp{B568}{2000}{263}\semi
W.L. van Neerven and A. Vogt, \NP \vyp{B588}{2000}{345}\semi
W.L. van Neerven and A. Vogt, \PL \vyp{B490}{2000}{111}.} based on 
calculation of moments in \ref\moments{S.A. Larin, P. Nogueira, 
T. van Ritenberg, and J.A.M. Vermaseren, \NP \vyp{B492}{1997}{338}\semi
A. Retey and J.A.M. Vermaseren, {\tt hep-ph/0007294}.}.}

Nevertheless, the BFKL equation did not seem to help these problems. 
The original LO BFKL prediction of a behaviour of the form $x^{-\lambda}$ 
for structure functions and splitting functions at
small $x$, with $\lambda \sim 0.5$, was clearly ruled out long ago. A
combination of the two approaches, using the BFKL equation to
supplement the Altarelli-Parisi splitting functions with higher terms
of the form $\alpha_s^{n+1} \ln^n(1/x)$ was originally successful (so long as
one avoided factorization scheme ambiguities by working in physical quantities)
\ref\LORSC{R.S. Thorne, \PL \vyp{B392}{1997}{463}; \NP 
\vyp{B512}{1998}{323}.}, but this success is not possible
to sustain with the most recent data \ref\newhone{H1 collaboration: C. Adloff
{\it et al}., \EPJ \vyp{C13}{2000}{609}\semi
H1 collaboration: C. Adloff {\it et al}., {\tt hep-ex/0012052}\semi
H1 collaboration: C. Adloff {\it et al}., 
{\tt hep-ex/0012053}.}\ref\newzeus{ZEUS collaboration: S. Chekanov {\it et al.}, 
{\tt hep-ex/0105090}.}. Moreover, the subject
was thrown into confusion by the calculation of the NLO
correction to the BFKL equation \ref\NLOBFKLlf{V.S. Fadin and
L.N. Lipatov, \PL \vyp{B429}{1998}{127}, and
references therein.}\ref\NLOBFKLcc{G. Camici and M. Ciafaloni, 
\PL \vyp{B430}{1998}{349}, and references therein.}.  
The results of this calculation were not very encouraging.
Ignoring the running of the coupling at NLO, i.e. proceeding with
the same sort of calculations as at LO but including the scale-independent 
NLO correction to the kernel, one obtains the ``intercept'' for the 
splitting function power-like behaviour, $x^{-\lambda}$, shifted from  
$\lambda=4\ln 2\bar\alpha_s$ 
to $\lambda=4\ln 2\bar\alpha_s(1-6.5\bar\alpha_s)$. 
This is clearly a huge correction,
and implies the breakdown of the perturbative expansion for this
quantity. More serious than this intercept is the power series for the 
splitting function, which may be calculated even taking into account the
renormalization and scale dependence introduced at NLO. Expanding 
this out formally to NLO in $\ln(1/x)$ one finds that it is is
dominated by the NLO corrections at all values of $x$ below about
$x=0.01$. For example, using the formulae in \NLOBFKLlf\
the first few terms in the power series for $P(x)$ go like
\eqn\unstab{\eqalign{xP(x,Q^2)=&\bar\alpha_s+2.4\bar\alpha_s^4
\xi^3/6+2.1\bar\alpha_s^6\xi^5/120+ \cdots\cr
&\hskip -0.3in -\bar\alpha_s(0.43\bar\alpha_s
+1.6\bar\alpha_s^2\xi+11.7\bar\alpha_s^3
\xi^2/2+13.3\bar\alpha_s^4\xi^3/6+39.7\bar\alpha_s^5\xi^4/24
+169.4\bar\alpha_s^6\xi^5/120+\cdots),\cr}}
where $\xi=\ln(1/x)$ and $\alpha_s \equiv \alpha_s(Q^2)$.
Clearly, the size of the coefficients more than compensates for the extra
power of $\alpha_s(Q^2)$, particularly at low $Q^2$ where 
the perturbative analysis of structure function evolution often takes place.
      
Hence, this NLO correction left open the whole question of how to 
address the evolution of structure functions at small $x$.
There has been considerable progress on the stability of the solutions to the 
BFKL equation in the intervening time. One major development was the 
observation that the resummation of double logarithmic terms in the 
transverse momentum $k^2$ is necessary in order to eliminate collinear 
divergences. This renders the intercept of the BFKL equation stable 
\ref\gavin{G.P. Salam, JHEP \vyp{9807}{1998}{19}.}, even when ignoring the 
renormalization scale dependence. 
This initial idea has been further
developed in \ref\ciafpcol{M. Ciafaloni 
and D. Colferai, \PL \vyp{452}{1999}{372}.}\ref\ciafsalam{M. Ciafaloni, 
D. Colferai and G.P. Salam,
\PR\vyp{D60}{1999}{114036}.}\ref\ciafcolsalam{M.Ciafaloni, 
D. Colferai and G.P. Salam, JHEP \vyp{9910}{1999}
{017}.}, where the effect of running coupling is also
considered in these later papers. This development is particularly 
important for the case of so-called ``single scale'' processes where both 
ends of the gluon Green's 
function are at high scales (not necessarily the same) where without this 
collinear resummation, all calculations are badly behaved over the full 
range of energy, not just in the asymptotic limit. 

However, for the type of situation embodied by DIS, where one end of the 
gluon Green's function is at some low non-perturbative scale, the factorization
theorem simplifies the problem. Although the growth of the coupling at low 
scales actually renders the solution of the BFKL equation formally 
divergent when the renormalization of the coupling is encountered, as 
realized as long ago as \ref\jan{J. Kwiecinski, \ZP \vyp{C29}{1985}{561}.}%
\nref\janpcoll{J.C. Collins and J. Kwiecinski, \NP \vyp{B316}{1989}{307}.}
and studied in detail in \ref\armesto{N. Armesto, J. Bartels and M.A. Braun, 
\PL \vyp{B442}{1998}{459}.}, all the uncertainty and indeed all the 
effects of the low $Q^2$ region are absorbed into the overall normalization
of the gluon, leaving the evolution and coefficient functions for hard 
scattering cross-sections calculable. However, these perturbatively 
calculable quantities are affected by the running of the coupling, and it
was argued in \ref\xscale{R.S. Thorne, \PR \vyp{D60}{1999}{054031}.} that
the effective result was as if the usual LO BFKL splitting functions should
be evaluated at an $x$-dependent scale, which grows with decreasing $x$,
due to increasing diffusion into the ultraviolet, 
leading to a decrease in the coupling. Hence, the effect of running coupling 
totally transforms the more simplistic LO BFKL results, making overall 
normalization of quantities incalculable, but moderating the effect of
those governing the evolution in $Q^2$. This moderation of the LO quantities 
also translated into a moderation of the effects of NLO corrections, leading 
to a much improved stability of the perturbative expansion, even without 
recourse to the type of resummation in \gavin-\ciafsalam.  
Indeed, for this case of deep inelastic scattering further resummation of 
this type is redundant. These modified BFKL contributions to the 
splitting functions, when combined with the conventional LO-in-$\alpha_s$
contributions, also led to improved fits compared to the usual DGLAP approach
\xscale\ and a more sensible prediction for $F_L(x,Q^2)$. This concept was 
put on a firmer footing in \ref\letter{R.S. Thorne,
\PL \vyp{B474}{2000}{372}.} where an explicit calculation of the 
BFKL splitting functions in powers of $\beta_0\alpha_s(Q^2)$, i.e. a
resummation of running coupling contributions, was outlined, and it was 
seen that over a wide range of the $x-Q^2$ range (including the HERA range)
the previous hypothesis was largely correct, and precise results were also
obtained outside this range.   

The purpose of this paper is to explain in detail, and expand upon the 
results of this previous letter, i.e. to present in full the calculation of 
splitting functions and coefficient functions for deep inelastic scattering 
obtained from the BFKL equation (both LO and NLO) and incorporating running
coupling contributions to all orders. Explicitly, while the usual BFKL 
equation presents an expression for these quantities which sums the leading 
power of $\xi$ at each power in $\alpha_s$, I will extend this  by 
producing expressions which
also include the leading power of $\beta_0$ at each power of $\alpha_s(Q^2)$ 
and $\xi$, e.g.
\eqn\expansion{xP_{gg}(x,Q^2) = \sum_{n=1}^{\infty} \sum_{m=0}^{n-1} 
a_{nm} \alpha_s^n(Q^2)\xi^{n-1-m} \beta_0^m,}  
though the formal divergence of the series will complicate this form 
a little. This presentation will begin, in section 
2, a brief review of the standard solution to the BFKL equation at LO, and 
then a detailed presentation of the solution at LO with running coupling. 
This will result in a solution for the gluon splitting function in an 
analytic form up to a small, unambiguous, correction of the form 
$\Lambda^2/Q^2$ 
(which is {\it not} higher twist) which may be modelled by an analytic 
function to excellent accuracy. Despite the integration over the infrared 
region when solving the running coupling BFKL equation there is no 
ambiguity in this splitting function.
Next, in section 3, will follow a discussion 
of some possible higher twist contributions at small $x$. It is argued that 
these are may be much smaller than generally supposed,
though the possibility of some large power-suppressed corrections 
(not necessarily higher twist) is left open. 
In section 4 I discuss the solution 
of the BFKL equation at NLO, defining precisely what I mean by the ``NLO
BFKL splitting function'', and showing that the NLO corrections for the 
gluon splitting function are moderate. 	In section 5 I consider real 
physical quantities, i.e the structure functions. Firstly, I calculate the 
quark-gluon splitting function and coefficient functions, 
and then consider the 
rather more direct physical splitting functions \ref\physanom{S. Catani,
in Proceedings of DIS 96, Rome, 1996, p. 165, {\tt hep-ph/9608310}; 
S. Catani \ZP \vyp{C75}{1997}{665}.}. I also consider how far one can 
calculate to NLO, defining a ``nearly NLO'' physical splitting function 
$P_{LL}(x, Q^2)$. The stability of the perturbative expansion is examined 
in detail, and seen to be very good. 
Finally, in section 6 phenomenology is briefly touched upon, 
and I present a summary and my conclusions.          

\newsec{BFKL Equation at LO.}

The BFKL equation for zero momentum transfer is an integral equation for 
the 4-point, transverse momentum-dependent gluon Green's function for
forward scattering in the high energy limit, $f(k_1, k_2, \alpha_s, N)$, 
where $N$ is the Mellin conjugate variable to energy. In the case of DIS
the second momentum $k_2$ is put equal to some non-perturbative scale 
$Q_0$, we let $k_1=k$, and $N$ becomes conjugate to $x$.
In order to obtain a structure function we attach the non-perturbative 
bare gluon distribution $g_B(N, Q_0^2)$ to the non-perturbative end of the 
gluon Green's function and convolute a hard scattering cross-section 
$h(Q^2/k^2,\alpha_s, N)$ to the perturbative end. 
   
In this section I will illustrate the effect that introducing the running 
coupling into the BFKL equation has. In order to do this I will first
begin with a brief presentation of the fairly simple traditional case of
fixed coupling before moving to the far more complicated case of running 
coupling. As will be seen, the introduction of renormalization, and hence 
running of the coupling, which is necessary except in the artificial model
of no consideration beyond LO, completely changes not only the detail of the
information one is able to extract from the BFKL equation, but also what
type of information one is able to extract.   

\subsec{Fixed Coupling.}

We simplify matters by working in moment
space, i.e. defining the moment of a structure function by
\eqn\melltranssf{{\cal F}(N,Q^2)= \int_0^1\,x^{N-1} F(x,Q^2)dx,}
and similarly for the parton distributions (scaled by $x$).
Doing this the BFKL equation is 
\eqn\bfkli{f(k^2, \bar\alpha_s/N)=f_I(k^2, Q_0^2)+{\bar
\alpha_s \over N}
\int_{0}^{\infty}{dq^2 \over q^2}K_{0}(q^2,k^2)f(q^2),}
where $f(k^2, \bar \alpha_s/N)$ is the unintegrated gluon four-point
function, $f_I(k^2,Q_0^2)$ is the zeroth order input, $\bar \alpha_s =(3/\pi)
\alpha_s$, and the LO kernel is defined by 
\eqn\kzero{K_0(q^2,k^2)f(q^2)= k^2
\biggl( {f(q^2)-f(k^2) \over \mid k^2-q^2\mid}
+{f(k^2) \over (4q^4+k^4)^{\half}}\biggr).}
It is convenient to define the input by $f_I(k^2,Q_0^2) = \delta(k^2-Q_0^2)$.
In fact in the leading twist factorization theorem this is the unique 
definition, and $Q_0^2$ is really just a regularization which we let $\to
0$ ultimately. Going beyond this approximation the dependence on $Q_0^2$ 
tells us about the higher twist due to the intrinsic transverse momentum 
of the gluon, and we will discuss this in section 3.
The ``gluon structure function'' is now given by          
\eqn\gluondef{{\cal G}(Q^2,N)=\int_{0}^{Q^2}{dk^2\over k^2} f(N,k^2,Q_0^2)
\times g_B(N,Q_0^2),}
where $g_B(N,Q_0^2)$ is the bare  gluon density in the
proton which implicitly absorbs the collinear divergences in $f(k^2)$. 
The BFKL equation is most easily solved by taking the Mellin
transformation to $\gamma$-space, i.e. 
\eqn\mellin{\tilde f(\gamma,N)=
\int_{0}^{\infty}d k^2 (k^2)^{-1-\gamma} f(k^2, N),}
where it reduces to 
\eqn\bfklii{\tilde f(\gamma,N)=\tilde f_I(\gamma, Q_0^2)+
(\bar \alpha_s/N)  \chi_0(\gamma)
\tilde f(\gamma, N), }
where $\tilde f(\gamma, Q_0^2)=\exp(-\gamma \ln (Q_0^2))$
and $\chi(\gamma)$ is the characteristic function
\eqn\kergam{\chi_0(\gamma)=2\psi(1)-\psi(\gamma)-\psi(1-\gamma).}
A little simple manipulation leads to the expression
\eqn\invmell{{\cal G}(Q^2,N)={1\over 2\pi i}
\int_{\half-i\infty}^{\half+i\infty}
d\gamma \exp(\gamma \ln(Q^2/Q_0^2)) {g_B(N,Q_0^2) \over \gamma
(1-(\bar\alpha_s/N) \chi_0(\gamma))}.}
This inverse transformation has a leading twist component given by 
the contribution of the leading pole at
$1-(\bar\alpha_s/N)
\chi_0(\gamma)=0$, and the solution is
\eqn\soliv{{\cal G}(Q^2,N)= {1 \over -(\bar\alpha_s/N)\gamma_0
\chi'_0(\gamma_0)}\biggl({Q^2\over
Q_0^2}\biggr)^{\gamma_0}g_B(N,Q_0^2).}
The anomalous dimension $\gamma_0(\bar\alpha_s/N)$ may be transformed
to $x$-space as a power series in $\bar \alpha_s\ln(1/x)$, and has a
branch point at $N=\lambda=4\ln 2 \bar\alpha_s$ (at which $\gamma\to \half$) 
leading to asymptotic small $x$ behaviour for the splitting function
\eqn\split{xP_{gg}^0(x,\bar\alpha_s) \to {0.07\bar\alpha_s x^{-\lambda.}
\over (\bar\alpha_s \xi)^{3/2}},}
In a similar fashion, assuming that the leading small
$x$ behaviour is dominated by the perturbative physics rather than by 
$g_B(Q_0^2, N)$, one can transform to $x$-space the normalization 
${1 \over -(\bar\alpha_s/N)\gamma_0\chi'_0(\gamma_0)}$ finding that this 
leads to a gluon normalization $xg(x)\propto {\bar\alpha_s x^{-\lambda.}
\over (\bar\alpha_s \xi)^{1/2}}$. 

\subsec{Running Coupling.}

Beyond strict leading order it is impossible to ignore the running of the
coupling. At NLO ultraviolet
regularization is required, resulting in a correction to the LO kernel of 
the form $-\beta_0\alpha_s(\mu_R^2)\ln(k^2/\mu_R^2)K_0(q^2,k^2)$, 
where $\mu_R$ is the renormalization scale which must now be introduced. 
Hence, it is unrealistic to simply use the LO kernel without considering the 
influence of such a correction. An obvious way in which to incorporate such
a term is to simply use the running coupling constant evaluated at the scale
$k^2$ in the previous LO BFKL equation. Since this, or something similar,  
is unavoidably forced upon us at NLO, it seems
sensible to consider the fixed coupling LO BFKL equation as just a
model which would apply in a conformally invariant world, and more
realistically to work with the BFKL equation with running coupling
\ref\glr{L.V. Gribov, E.M. Levin and 
M.G. Ryskin, \PRep \vyp{100}{1983}{1}.} 
\ref\liprun{L.N. Lipatov, {\it Sov. Phys. JETP} \vyp{63}{1986}{904}.}
\jan\janpcoll\ from the beginning. Doing this we obtain 
\eqn\bfklruni{f(k^2, Q_0^2, \bar\alpha_s(k^2)/N)=f_I(k^2, Q_0^2)+
{\bar \alpha_s(k^2) \over N}
\int_{0}^{\infty}{dq^2 \over q^2}K_{0}(q^2,k^2)f(q^2),}
where
\eqn\coupdef{\alpha_s=1/(\beta_0\ln(k^2/\Lambda^2)),}
$\beta_0 = (11-2N_f/3)/(4\pi)$, and $N_f$ is the number of active 
flavours. 

One can solve this equation in the 
same type of way as for the fixed coupling case, i.e. take the Mellin 
transformation, but now with respect to $(k^2/\Lambda^2)$. 
It is most convenient first to multiply through by 
$\ln(k^2/\Lambda^2)$, in which case one obtains 
\eqn\bfklrunii{{d\tilde f(\gamma,N)\over d \gamma}={d \tilde 
f_I(\gamma, Q_0^2) \over
d\gamma}-{1\over \bar\beta_0 N} \chi(\gamma)
\tilde f(\gamma, N),}
where $\bar \beta_0=(\pi\beta_0/3)$. 
Hence, the inclusion of the running coupling has completely changed
the form of 
our double Mellin space equation, turning it into a 
first order differential equation. This has a profound effect on the
form of the solutions. 
The equation may easily, if formally, be solved 
giving,
\eqn\solruni{\tilde f(\gamma,N)=\exp(-X_0(\gamma)/(\bar\beta_0 N))
\int_{\gamma}^{\infty}
{d \tilde f_I(\tilde \gamma,N,Q_0^2)
\over d\tilde \gamma}\exp(X_0(\tilde \gamma)/(\bar\beta_0 N))d\tilde\gamma,}
where
\eqn\solrunii{X_0(\gamma)=
\int_{\half}^{\gamma}\chi_0(\hat\gamma)d\hat\gamma
\equiv \biggl(2\psi(1)(\gamma-\half)-
\ln\biggl({\Gamma(\gamma) \over \Gamma(1-\gamma)}\biggr)\biggr).}
$X_0(\gamma) \to \ln(\gamma)$ at $\gamma=0$ and hence 
$\exp(-X_0(\gamma)/(\bar\beta_0 N))$ has a branch point at $\gamma=0$ 
($\exp(-X_0(\gamma)/(\bar\beta_0 N))\to \gamma^{-1/\bar\beta_0 N}$) 
with similar branch points at all negative integers. It is easiest 
to choose each of the cuts along the negative real axis.
$\exp(X_0(\gamma)/(\bar\beta_0 N))$ has similar branch points at every 
positive integer, and it is easiest to choose these cuts along the 
positive real axis. This means that the integral in \solruni\ is ambiguous
due to the available choice in avoiding the cuts. This ambiguity can only
really be removed by regulating the Landau pole in the definition of
the coupling. However, this introduces model dependence, and also makes 
analytic progress rather more difficult, so I simply accept this ambiguity
for this function.\foot{The problem due to the Landau pole is illustrated 
using an alternative method of solution in \armesto. In this paper the 
solution of the equation where the NLO coupling effect is left
simply as $-\beta_0\alpha_s(\mu_R^2)\ln(k^2/\mu_R^2)K_0(q^2,K^2)$ rather 
than resummed is also 
considered. This does not improve the situation, i.e. an ambiguity in 
the solution remains even in this case.}      

In order to simplify \solruni, and introduce factorization we trivially 
rewrite it as 
\eqn\solruninew{\tilde f(\gamma,N)=\exp(-X_0(\gamma)/(\bar\beta_0 N))
\biggl[\int_{0}^{\infty}-\int_{0}^{\gamma}\biggr]
{d \tilde f_I(\tilde \gamma,N,Q_0^2)
\over d\tilde \gamma}\exp(X_0(\tilde \gamma)/(\bar\beta_0 N))d\tilde\gamma,}
In the region of $\gamma=0$ the integrand in \solruninew\ 
is $\propto \gamma^{1/\bar\beta_0 N}$, so the integral of this from $0\to
\gamma$ is $\propto \gamma^{1+1/\bar\beta_0 N}$.      
Hence, the leading singularity in the $\gamma$ plane for
$\exp(-X_0(\gamma)/(\bar\beta_0 N))$, is 
cancelled by the integral from $0 \to \gamma$ of this integrand \janpcoll, 
and the new leading singularity is at $\gamma = -1$. Since ${\cal G}(Q^2,N)$
is obtained by an inverse Mellin transformation with respect 
to $Q^2/\Lambda^2$, the part of \solruninew\ coming from the integral from 
$0$ to $\gamma$ will behave like $\Lambda^2/Q^2$ (actually 
$Q_0^2/\lambda^2$ as we will see later). Hence, disguarding 
this power-suppressed correction, which will be considered in some 
detail in section 3, we keep only the first term in \solruninew,
obtaining for the gluon distribution 
\eqn\solruniv{\eqalign{{\cal G}(Q^2,N)&={1\over 2\pi i}
\int_{\half -i\infty}^{\half+i\infty}
{1\over \gamma}
\exp(\gamma\ln(Q^2/\Lambda^2)-X_0(\gamma)/(\bar\beta_0 N))
d\gamma
\cr
&\hskip 1.5in \times\int_{0}^{\infty}
\exp(-\tilde\gamma\ln(Q_0^2/\Lambda^2)+X_0(\tilde \gamma)/
(\bar\beta_0N))d\tilde\gamma
\,g_B(Q_0^2,N)\cr
&={\cal G}_E(Q^2,N){\cal G}_I(Q_0^2,N)g_B(Q_0^2,N).\cr}}

Therefore, we have factorization up to well-defined corrections of ${\cal O}
(Q_0^2/Q^2)$,  which genuinely do vanish as $Q_0^2 \to 0$ (see section 3).
As mentioned, 
$\exp(X_0(\gamma)/(\bar\beta_0N))$, contains singularities at all positive 
integers, and ${\cal G}_I(Q_0^2,N)$ is
not properly defined, since the integrand has singularities 
lying along the line of integration. However, since this factor is 
independent of $Q^2$, it does not contribute at all to the evolution of the
structure function. 
It is also divergent as $Q_0^2 \to 0$, and as usual in the factorization 
theorem these divergences are implicitly cancelled by $g_B(Q_0^2,N)$,
and we can imagine the ambiguity to be cancelled in the same manner. 
So the overall normalization is incalculable, but there is a  
calculable function ${\cal G}_E(Q^2,N)$ 
whose form is determined by the singularities of 
$\exp(-X_0(\gamma)/(\bar\beta_0 N))$ in the $\gamma$ plane.
This also leads to a fundamental difference between the cases of the fixed and 
running couplings. Whereas previously the leading singularity was a pole 
at $(\bar\alpha_s/N)\chi(\gamma)=1$, i.e. at $\gamma\to \half$ as $N \to 
4\ln 2\bar\alpha_s$, now the leading singularity is an cut at
$\gamma=0$: there is no power-like behaviour in $Q^2$. Similarly, the 
branch point in the $N$ plane at $4\ln 2\bar\alpha_s$ has become an essential
singularity at $N=0$: there is no power-like behaviour in $x$ in the 
evolution factor for the gluon. The introduction
of the running of the coupling has changed the character of the
solution completely. 

One can now proceed with the solution to the LO BFKL equation by 
acknowledging that the only real information contained in ${\cal G}_E(N,Q^2)$
is on the evolution of the structure function, i.e. defining 
\eqn\evol{{d \ln{\cal G}(N,Q^2) \over d\ln(Q^2)} =
{d\ln {\cal G}_E(N,Q^2) \over d \ln(Q^2)}\equiv \gamma_{gg}(N,Q^2).}
${\cal G}_E(N,Q^2)$ gives us an entirely perturbative effective
anomalous dimension governing the evolution of the gluon structure function. 
The usual technique for solving for ${\cal G}_E(N,Q^2)$ 
is to expand the
integrand in \solruniv, about the saddle-point. This results in a
contour of integration parallel to the imaginary axis, with real part
$\to \half$ for the small $x$ solutions, see \fig\contourfig{The branch points 
and cuts associated with $\exp(-X_0(\gamma)/(\bar\beta_0N))$ and the 
saddle-point contour, the Gamma-function contour and the numerical integration 
contour.}. Using this results in an anomalous dimension
\eqn\nloevol{\gamma_{gg}(N,Q^2)=\gamma_0(\bar\alpha_s(Q^2)/N)
+\sum_{n=1}^{\infty}(-\beta_0\alpha_s(Q^2))^n
\tilde\gamma_n(\bar\alpha_s(Q^2)/N),}
i.e., the effective anomalous dimension is the naive leading-order result
with coupling at scale $Q^2$ plus an infinite  series of corrections in 
increasing powers
of $-\beta_0\alpha_s(Q^2)$ \xscale. However, each of the $\tilde
\gamma(\bar\alpha_s(Q^2)/N)$ is singular at $N=\lambda(Q^2)$, and the
power of the singularity increases with increasing $n$. Hence,
although the series for the resulting splitting function is in the
small quantity $\alpha_s(Q^2)\beta_0$, the accompanying coefficients are
progressively more singular as $x\to 0$. The saddle-point approximation
is therefore not a reliable result as $x\to 0$ and explicit
investigation reveals that it is only really quantitatively useful when
$\bar\alpha_s(Q^2)\ln(1/x)$ is so small that the effective anomalous
dimension is effectively the LO in $\alpha_s$ part,
$xP_{gg}(x)=\bar\alpha_s(Q^2)$\xscale. 
This translates into $x \gsim 0.01$ in the HERA range. Therefore the 
calculations of
the anomalous dimension which rely on an expansion about the
saddle-point, i.e. the conventional expansion in decreasing powers of 
$\ln(1/x)$ at fixed power of $\alpha_s$, leads to very inaccurate and 
misleading results for small $x$. 
This instability is not surprising. If one examines the integrand along
the saddle-point contour of integration one finds that it is very
different from the Gaussian form the saddle-point method assumes \xscale.  
Also this is an expansion obtained from approaching $\gamma=\half$
and in terms of functions of $N$ which are singular at
$N=\lambda(Q^2)$, whereas we know that the full solution no longer
sees these points as anything special. In fact, the known
singularity structure of the integrand implies that $\gamma=0$ is
the point on which to concentrate. 

This suggests an alternative method 
of solution for the anomalous dimension. In order to concentrate on this 
leading singularity we may
move the contour of integration to the left and simultaneously use the
property that the integrand dies away very quickly at infinity (for $\Real 
\gamma \leq 1/2$) to
close the contour so that it simply encloses the real axis for
$\gamma<0$ \contourfig. It is then useful to express $\chi_0(\gamma)$ 
in the form  
\eqn\expan{\chi_0(\gamma)=1/\gamma+\sum_{n=1}^{\infty}2\zeta(2n+1)\gamma^{2n},}
which is, however, only strictly valid only for $|\gamma|<1$. 
Doing this we may write
\eqn\expX{X_0(\gamma)= \ln(\gamma)+\gamma_E+
\sum_{n=1}^{\infty}2{\zeta(2n+1)\over
2n+1}\gamma^{2n+1},}
and the integrand for ${\cal G}_E(N,Q^2)$ becomes
\eqn\expint{\gamma^{-1/(\bar\beta_0N)-1}\exp\biggl(\gamma t
-{1\over(\bar\beta_0N)}(\gamma_E+\sum_{n=1}^{\infty}a_n\gamma^{2n+1})\biggr),}
where $t=\ln(Q^2/\Lambda^2)$ and $a_n=2\zeta(2n+1)/(2n+1)$. The
contribution to the integral from $0 \to -\infty +i\epsilon$ is now
the same as
that from $-\infty-i\epsilon \to 0$ up to a phase factor, and we may write
\eqn\contour{{\cal G}_E(N,t)=-\sin\biggl({\pi\over(\bar\beta_0N)}\biggr)
\exp\biggl(-{\gamma_E\over(\bar\beta_0N)}\biggr)
\int_{-\infty}^0 
\gamma^{-1/(\bar\beta_0N)-1}\exp\biggl(\gamma t
-{1\over (\bar\beta_0N)}\sum_{n=1}^{\infty}a_n\gamma^{2n+1}\biggr)d \gamma,}
where the integral has to be understood as an analytic continuation,
since there are singularities along the real axis, and strictly
speaking the integrand is well defined only for $\gamma > -1$. Since
the factor of $\exp(\gamma t)$ is present this latter point leads,
in principle, to an error of order $\exp(-t)$, i.e. ${\cal O}(\Lambda^2/Q^2)$ 
into the value of ${\cal G}_E(N,t)$. This will be discussed in more detail 
below.

In order to evaluate the above integral it is convenient to let
$y=\gamma t$, resulting in 
\eqn\contouri{\eqalign{{\cal G}_E(N,t)=
-\sin\biggl({\pi\over(\bar\beta_0N)}\biggr)
&\exp\biggl(-{\gamma_E\over(\bar\beta_0N)}\biggr)\cr
&\hskip -0.8in t^{1/(\bar\beta_0N)}
\int_{-\infty}^0 
y^{-1/(\bar\beta_0N)-1}\exp(y)\exp\biggl(
-{1\over(\bar\beta_0N)}\sum_{n=1}^{\infty}a_n(y/t)^{2n+1}\biggr)d y.\cr}}
The latter exponential may be expanded as a power series in $y/t$ and
the each term in the integral then precisely evaluated using the standard 
result that  
\eqn\gamdef{(-1)^n\Gamma(-1/(\bar\beta_0N)+n)=\int_{-\infty}^0 
y^{-1/(\bar\beta_0N)-1}\exp(y)y^{n}d y.}
Hence, we may formally write  
\eqn\result{\eqalign{{\cal G}_E(N,t)&=-\sin\biggl({\pi\over(\bar\beta_0N)}
\biggr)\exp\biggl(-{\gamma_E\over(\bar\beta_0N)}\biggr)
\Gamma(-1/(\bar\beta_0N))\cr
&\hskip 1in t^{1/(\bar\beta_0N)}
\biggl(1+\sum_{n=3}^{\infty}A_n(1/(\bar\beta_0N))t^{-n}(-1)^n
{\Gamma(-1/(\bar\beta_0N)+n)\over \Gamma(-1/(\bar\beta_0N))}\biggr),\cr}}
plus an error of ${\cal O}(\Lambda^2/Q^2)$.
We note that we could have reached this final expression \result\ in a 
slightly more rigorous manner. After performing the expansion of 
$X_0(\gamma)$ in \expX\ we could have produced a well-defined integral in 
\contour by taking the lower limit of integration to be $-1+\epsilon$ so that
the expansion is valid over the region of integration. This would mean that 
there is region of integration $\gamma \leq -1$ absent, which due to the
factor of $\exp(\gamma t)$ would mean a missing contribution of 
${\cal O}(\Lambda^2/Q^2)$. This new limit of integration would result in the 
lower limit of $-t$ in \contouri\ and \gamdef\ and consequently we would 
obtain incomplete gamma functions $\gamma(-1/(\bar\beta_0N)+n,t)$ rather than 
$\Gamma(-1/(\bar\beta_0N)+n)$. However, $\gamma(-1/(\bar\beta_0N)+n,t)=
\Gamma(-1/(\bar\beta_0N)+n)+{\cal O}(\Lambda^2/Q^2)$, so disguarding the 
contributions of ${\cal O}(\Lambda^2/Q^2)$ we regain \result, which is 
formally equivalent to \contour, but we have seen explicitly the origin 
of the intuitively obvious ${\cal O}(\Lambda^2/Q^2)$ corrections to \result.
        
The result \result\ was first noted in \ref\alanjan{A.D. Martin and 
J. Kwiecinski, \PL \vyp{B353}{1995}{123}.}, and was simplified
by using the relationship that as $N\to 0$, 
$(\Gamma(-1/(\bar\beta_0N)+n)/ \Gamma(-1/(\bar\beta_0N))) \to
(-1/(\bar\beta_0N))^{n}$. However, it is important to notice the more
general result that for all $N$ 
\eqn\deltadef{(-1)^n{\Gamma(-1/(\bar\beta_0N)+n)\over 
\Gamma(-1/(\bar\beta_0N))}=\Delta_n(-1/(\bar\beta_0N)),}
where 
\eqn\deltadefi{\Delta_n(-1/(\bar\beta_0N))=\sum_{m=0}^{n-1} (-1)^{m}d_{mn} 
(\bar\beta_0N)^{-n+m},}
and $d_{mn}$ are positive coefficients and $d_{0n}=1$. 
Explicitly the first few $\Delta_n(-1/(\bar\beta_0 N))$ are
\eqn\explicit{\eqalign{\Delta_1(-1/(\bar\beta_0N)) &= \biggl({1 \over 
(\bar\beta_0N)}\biggr) \cr
\Delta_2(-1/(\bar\beta_0N)) &= \biggl({1 \over (\bar\beta_0N)}\biggr)^2
-\biggl({1 \over (\bar\beta_0N)}\biggr)\cr
\Delta_3(-1/(\bar\beta_0N)) &= \biggl({1 \over (\bar\beta_0N)}\biggr)^3
-3\biggl({1 \over (\bar\beta_0N)}\biggr)^2+2\biggl({1 \over (\bar\beta_0N)}
\biggr)\cr
\Delta_4(-1/(\bar\beta_0N)) &= \biggl({1 \over (\bar\beta_0N)}\biggr)^4
-6\biggl({1 \over (\bar\beta_0N)}\biggr)^3+11\biggl({1 \over (\bar\beta_0N)}
\biggr)^2 -6\biggl({1 \over (\bar\beta_0N)}
\biggr)\cr
\Delta_5(-1/(\bar\beta_0N)) &= \biggl({1 \over (\bar\beta_0N)}\biggr)^5
-10\biggl({1 \over (\bar\beta_0N)}\biggr)^4+35\biggl({1 \over (\bar\beta_0N)}
\biggr)^3 -50\biggl({1 \over (\bar\beta_0N)}
\biggr)^2+24\biggl({1 \over (\bar\beta_0N)}
\biggr).\cr}}
These functions oscillate a great deal and only approach the 
asymptotic values of $1/(\bar\beta_0N)^n$ at low values of $N$ which 
decrease with increasing $n$. The comparison of  $\Delta_4(-1/(\bar\beta_0N))$
with $1/(\bar\beta_0N)^4$ is shown in \fig\gammas{The  
expression $\Delta_4(-1/(\bar\beta_0N))$ as a function of $N$ compared to
$1/(\bar\beta_0N)^4$.}, and illustrates this feature clearly.

Ignoring the common factor of
$-\sin(\pi/(\bar\beta_0N))\Gamma(-1/(\bar\beta_0N))
\exp(-\gamma_E/(\bar\beta_0N))$, which has no $t$
dependence, and is irrelevant for the calculation of the anomalous dimension, 
\eqn\resulti{{\cal G}_E(N,t)=t^{1/(\bar\beta_0N)}\biggl(1+
\sum_{n=3}^{\infty}A_n(1/(\bar\beta_0N))t^{-n}\Delta_n(-1/(\bar\beta_0N))
\biggr)}
where the $A_n$ are simply calculable from the expansion of $\exp\biggl(
-1/(\bar\beta_0N)\sum_{n=1}^{\infty}a_n(y/t)^{2n+1}\biggr)$. 
The common factor of
$t^{1/(\bar\beta_0N)}$ is the well-known double-leading-log result
coming from just the LO $\alpha_s(Q^2)/N$ part of the anomalous
dimension. Multiplying this we have an expansion as a power series in $1/t$ or
equivalently in $\alpha_s(Q^2)$. In fact 
\eqn\expanbit{t^{-n}\Delta_n(-1/(\bar\beta_0N))=(\bar\alpha_s(Q^2)/N)^n
\sum_{m=0}^{n-1}d_{mn}(-\beta_0\alpha_s(Q^2))^{m}(\bar\alpha_s(Q^2)/N)^{-m}.}
This explicitly
demonstrates that we obtain a set of running coupling corrections to a
LO result, i.e. in solving the BFKL equation we are now obtaining not 
only the leading power in $1/N$ (corresponding to the leading power 
of $\ln(1/x)$) at each order in $\alpha_s(Q^2)$, but we also obtain the 
leading power in $\beta_0$ at each power of $\alpha_s(Q^2)$ and $1/N$. 
Substituting this type of expansion into \resulti, putting 
the resulting expression for
${\cal G}_E(N,t)$ in \evol\ and expanding in inverse powers of $t$, 
one obtains an expression for the anomalous
dimension as a power series in $\alpha_s(Q^2)$, where at each order we
have the leading divergence in $1/N$ plus a sum of running
coupling correction type terms.
With a little work one may regain the whole leading
$\gamma_0(\alpha_s(Q^2)/N)$ (though it is necessary to keep some
subleading terms in the $\Delta_n$ to do this), along with a tower of
terms which are subleading in powers of $\beta_0\alpha_s(Q^2)$ to this
leading anomalous dimension -- one obtains all the corrections to
this naive LO anomalous dimension due to the running of the coupling,
i.e. the whole of \nloevol\ is regained, but ordered in powers of 
$\alpha_s(Q^2)$ rather than in $\beta_0\alpha_s(Q^2)$.

The general features of this full, running coupling BFKL gluon Green's 
function and consequent anomalous dimension may be appreciated quite easily.
The important fact to note is that although the
$\Delta_n(-1/(\bar\beta_0N)))
\to (1/(\bar\beta_0N))^n$ as $N\to 0$, the function oscillates a great deal 
with $1/(\bar\beta_0N)$, and remains much smaller in magnitude than this
asymptotic form until very small $N$, roughly until $1/N>n$. This
coupled with the accompanying factor of $t^{-n}$ means that for
reasonable $t$, i.e $t\gsim 4-5$ ($Q^2\gsim 1\Gev^2$), only the first $5$ or
so terms in \resulti\ make a significant contribution for $N>0.25$. 
Hence, to a very good approximation    
\eqn\resultii{{\cal G}_E(N,t)=t^{1/(\bar\beta_0N)}\biggl(1-
{2\zeta(3)\over 3(\bar\beta_0N)t^3}\Delta_3(-1/(\bar\beta_0N))
-{2\zeta(5)\over 5(\bar\beta_0N)t^5}\Delta_5(-1/(\bar\beta_0N))\biggr),}
and in fact the smallness of the coefficient makes even the $t^{-5}$
term almost negligible in this case. 
${\cal G}_E(N,t)$ initially grows as $N$ falls due to
the $t^{1/(\bar\beta_0N)}$ term. However, for $N\sim 0.6$ the negative
contribution from the $t^{-3}$ term starts to become significant
and ultimately drives the gluon structure function 
to negative values. The result is shown in
\fig\gluon{The $Q^2$-dependent part of the gluon structure function,
${\cal G}_E(N,t)$, and of $d{\cal G}_E(N,t)/dt$ as a function of $N$ for
$t=6$ ($Q^2 \sim 6\Gev^2$). The $Q^2$-independent factor of
$-\sin(\pi/(\bar\beta_0N))\Gamma(-1/(\bar\beta_0N))
\exp(-\gamma_E/(\bar\beta_0N))$ is included in both in order to 
produce a
smoother $N$-dependent normalization of the functions.}. 
$d{\cal G}_E(N,t)/dt$ may simply be
evaluated also using \resulti, and shows the same general shape, but
does not become negative until a slightly lower value of $N$ as also
seen in \gluon.  
Hence the anomalous dimension develops a leading pole at a finite value of
$N$, given by 
\eqn\pole{t^3=
{2\zeta(3)\over
3(\bar\beta_0N)}\biggl({1\over(\bar\beta_0N)^3}-{3\over(\bar\beta_0N)^2} 
+{2\over(\bar\beta_0N)}\biggr).}
This result is accurate to better than $10\%$ even at $Q^2 \sim 1 \Gev^2$, 
and is much better at higher $Q^2$, the right-hand-side 
receiving corrections formally
of ${\cal O}(1/(t^2 \bar \beta_0^5 N^5))$, but which are numerically small. 
The value of $N$ for this leading pole
is shown as a function of $t$ in \fig\intercepts{The positions of the
leading poles in the anomalous dimensions for the gluon structure
function at LO and NLO, and for $F_L$ at LO and NLO.},
and for the sort of values of $t$ relevant at HERA is $\sim 0.25$. 
Going to $N<0.25$ higher order terms in \resulti\ become
important, and the positive $1/((\bar\beta_0N)^2t^6)
\Delta_6(-1/(\bar\beta_0N))$ term absent in \resulti\ 
pulls ${\cal G}_E(N,t)$ back to positive
values, and another pole, with opposite sign residue, appears in
$\gamma_{gg}(N,t)$. At even lower $N$ the analytic expression eventually
breaks down, as discussed below, but numerical results show a series of 
poles becoming closer together. Nevertheless, the position of the leading 
pole is essentially determined by the first handful of terms in the
power series in $\alpha_s(Q^2)$ for ${\cal G}_E(N,t)$, and hence so is the
asymptotic behaviour of the small $x$ splitting function, i.e. 
$P_{gg}(x,t)\sim x^{-0.25}$. So we see that the
introduction of the running coupling has a dramatic effect on the
singularity structure of the LO BFKL anomalous dimension, turning the
cut into a series of poles, and changing the
position of the rightmost singularity by a factor of $\sim 0.4$.  
This result of the pole in the anomalous dimension was previously 
proved in detail in \ciafsalam\ using numerical techniques and in the 
context of the collinearly resummed NLO kernel, and also 
indicated here using an approximate analytical solution first suggested 
in \liprun. 
However, in this paper I particularly stress the quantitative result of the 
huge modification of the naive LO BFKL anomalous dimension due to the running 
coupling contributions alone. This is apparent over a wide range of $N$, and 
in \fig\anom{a. The anomalous dimension for the gluon structure
function at LO plotted as a function of $N$ for
$t=6$ ($Q^2 \sim 6\Gev^2$). Also shown is the ${\cal
O}(\alpha_s(Q^2))$ contribution $\bar\alpha_s(Q^2)/N$, and the full naive 
LO BFKL anomalous dimension.
b. The anomalous dimensions for the gluon
at LO and at NLO plotted as functions of $N$ for
$t=6$.}.a I show the 
anomalous dimension
as a function of $N$ for all values right of the leading
singularity. As one sees, it is much closer to the simple
$\alpha_s(Q^2)/N$ expression than to the naive BFKL result. 
 
Before going into more precise detail and more general situations there 
are two important points I should address. These are the choice of the scale 
of the running coupling in \bfklruni\ as $k^2$ and the fact that the 
expansion of $\chi_0(\gamma)$ in powers of $\gamma$ is not convergent 
over the whole range of the contour of integration. The former of these 
is the simpler, so firstly I shall address the choice of scale. 
It was known in \ref\ciafrun{G. Camici and M. Ciafaloni, 
\PL\vyp{B386}{1996}{341}.} that the correct scale seemed as if it were 
really the symmetric choice $(k-q)^2$, but that $k^2$ could
be used instead, leading to contribution to the NLO kernel which is 
proportional to
$\beta_0$. In practice it is much easier to obtain analytic results using 
$k^2$, and this $\beta_0$-dependent NLO term leads to a contribution to 
the Mellin transformation of the NLO kernel, $\chi_1(\gamma)$, of the
form $\half\bar\beta_0(\chi_0^2(\gamma) +\chi'_0(\gamma))$. Including this
in the integrand for the expression for ${\cal G}_E(N,t)$ at NLO
(to be discussed in detail in section 4) leads to a multiplicative 
contribution of the form $\exp(\half(\ln(\chi_0(\gamma))+X_0(\gamma)))
\equiv f^{\beta_0}(\gamma)$. 
This can be expanded as a power series which at low orders is 
\eqn\serfbet{f^{\beta_0}(\gamma) =1+1.60\gamma^3+1.24\gamma^5 -0.163\gamma^6
+1.15\gamma^7 +\cdots.}
Including this additional factor in \contour\ modifies \resultii\ to  
\eqn\resultiialt{\eqalign{{\cal G}_E(N,t)=t^{1/(\bar\beta_0N)}\biggl(1-
{(2/3\zeta(3)-1.60(\bar\beta_0N))\over (\bar\beta_0N)t^3}
&\Delta_3(-1/(\bar\beta_0N))\cr
&-{(2/5\zeta(5)-1.24(\bar\beta_0N))\over (\bar\beta_0N)t^5}
\Delta_5(-1/(\bar\beta_0N))\biggr).\cr}}
For a given power of $\alpha_s(Q^2)$ these new contributions 
produce terms a power of 
$\bar\beta_0 N$ up on the other terms and hence, not
surprisingly, result in additional running coupling corrections to the
gluon and anomalous dimension. However, the new terms in the series 
in powers of $\gamma$ do not start
until third order and have rather small coefficients. The resulting change 
in the anomalous dimensions, both for general values of $N$ and for the 
position of the leading pole is very minor.   
Therefore, the correction for my original
``incorrect'' choice of scale is very small. However, in principle it seems as
though the factor just considered should really be taken as part of the LO
result since it just gives running coupling corrections.  
I will adopt this convention and the LO anomalous dimensions
and splitting functions presented in this paper will explicitly 
contain the corrections from this factor, and in fact the results already 
presented in \gluon, \intercepts\ and \anom\ include these (very small) 
effects. In principle one could sum the corrections needed due to the simple
choice of $k^2$ in the coupling, rather than $(k-q)^2$, by including 
contributions induced in the kernel at NNLO and beyond. In practice, beyond
NLO the change seems too tiny for one to be concerned.    
    
I should also comment on the limit of applicability of
the analytic expression \resulti. 
As noted, it is obtained via a series expansion which is 
not valid over the whole 
contour of integration. This is reflected in the 
error of ${\cal O}(\Lambda^2/Q^2)$ we discovered for this expression but also 
in the fact that the overall
magnitude of the $\Delta_n(-1/(\bar\beta_0N))$ actually increases like 
$n!$ in general. This latter point means that the series in 
\resulti\ is actually
asymptotic. It turns out that it contains both infrared and ultraviolet 
renormalon contributions, and hence it must be truncated to obtain sensible 
results. The greatest accuracy may be obtained from \resulti\ by
truncating the series at order $n_0\sim t$, the precise value 
depending on the size of the coefficients in the series expansion.
For the LO gluon these are small and one could use $n_0 \sim 10$,
but from experience with other variables (see later)
and the desire to go down to $Q^2 \sim 1\Gev^2$, i.e. $t \sim 4-5$, 
in practice I always 
use $n_0=5$. (For the LO gluon the contribution from $n=6\to 10$ is practically
negligible.) Using the truncated expression for ${\cal G}_E(N,t)$ in the manner
already discussed, then
results in an infinite series in $\alpha_s(Q^2)$ for $\gamma_{gg}(N,t)$
which is convergent for any $N$ right of the leading pole, but different to
the real, divergent series beyond $6$th order in $\alpha_S(Q^2)$. 

It is vital to note that although the formal expression for the gluon, and 
hence anomalous dimension, as a power series in $\alpha_s(Q^2)$ \resulti\
contains infrared renormalons\foot{In unphysical regularization schemes,
such as ${\overline{\rm MS}}$, the anomalous dimensions are not expected to 
contain renormalons (see section 3.4 of \ref\beneke{M. Beneke, \PRep 
\vyp{317}{1999}{1}.} for a discussion), 
these being confined to the coefficient functions 
relating the parton distributions to physical quantities. However, by 
regularizing via a finite $Q_0$, and defining the gluon density as the bare
density convoluted with the gluon Green's function we have implicitly chosen 
a more physically motivated factorization scheme which allows the presence
of renormalons.}, and hence has an ambiguity of ${\cal O}
(\Lambda^2/Q^2)$, the integral in \solruniv, which properly defines
the leading twist gluon and anomalous dimension, does exist and produces 
well-defined results. The ambiguity of ${\cal O}
(\Lambda^2/Q^2)$ in \resulti\ cancels with an ambiguity in the ${\cal O}
(\Lambda^2/Q^2)$ correction to this power-series expansion which we 
discovered in the derivation of \resulti.
The accuracy of the (truncated) analytic
expression can be found by comparing with results obtained from
evaluating \solruniv\ using numerical integration along the contour shown in 
\contourfig. For the gluon
structure function for $N$ to the right of the leading pole the analytic
approximation to the anomalous dimension is found to
be a fraction of a percent for $t=6$, and falls like $\exp(-t)$.
Strictly speaking there is an $\exp(-t)$ contribution from the 
correction to \resulti\ (with the renormalon ambiguity removed) plus a 
$1/t^7$ correction due to the truncation. However, $1/t^7$ is similar 
to $\exp(-t)$ in the range considered. 
Hence, we have a power-like correction to the power series in $\alpha_s(Q^2)$
obtained from the truncated expression which is completely well-defined. This 
illustrates that the presence of infrared renormalons in a physical quantity 
is not necessarily due to an inherent ambiguity in the quantity itself 
(due, for example,
to the Landau pole in the coupling) as is commonly thought, but rather due 
to the impossibility of completely expressing the physical quantity as a 
power series in $\alpha_s(Q^2)$ \ref\renormexp{G. Grunberg, \PL 
\vyp{B372}{1996}{121} \semi Yu.L. Dokshitzer and N.G. Ultrasev, \PL 
\vyp{B380}{1996}{141}.}. In truncating the power-series expansion in 
\resulti\ I simply choose to split the expression for the gluon 
as some general function of $N$ and $Q^2$ into a perturbatively calculable 
part as a power-series in $\alpha_s(Q^2)$ and a remainder which is 
approximately of order ${\cal O}(\Lambda^2/Q^2)$. The point of truncation is
then chosen empirically so as to make this remainder term as small as 
possible. This seems to be the way to obtain the most accurate analytic
results. It is important to note that the remainder term, although 
power-suppressed, is not in any way higher twist, since it is obtained 
from the leading twist part of the solution to the BFKL equation.        

Having got these two points out of the way we can now begin to discuss 
the quantitative results of the running coupling BFKL equation. 
In order to investigate the real effect of the BFKL anomalous
dimension on structure function evolution it is necessary to calculate
the BFKL splitting
function as a function of $x$. This is where an analytic expression for
the anomalous dimension is particularly useful. A series of
numerically obtained values of $\gamma_{gg}(N,t)$ allows an approximate
determination of $P(x,t)$, but it is extremely difficult to be accurate,
especially for the wildly oscillating functions of $1/N$ which do in
fact make up ${\cal G}_E(N,t)$. However, I now
have an explicit series for $\gamma_{gg}(N,t)$ in powers of
$\alpha_s(Q^2)$, obtained from the truncated expression for 
${\cal G}_E(N,t)$. 
The $N$-dependent functions at each power of 
$\alpha_s(Q^2)$
become larger at small $N$ as the series progresses, of course, 
and to reach small
enough $x$ more and more terms are needed. However, at a fixed value of
$N$ there is no such growth, and the same is therefore true for fixed
$x$. Hence,
one only needs to work to a finite order. Limiting oneself to
$x>10^{-5}$ and $t>4.5$, i.e. $Q^2 \gsim 1 \Gev^2$, the suppression of the
$\Delta_n(-1/(\bar\beta_0N))$
is quite significant and seventh order in $\alpha_s(Q^2)$ is easily
sufficient. This results in a power-series contribution to the splitting 
function 
\eqn\pgglopow{\eqalign{ xP^{LO}_{gg}(\xi,\alpha_s(Q^2)) &= \bar\alpha_s(Q^2)
+\bar\alpha^4_s(Q^2)\biggl(2.4 {\xi^3 \over 3!}-12.01\bar\beta_0{\xi^2 \over 2}
+9.206\bar\beta_0^2\xi-9.60 \bar\beta_0^3\biggr)\cr
&\hskip -1.1in +\bar\alpha^6_s(Q^2)\biggl(2.08{\xi^5 \over 5!}
-26.95\bar\beta_0{\xi^4 \over 4!}+ 134.6\bar\beta_0^2{\xi^3 \over 3!}
-320.7\bar\beta_0^3{\xi^2 \over 2}
+359.8\bar\beta_0^4\xi-148.8 \bar\beta_0^5\biggr)\cr
&\hskip -1.1in +\bar\alpha^7_s(Q^2)\biggl({1.92\over \bar\beta_0}
{\xi^7\over 7!} -19.23
{\xi^6 \over 6!}+78.94\bar\beta_0{\xi^5 \over 5!}
-169.2\bar\beta_0^2{\xi^4 \over 4!}+ 199.8\bar\beta_0^3{\xi^3 \over 3!}
-122.9\bar\beta_0^4{\xi^2 \over 2}
+30.72\bar\beta_0^5\xi\biggr).\cr}}
This contribution to the splitting function for $t=6$ and is shown 
in \fig\splitf{a. The splitting function $xP^{LO}_{gg}(x)$ and its 
power series and power-suppressed contributions
plotted as functions of $x$ for
$t=6$. b. The splitting function $xP^{LO}_{gg}(x)$ 
plotted as a function of $x$ for
$t=6$ ($Q^2 \sim 6\Gev^2$). Also shown is the ${\cal O}(\alpha_s(Q^2))$ 
contribution $\bar\alpha_s(Q^2)$, and the naive LO BFKL splitting
function with coupling $\alpha_s(Q^2)$.}.a. Note that because of the 
truncation of ${\cal G}_E(N,t)$, beyond 6th order the expression for 
$P^{LO}_{gg}(\xi,\alpha_s(Q^2))$ is not what one would really get from 
the true power-series. In particular there are higher powers of $\xi$ than
strictly allowed. Nevertheless, it represents a very accurate approximation 
to the full result whereas the correct series would simply diverge.      

We also have to consider the power-suppressed contribution. 
Although this is only calculated numerically in $N$-space it is only 
a small correction of order $0.05\%$ for $\gamma^{LO}_{gg}(N,t)$ a t $t=6$, 
and can also be calculated for a wide variety of values of 
$N$ and $t$ without too much work. It can then be modelled by an analytic
function which may easily be converted to $x$-space.
Hence, I choose to calculate it for
$t=4.5$ ($Q^2\sim 1\Gev^2$) and $t=6$ ($Q^2\sim 6\Gev^2$) and $N$-values
$0.4, 0.5, 0.6, 0.7, 0.8, 0.9, 1.0, 1.5, 2, 3, 5, \infty$. The lower $t$ 
value is the lower limit at which we will trust this LO perturbative result,
and for $t$ above $6$ the power-suppressed effect is very small. The $N$ 
values go low enough to correspond safely to $x>0.00001$ and are sufficient
that very accurate modeling can be done. The values are fit to a function 
of the form 
\eqn\model{a_0\exp(-b_0t)+\exp(-t)\biggl(\sum_{n=1}^7 a_n 
\biggl({\alpha_s(t)\over \alpha_s(t=4.5)}\biggr)^{b_n}{1\over N^n}
\biggr).}
Introducing further degrees of freedom beyond this does not seem to change 
the results. This expression can then be trivially converted to $x$-space.
Performing this procedure
in the case of the power-suppressed contributions to the 
LO gluon anomalous dimension I obtain the explicit result
\eqn\modellogg{\eqalign{4.92\exp(-1.62t)&\delta(1-x)+
\exp(-t)\biggl( 1.068 
\biggl({\alpha_s(t)\over \alpha_s(4.5)}\biggr)^{1.98}
+5.257 \biggl({\alpha_s(t)\over \alpha_s(4.5)}\biggr)^{3.06}\xi\cr
&-18.73 \biggl({\alpha_s(t)\over \alpha_s(4.5)}\biggr)^{2.90}{\xi^2 \over 2!}
+21.56 \biggl({\alpha_s(t)\over \alpha_s(4.5)}\biggr)^{2.90}{\xi^3 \over 3!}
-11.60 \biggl({\alpha_s(t)\over \alpha_s(4.5)}\biggr)^{2.79}{\xi^4 \over 4!}\cr
&+3.00 \biggl({\alpha_s(t)\over \alpha_s(4.5)}\biggr)^{2.55}{\xi^5 \over 5!}
-0.301 \biggl({\alpha_s(t)\over \alpha_s(4.5)}\biggr)^{2.17}{\xi^6 \over 6}
\biggr).\cr}}
This power-suppressed correction is shown along with the power-series 
part and the full LO splitting function in \splitf.a.  
Although the power-suppressed contribution in $x$ space
turns out to be a larger fraction of the total than in $N$-space,
it still only makes a very small correction to the evolution. 
However, one notices than the logarithmic terms in \modellogg\ are such that
it falls more quickly than $(\Lambda^2/Q^2)$, or alternatively, grows more
quickly than this as $Q^2$ falls. This may be due to the presence of a 
significant $(\Lambda^4/Q^4)$ term in practice. 

The full LO splitting function is shown in \splitf.b
along with the purely order $\alpha_s(Q^2)$ contribution and the naive BFKL 
splitting function. One
sees that it is hugely suppressed compared with the naive LO BFKL
splitting function, and is even lower than the ${\cal O}(\alpha_s(Q^2)$)
contribution for $x$ between about $0.1$ and $0.001$. Finally I note that 
the LO running coupling BFKL equation has also been calculated in 
\ref\salamlo{M. Ciafaloni, D. Colferai and G.P. Salam, JHEP \vyp{0007}{2000}
{054}.}, but numerically, with coupling scale equal to $(k-q)^2$, and with 
the coupling frozen below a particular scale and $Q_0$ taken to be a finite 
value. The results are displayed for high $t$ (where my power-series 
is essentially exact) and despite the above differences seem to be in very 
good agreement with the results in \letter\ and this paper. The freezing of 
the coupling and the finite $Q_0$
introduce choice-dependent non-perturbative effects which become important
at extremely low values of $x$, which in general become lower as $Q_0$ and 
the scale of
freezing decrease. This seems to support the results obtained by 
my method of formally
factorizing the non-perturbative effects into ${\cal G}_I(Q_0^2,N)$ and
extracting as much information as possible in an analytic model-independent
manner.        

\newsec{Higher Twist at Small $x$.}

In this section I will show that as far as the information from the BFKL 
equation is concerned calculable higher twist contributions are small.
I will also suggest that some other powerlike corrections at small $x$ 
may perhaps be less significant than often claimed. 
As a first point I note that it has been claimed that there are likely to
be large infrared renormalon contributions to structure functions at small 
$x$ \ref\renorm{E. Stein, M. Maul, L. Mankiewicz and A. Sch\"afer, 
\NP \vyp{B536}{1998}{318}\semi G.E. Smye, \NP \vyp{B546}{1999}{315}\semi
G.E. Smye, {\tt hep-ph/0105015}.}. 
As shown in the previous section for the case of the gluon both infrared
and ultraviolet renormalons do show up in the solution to the BFKL equation 
if one insists upon trying to express results entirely in terms as a 
power-series in $\alpha_s(Q^2)$ and uses the whole of \resulti\ rather than 
truncating. Presumably these are an extension of the small $x$ divergent
contribution to the renormalons in \renorm. 
However, these renormalons are circumvented if one considers the full solution 
to the $Q^2$-dependent part of the BFKL equation. Precisely the same argument
works for the case of real structure functions, as will be shown explicitly 
in section 5.  
This is not to say that there are not relatively large power-suppressed
corrections to the (truncated) perturbative-series. We have already seen a 
non-negligible contribution to $P^{LO}_{gg}(x,Q^2)$, and the power-suppressed
contributions turn out to be larger for physical quantities. However, these
contributions are calculable and unambiguous. Hence, solution of the BFKL 
equation, which provides results more general than a power series in 
$\alpha_s(Q^2)$ avoids the renormalon ambiguity. This means that  
renormalons obtained from unresummed (in $\ln(1/x)$) calculations require 
not only a $\ln(1/x)$ resummation but also the consideration of results beyond 
the power-series expansion. This implies they do not really tell us anything 
truely quantitative about power corrections in practice.        

Now let us consider genuine higher twist effects. Some of these are contained
within the BFKL equation, since if $Q_0^2$ is allowed to be non-zero a series
in powers of $(Q_0^2/Q^2)$ is obtained which tells us about the higher twist
contributions due to the intrinsic transverse momentum in the two-gluon
operator. This is the only information, however, and we learn nothing 
about the other three contributions to next-to-leading twist (discussed,
for example in \ref\htwist{J. Bartels and C. Bontus, \PR 
\vyp{D61}{2000}{034009}.}), in particular those due to the four-gluon operator
and hence possible saturation effects. However, it is possible to obtain some
useful and interesting results. 

Let us first consider the fixed coupling BFKL equation. When solving \invmell\
it is straightforward to also calculate the higher twist contributions by 
picking up the non-leading poles in $\gamma$. The easiest way to proceed is
to obtain $G(Q^2,x)$ by first taking the exact inverse Mellin 
transformation back to $x$-space by picking up the simple pole at 
$N=\bar\alpha_s\chi_0(\gamma)$ resulting in 
\eqn\invmellht{ xG(Q^2,x)\propto {1\over 2\pi i}
\int_{\half-i\infty}^{\half+i\infty}
d\gamma \exp(\gamma \ln(Q^2/Q_0^2)) \exp(\xi\bar\alpha_s\chi_0(\gamma)).}
This can now accurately be evaluated in the asymptotic small $x$ limit 
using the saddle-point technique, i.e. 
integrating along the contour determined by the condition 
$(d\chi_0(\gamma)/d\gamma)=0$ which defines $\gamma_0$. At leading twist,
$0\geq\Real \gamma \geq 1$, 
$\gamma_0=1/2$ and $\chi_0(\gamma_0)=4\ln(2)$, leading to the usual 
power-like growth at small $x$. However, looking for the solutions to  
$(d\chi_0(\gamma)/d\gamma)=0$ for $-1\geq\Real \gamma\geq 0$, i.e. examining 
the higher twist operator and its anomalous dimension, one finds
\eqn\nlotwistsp{\gamma^{HT}_0=-0.425 \pm 0.474i \qquad\qquad 
\chi_0(\gamma^{HT}_0) = -2.64 \pm 2.393i.}
Hence, the features of the saddle-point are completely different at 
next-to-leading twist. Not only are there complex conjugate saddle-points 
leading to an oscillatory behaviour, but the real part of 
$\chi_0(\gamma^{HT}_0)$ is negative rather than positive. Inserting 
\nlotwistsp\ into \invmellht\ one obtains 
\eqn\htsolnorm{xG^{HT}(Q^2,x) \propto x^{2.64\bar\alpha_s}
\cos(2.393\bar\alpha_s\xi),}
i.e. a valence-like gluon rather than one growing at small $x$. The 
corresponding higher twist splitting function has the same general behaviour 
as the gluon as $x \to 0$. 

One can also find the splitting function by solving $1-(\bar\alpha_s/N)
\chi_0(\gamma)$ as a power-series in $(\bar\alpha_s/N)$ for the 
next-to-leading twist solution. This results in the explicit series
\eqn\nlotwistanom{\eqalign{\gamma^{HT}_{0}(\bar\alpha_s/N)+1&=
\biggl({\bar\alpha_s \over N}\biggr)-2\biggl({\bar\alpha_s \over N}\biggr)^2
+2\biggl({\bar\alpha_s \over N}\biggr)^3
+4.4\biggl({\bar\alpha_s \over N}\biggr)^4
-29.2\biggl({\bar\alpha_s \over N}\biggr)^5
+80.2\biggl({\bar\alpha_s \over N}\biggr)^6\cr
&-90.6\biggl({\bar\alpha_s \over N}\biggr)^7
-298\biggl({\bar\alpha_s \over N}\biggr)^8
+2084\biggl({\bar\alpha_s \over N}\biggr)^9
-6446\biggl({\bar\alpha_s \over N}\biggr)^{10}\cr
&+9157\biggl({\bar\alpha_s \over N}\biggr)^{11}
+20919\biggl({\bar\alpha_s \over N}\biggr)^{12}
-187924\biggl({\bar\alpha_s \over N}\biggr)^{13}
+666008\biggl({\bar\alpha_s \over N}\biggr)^{14}\cr
&-1.2\times10^6\biggl({\bar\alpha_s \over N}\biggr)^{15}
+1.3\times10^6\biggl({\bar\alpha_s \over N}\biggr)^{16}
+1.9\times10^7\biggl({\bar\alpha_s \over N}\biggr)^{17}
-7.7\times10^7\biggl({\bar\alpha_s \over N}\biggr)^{18}\cr  
&-1.7\times10^8\biggl({\bar\alpha_s \over N}\biggr)^{19} 
-2.1\times10^7\biggl({\bar\alpha_s \over N}\biggr)^{20}
-2.0\times10^9\biggl({\bar\alpha_s \over N}\biggr)^{21}
+\cdots,\cr}}
which can be easily converted to $x$-space. The corresponding splitting 
function is plotted for $\bar\alpha_s=0.2$ in \fig\htwistsp{The 
next-to-leading twist splitting function for $\bar\alpha_s=0.2$.}, and
it clearly fits the expectation that $xP^{HT}_{gg}(x,\bar\alpha_s) \sim
x^{0.5}\cos(0.5\xi)$ as $x \to 0$.\foot{Unfortunately, because 
of large cancellations, the first 21 terms in the series for 
$xP^{HT}_{gg}(x,\bar\alpha_s)$ are needed for $x\geq 0.00001$.}
Hence, although the first term in the series is the same as at leading twist,
and implies a growth at small x, the summation of the series is extremely
different, and the next-to-leading twist contributions from the BFKL
equation are not only suppressed by $(Q_0^2/Q^2)$, but also become negligible
at small $x$. This can also be shown to be true for the even higher twist 
contributions using the same techniques.
This highlights the danger of using low order terms in 
the series for the splitting functions to estimate higher twist corrections,
as in \htwist\ - the summation of leading $\ln(1/x)$ terms may be very
important, in this case of the two-gluon operator leads to a complete
change of conclusion on the import of higher twist. Unfortunately, there is
no knowledge at all of the corresponding series for the four-gluon operators. 

Given that the results from the fixed coupling BFKL equation were altered so
dramatically at leading twist by the inclusion of the running coupling, 
we should see what happens at higher twist. As already mentioned, the higher
twist contribution to the running coupling BFKL equation is given by
\eqn\solrunht{\eqalign{{\cal G}^{HT}(Q^2,N)&={1\over 2\pi i}
\int_{-\epsilon -i\infty}^{-\epsilon+i\infty}
{1\over \gamma}
\exp(\gamma\ln(Q^2/\Lambda^2)-X_0(\gamma)/(\bar\beta_0 N))
d\gamma
\cr
&\hskip 1in \times\int_{\gamma}^{0}
\exp(-\tilde\gamma\ln(Q_0^2/\Lambda^2)+X_0(\tilde \gamma)/
(\bar\beta_0N))d\tilde\gamma
\,g_B(Q_0^2,N),\cr}}
where the contour in the first integral has been moved to the left since the 
leading singularity at $\gamma=0$ is eliminated by the second integral. 

Let us consider first the case where $t =\ln(Q^2/\Lambda^2) \gg t_0 =
\ln(Q_0^2/\Lambda^2)$, which would be the case for deep inelastic scattering.
Let us also, without justification for the moment, let the lower limit on 
the second integral be a constant, $k \sim -1$, so that we have factorization
imposed. In this case we can evaluate the two integrals separately.
Both the integrals can be calculated accurately using the saddle-point 
method. Thus, using the type of steps outlined in (4.1)--(4.5) of \xscale\
one obtains
\eqn\htsolt{{\exp\biggl(\int^{Q^2} \gamma^{HT}_0(\bar\alpha_s(q^2)/N)d 
\,\ln q^2\biggr) \over \gamma^{HT}_0(\bar\alpha_s(Q^2)/N)[-\chi_0'(
\gamma^{HT}_0(\bar\alpha_s(Q^2)/N))]^{\half}},}
for the first integral and         
\eqn\htsolto{{\exp\biggl(-\int^{Q_0^2} \gamma^{HT}_0(\bar\alpha_s(q^2)/N)d 
\,\ln q^2\biggr) \over [-\chi_0'(
\gamma^{HT}_0(\bar\alpha_s(Q_0^2)/N))]^{\half}},}
for the second. It can be verified numerically that these expressions are 
indeed good approximations to the precise results. Combining these we get
the full next-to-leading twist gluon Green's function.
\eqn\htsolfull{{\exp\biggl(\int_{Q_0^2}^{Q^2} \gamma^{HT}_0
(\bar\alpha_s(q^2)/N)d 
\,\ln q^2\biggr) \over \gamma^{HT}_0(\bar\alpha_s(Q^2)/N)[-\chi_0'(
\gamma^{HT}_0(\bar\alpha_s(Q^2)/N))]^{\half}[-\chi_0'(
\gamma^{HT}_0(\bar\alpha_s(Q_0^2)/N))]^{\half}}.}
Hence, the anomalous dimension for the higher twist operator 
is simply that obtained for the fixed 
coupling, but with the coupling constant allowed to run with the scale, while
the normalization is (roughly) the root of the fixed coupling normalization 
evaluated for $\alpha_s(Q^2)$ multiplied by the same for $\alpha_s(Q_0^2)$.
Hence, the result is much the same as for the fixed coupling case, with 
both the splitting function and the normalization decreasing and oscillating 
as $x \to 0$. 

It order to justify this conclusion it is only necessary to explain why we
could assume the factorization. To do this we note that the saddle-point for 
the first integrand is at $t=(1/\bar\beta_0N)\chi_0(\gamma^{HT}_0(t))$ and 
similarly for the second integrand with $t \to t_0$. However, since $t\gg t_0$
$\gamma^{HT}_0(t_0)$ is significantly to the right of $\gamma^{HT}_0(t)$.
The value of $\exp(-\tilde\gamma t_0+X_0(\tilde \gamma)/(\bar\beta_0N))$ along
the real axis along with $\tilde \gamma = \gamma_0^{HT}(t_0), 
\gamma_0^{HT}(t)$ is shown in \fig\sadpoints{The value of 
$\exp(-\tilde\gamma t_0+X_0(\tilde \gamma)/(\bar\beta_0N))$, along
the real axis for $N=0.4$ and $t_0=2$, along with $\tilde \gamma = 
\gamma_0^{HT}(t_0), 
\gamma_0^{HT}(t)$ for $t\gg t_0$.}. It is simple to rewrite \solrunht\
in the equivalent form
\eqn\solrunhtalt{\eqalign{{\cal G}^{HT}(Q^2,N)&={1\over 2\pi i}
\int_{\gamma^{HT}_0(t) -i\infty}^{\gamma^{HT}_0(t)+i\infty}
{1\over \gamma}
\exp(\gamma\ln(Q^2/\Lambda^2)-X_0(\gamma)/(\bar\beta_0 N))
d\gamma\cr
&\times\biggl[\int_{\gamma^{HT}_0(t)}^{0}
\exp(-\tilde\gamma\ln(Q_0^2/\Lambda^2)+X_0(\tilde \gamma)/
(\bar\beta_0N))d\tilde\gamma \cr
&\hskip 1in +\int_{\gamma}^{\gamma^{HT}_0(t)}
\exp(-\tilde\gamma\ln(Q_0^2/\Lambda^2)+X_0(\tilde \gamma)/
(\bar\beta_0N))\biggr]d\tilde\gamma
\,g_B(Q_0^2,N).\cr}}
Using \sadpoints, and remembering that the saddle-point integral for 
the first integral is parallel to the imaginary axis, and that the 
integrand very quickly decreases away from $\gamma^{HT}_0(t)$, we conclude 
that the value of  the second integral in the second line of \solrunhtalt\
is negligible compared with the first. Also noting from \sadpoints\ 
that there is little 
change if we alter the lower limit of the first integral in the second line 
to $k\sim -1$, we obtain the factorization assumed above. Hence, in this 
$t\gg t_0$ limit we find that we obtain factorization of the next-to-leading
twist solution and that as for the fixed coupling case this is negligible 
as $x\to 0$.     

Even if $t_0$ approaches $t$ the results can be shown to be similar by 
numerical calculation. For example, in the extreme limit of $t=t_0$ the first 
integral in the second line of \solrunhtalt\ gives only half the 
saddle-point contribution, but one can check that the previously negligible 
second integral now gives a roughly equal contribution for all $N$. However,
factorization is now clearly broken. Detailed numerical investigation shows 
that for $t_0$ not much smaller than $t$ we can write the higher twist 
contribution in the form $(Q_0^2/Q^2)f(Q^2,Q_0^2,N)$ where the total  
is a function of $N$ which grows slowly with $N$, approaching a constant 
as $N \to 0$. This is consistent with the form $x^a \cos(b\ln(1/x))$
which we get for the factorized next-to-leading twist solution (the 
Mellin transformation of which is $(N+a)/((N+a)^2+b^2)$), and certainly 
confirms that the gluon Green's function is falling as $x \to 0$.         

Therefore, the higher twist operators and their anomalous dimensions 
derived from
either the fixed coupling or running coupling BFKL equation are negligible 
at small $x$, and for these higher twist contributions the use of the
running coupling equation does not qualitatively change anything. However, we
are currently not able to say anything about the contributions from the 
four-gluon operators, and hence about shadowing corrections etc., beyond 
relatively simple results, e.g. anomalous dimensions in the small $x$ limit 
at LO in $\alpha_s$. There have been various suggestions that such shadowing
corrections are large, but I feel that these estimates may well be severely
exaggerated by the use of the approximation of this LO in $\alpha_s$ anomalous
dimension, and also by the fact that the even more restrictive double-leading
logarithmic approximation is often used. This often seriously overestimates 
the size 
of the anomalous dimensions, coefficient functions, and also the gluon
distribution.  I hope I have 
demonstrated that for the evolution of the higher twist two-gluon operator
the LO-in-$\alpha_s$ double-leading-log approximations is indeed totally 
misleading.
It is also interesting to note that a more complete
calculation of the higher twist coefficient functions for the evolution of 
$F_2(x,Q^2)$ due to the four-gluon operators \ref\blumht{J. Bl\"umlein,
V. Ravindran, J. Ruan and W. Zhu, \PL \vyp{B504}{201}{235}.} 
implies that the double leading log approximation is a vast overestimate.
Even using very small values of the screening length ($R = 2\Gev^{-2}$ rather 
than the more usual $R \sim 10\Gev^{-2}$) and the very large LO GRV gluon 
distribution
\ref\grv{M. Gl\"uck, E. Reya and A. Vogt, \EPJ \vyp{C5}{1998}{461}.},
it seems that the shadowing correction is almost negligible in the 
perturbative HERA range. Saturation effects will
no doubt eventually set in for low enough $x$ and $Q^2$, but presently I feel 
the technology is not such as to predict where with any real accuracy. 
Certainly, 
resummations in $\ln(1/x)$ tend to decrease the size of the gluon extracted 
from data, and this combined with the above considerations suggests
a much smaller saturation effect, and total higher twist effect, 
than often supposed. Certainly the model-independent ``rule of thumb'' 
for strong saturation contributions that $d F_2(x,Q^2)/d \ln Q^2 \approx 
Q^2 \sigma(x)$ and hence $d \ln(F_2(x,Q^2))/d \ln Q^2\approx 1$ is not
even closely approached for any HERA data with $Q^2 \geq 1 \Gev^2$.    
     
However, I note that in my examination of higher twist I have not
examined the mixing between leading twist and higher twist operators or 
included any nonperturbative contributions due to, for example, the behaviour 
of the coupling constant at low scales. These two effects are related to 
each other. Such questions have been considered 
for toy models in \ciafsalam\ and \salamlo, and numerically for the full LO
running coupling BFKL equation \salamlo. These papers have considered the 
full anomalous dimension defined by $d \ln({\cal G}(Q^2,N))/dt$, 
and the way this is
affected by the higher twist corrections, rather than just 
$d \ln({\cal G}^{HT}(Q^2,N))/dt$ considered above. 
They demonstrate that there are
potentially serious modifications to the leading twist anomalous 
dimension due to the higher twist corrections introducing sensitivity to
the form of the normalization factor ${\cal G}_I(Q_0^2,N)$ which depends on
the regularization of the coupling at low scales and on the $Q_0^2$ 
dependence. Depending on the assumptions about the nonperturbative physics, 
these contributions can be be important at extremely small $x$, generally 
changing the precise form of the 
power-like behaviour, and for more severe imposition of nonperturbative 
effects, i.e. letting them set in at higher scales, introducing a completely
different asymptotic behaviour. Unfortunately, within the framework of my 
paper the formal divergence of ${\cal G}_I(Q_0^2,N)$ makes a similar study 
impossible and, as mentioned at the end of the previous section, I simply 
have to appeal to these alternative results, in particular the smallness of 
$x$ at which the
power-suppressed modifications set in, in order to support the reliability 
of my more formal calculations. However, I also note that the smallness of the 
higher twist operators and their anomalous dimensions calculated in this 
section suggest that whilst these contributions 
from non-perturbative sources only set in at low $Q^2$ or very
small $x$ indeed  it seems perfectly possible that they will give a 
comparable, or even larger contribution at low $x$ and low $Q^2$ than the 
genuine higher twist contributions.

\newsec{NLO Corrections.}   

In section 2 I demonstrated that using $\alpha_s(k^2)$ in the BFKL
equation, as in \bfklruni, has a profound effect on the form of the
solution both for the normalization and for the anomalous dimension. 
However, given the first conclusions regarding NLO corrections in the 
essentially fixed coupling case, it is particularly necessary to
check that the results presented are not severely modified by the
inclusion of the NLO kernel, i.e. that the perturbative calculations are
stable. The NLO kernel was
presented in \NLOBFKLlf\ and the way in which to solve at NLO with a running
coupling was presented in \ciafpcol. 
Writing the NLO equation as 
\eqn\NLOBFKLalti{f(k^2,Q_0^2)= f_I(k^2,Q_0^2)
+\biggl({\bar\alpha_s(k^2) \over N}\biggr)
\int_0^{\infty}{dq^2\over q^2}(K_0(q^2,k^2)
-\alpha_s(k^2)K_1(q^2,k^2))f(q^2),}
and using just the one-loop expression for the coupling\foot{Using the full
NLO expression for the running coupling would lead to a huge degree of 
complication, and this has never been attempted. Since, so long as $\Lambda$ 
is chosen appropriately, the one- and two-loop couplings are very similar,
I do not imagine any major errors in the results below.} leads to a 
2nd order differential equation in $\gamma$-space
\eqn\bfklrunnlotot{{d^2 \tilde f(\gamma,N)\over d \gamma^2}={d^2\tilde 
f_I(\gamma, Q_0^2) \over
d\gamma^2}-{1\over \bar\beta_0 N} {d (\chi_0(\gamma)\tilde 
f(\gamma,N))\over d\gamma}-{\pi\over 3\bar\beta^2_0 N}\chi_1(\gamma)
\tilde f(\gamma, N).}
This can be solved in a very similar way to LO,
i.e. it factorizes into the same form as \solruniv\
with $Q^2$-dependent part given by
\eqn\solrunnloii{{\cal G}_{E}^1(N,t)={1\over 2\pi i}
\int_{\half -i\infty}^{\half+i\infty}
{1\over \gamma}
\exp(\gamma t-X_{1}(\gamma,N)/(\bar\beta_0 N))
d\gamma.}
However, $X_{1}(\gamma,N)$ is rather more
complicated than the previous $X_0(\gamma)$.
It can still be expressed in the form 
\eqn\solrunnloiii{X_{1}(\gamma,N)=\int_{\half}^{\gamma}
\chi_{NLO}(\hat\gamma,N)d\hat\gamma,}
but now $\chi_{NLO}(\gamma,N)$ can be written as a power series in $N$ 
beginning at zeroth order with $\chi_0(\gamma)$. As seen in \ciafpcol,
though here ignoring resummations in $N$, the explicit form is 
\eqn\solrunnloiv{\chi_{NLO}(\gamma,N) =
\chi_0(\gamma)-N{\chi_1(\gamma) \over \chi_0(\gamma)} +{N^2 \over
\chi_0}\biggl(-\biggl({\chi_1(\gamma)\over
\chi_0(\gamma)}\biggr)^2-\beta_0 \biggl({\chi_1(\gamma)\over 
\chi_0(\gamma)}\biggr)'\biggr) +\cdots,}
where the currently unknown NNLO contribution to the kernel, $\chi_2(\gamma)$, 
would also appear at order $N^2$ in principle.  

As already discussed in section 2 there is a contribution to $\chi_1(\gamma)$
from the $\beta_0$-dependent terms induced by an ``incorrect'' choice of the 
scale for the coupling -- $k^2$ rather than $(k-q)^2$. Taking this 
contribution to
the term in \solrunnloiv\ which is linear in $N$, and combining with the 
LO expression we find the previously discussed result of 
only a minor change in the anomalous dimension and splitting function 
extracted. Hence, the choice of $\alpha_s(k^2)$ is reliable, and
is easily corrected for. In this section I consider the rest of the NLO
correction to the kernel, which is much larger, and henceforth I denote 
$\chi_1(\gamma)$ as the NLO kernel with the $\beta_0$-dependent part 
$\half\bar\beta_0(\chi^2_0(\gamma)+\chi'_0(\gamma))$ already extracted,
and include the multiplicative factor $f^{\beta_0}(\gamma)$ in the integrand 
in \solrunnloii. This still leaves a decision
as to precisely what I take ``the NLO calculation'' to mean. There are 
various possibilities. I could work at the level of the NLO correction 
to the kernel, and hence the BFKL equation, and solve \NLOBFKLalti, 
producing the infinite series in \solrunnloiv. Alternatively, I could 
truncate $\chi_{NLO}(\gamma,N)$ in \solrunnloiv\ after the second term.
However, doing this still leaves the question of whether to use the 
whole of $\exp(1/\bar\beta_0\int_{\half}^{\gamma}
(\chi_1(\hat\gamma)/\chi_0(\hat\gamma))d\hat\gamma)$ or just expand it
out to first order in $\bar\beta_0^{-1}$. 

There are particular problems associated with all
choices. If one solves using the full NLO corrected kernel then there 
is an infinite series in powers of $N$ to consider in \solrunnloiv, 
which turns out to be important in practice 
(see below). Also, the gluon Green's function and anomalous dimensions 
obtained from this solution contain many subleading terms beyond just  LO and 
NLO in $\ln(1/x)$ (and running coupling type corrections to these), as is 
essentially obvious from looking at \NLOBFKLalti\ - 
iteration of $f$ leads the last term producing NNLO then NNNLO and so on. 
Hence, this 
choice is disguarded. If one instead truncates \solrunnloiv\ at order $N$, one
still generates a subset of higher order terms beyond those one wishes,
though it is possible to
proceed in this case at least. One can see the explicit 
form of the solution by 
substituting the truncated \solrunnloiv\ into \solrunnloii\ and 
proceeding as in section 2. The contribution to $X_{1}(\gamma, N)$
coming from the second term, $-N(\chi_1(\gamma)/\chi_0(\gamma))$, leads to an 
expression of the same form as in \expX, i.e.  
\eqn\expXnlo{X_{1}(\gamma,N)= X_0(\gamma)-c_lN\ln(\gamma)-N c_0 -
N\sum_{n=1}^{\infty}c_n \gamma^{n},}
where the $c_n$ may be calculated easily by performing a power-series 
expansion of the known functions of $\gamma$, i.e.
\eqn\expnloser{\sum_{n=1}^{\infty} c_n\gamma^n = 0.424\gamma + 0.805\gamma^2
+0.521\gamma^3+2.290\gamma^4+1.287\gamma^5+2.980\gamma^6+\cdots.}
Hence, the integrand for ${\cal G}^1_E(N,Q^2)$ becomes 
\eqn\expintnlo{\gamma^{-(1-c_lN)/(\bar\beta_0N)-1}f^{\beta_0}(\gamma)
\exp\biggl(\gamma t
-{1\over(\bar\beta_0N)}(\gamma_E-c_0 N+\sum_{n=1}^{\infty}
(a_n\gamma^{2n+1}-Nc_n\gamma^{n}))\biggr).}
Performing precisely the same type of manipulations as in section 2 results 
in the expression 
\eqn\resultnlo{\eqalign{{\cal G}_E^1(N,t)&=-
\sin\biggl({\pi(1-c_lN)\over(\bar\beta_0N)}\biggr)
\Gamma(-(1-c_lN)/(\bar\beta_0N))
\exp\biggl(-{\gamma_E-c_0N\over(\bar\beta_0N)}\biggr)
t^{(1-c_lN)/(\bar\beta_0N)}\cr
&\times\biggr(1+
\sum_{n=1}\sum_{m=1}\biggr[\biggl(1+\tilde A_n(1/(\bar\beta_0N))
\biggr)\biggl(1+C_m(1/\bar\beta_0)\biggr)-1\biggr]t^{-n-m}
\Delta_{n+m}\biggl({-(1-c_lN)\over (\bar\beta_0N)}\biggr)\biggr),\cr}}
where 
\eqn\sumdef{1+\sum_{m=1}^{\infty}C_m(1/\bar\beta_0)\gamma^m=\exp\biggl(
{1 \over \bar\beta_0}\sum_{n=1}^{\infty}c_n\gamma^n\biggr),}
and the $\tilde A(1/(\bar\beta_0N))$ include the contributions from 
$f^{\beta_0}(\gamma)$, i.e. are of the form in \resultiialt.
The factoring of the terms independent of $t$ then results in the expression
\eqn\resultinlo{\eqalign{{\cal G}^1_E(N,t)&=t^{(1-c_lN)/(\bar\beta_0N)}
\biggl(1+
\sum_{n=1}^{\infty}\sum_{m=1}^{\infty}\biggl[\biggl(1+\tilde 
A_n(1/(\bar\beta_0N))\biggr)\cr
&\hskip 1.6in \times\biggl(1+C_m(1/\bar\beta_0)\biggr)-1\biggr]t^{-n-m}
\Delta_{n+m}\biggl({-(1-c_lN)\over (\bar\beta_0N)}\biggr)\biggr).}}
There are two sources of corrections beyond NLO in $\ln(1/x)$, other than 
running coupling corrections, in \resultinlo. Firstly, $C_n(1/\bar\beta_0)$
can be expanded as a power series in $1/(\bar\beta_0)$. Only the first term 
in this series is genuinely a NLO correction to the LO result. 
Terms of higher order lead to contributions to the anomalous
dimensions which are beyond NLO in $\ln(1/x)$ without compensating factors of 
$\beta_0$ which would enable them to be interpreted as running coupling
corrections. Secondly, when one 
expands terms of the form $((1-c_lN)/(\bar\beta_0N))^n$ which appear in the 
$\Delta_n$ in \resultinlo, one obtains a power-series of the form,
\eqn\delexpnlo{\biggl({(1-c_lN)\over (\bar\beta_0N)}\biggr)^n = 
\biggl({1\over (\bar\beta_0N)}\biggr)^n\biggl[1 -nc_lN
+{n(n-1)\over2}(c_lN)^2 +\cdots\biggr].}
The second term in this series gives the NLO in $\ln(1/x)$ 
correction while the remainder give higher corrections without compensating
powers of $\beta_0$. Therefore, both these power-series expansions, i.e of 
the $C_n$ in powers of $1/(\bar\beta_0)$, and the $\Delta_n$ in powers of 
$N$ should be stopped at first order in $\bar\beta_0^{-1}$ or $N$, and the 
cross-terms coming from first-order in both expansions, which are of overall 
second order, should be eliminated to obtain truly NLO results.\foot{Ignoring 
this 
requirement and using the whole of \resultinlo, it turns out that the 
resultant expression is very badly behaved - blowing up at large $N$. 
This is almost entirely due to the higher-order terms in
the expansion of the $\Delta_n$. Using the full $C_n(1/\bar\beta_0)$ does not
change things much in practice. This large $N$ instability 
translates into huge corrections in the splitting function at large $x$. 
Presumably this instability at large $N$ and $x$ is cured if one resums the 
whole series in \solrunnloiv. Including just the ${\cal O}(N^2)$ term does 
seem to improve matters.}

Ultimately I define NLO by appealing to the perturbative form of the gluon 
Green's function and anomalous dimension produced and hence
by choosing the NLO definition such that the Green's
function does receive only corrections which are no more than one power of
$\alpha_s(Q^2)$ (without compensating factors of $\beta_0$) down on the 
leading order one. This means using an 
expression for the gluon Green's function of the form 
\eqn\resultinloprop{\eqalign{{\cal G}^1_E(N,t)&=t^{(1-c_lN)/(\bar\beta_0N)}
\biggl(1+\sum_{n=1}^{\infty}\sum_{m=1}^{\infty}\biggl[\biggl(1+
\tilde A_n(1/(\bar\beta_0N))\biggr)\biggl(1+ 
c_n/\bar\beta_0\biggr)-1\biggr]\cr
& t^{-n-m}\Delta_{n+m}(-1/(\bar\beta_0N))
-{c_l \over \bar\beta_0} \sum_{n=1}^{\infty}\tilde 
A_n(1/(\bar\beta_0N)) t^{-n}{ d 
\,\Delta_n(-1/(\bar\beta_0N))\over d(-1/(\bar\beta_0N))}
\biggr),\cr}}
where the $c_n/\bar\beta_0$ are obtained by expanding the exponential
expression 
$\exp(1/\bar\beta_0\bigl(\int_{\half}^{\gamma}
(\chi_1(\hat\gamma)/\chi_0(\hat\gamma)+c_l+c_0\bigr))d\hat\gamma)$, 
out to just first order in $1/\bar\beta_0$. Implicitly there is 
also a factor of $-\sin\biggl({\pi(1-c_lN)\over(\bar\beta_0N)}\biggr)
\Gamma\biggl({-(1-c_lN)\over(\bar\beta_0N)}\biggr)
\exp(-\gamma_E/(\bar\beta_0N)+c_0/\bar\beta_0)$ which contributes 
to the normalization in \resultinloprop.

Now that we have this NLO expression for the gluon Green's function it is 
necessary to make one more decision regarding the definition of the 
anomalous dimension. This is obtained from $\gamma^{LO+NLO}(N,t)
= (d \ln({\cal G}^1_E(N,t))/dt)$. However, 
strictly speaking, in order to obtain only NLO contributions to the anomalous
dimension $({\cal G}^1_E(N,t))^{-1}$ in this expression should be 
expanded only to NLO. This leads to a formal problem already pointed out in
section 6 of \xscale. Using the whole of $({\cal G}^1_E(N,t))^{-1}$ in the
expression for the anomalous dimension we notice that the position of the 
first zero is changed from that at LO, leading to a shift, in fact a decrease,
in the leading pole for the anomalous dimension, and hence in the power of 
leading behaviour of the splitting function as $x \to 0$. So the $x \to 0$ 
behaviour of the splitting function becomes 
$P_{gg}(x)=\exp(\lambda_0\xi-\Delta\lambda\xi)$. However, since 
$\Delta\lambda$ 
is due to NLO corrections, the strict NLO expansion is just $P_{gg}(x)=
\exp(\lambda_0\xi)-\Delta\lambda\xi \exp(\lambda_0\xi)$. This definition 
does not 
explicitly retain the shift in the power-like behaviour, and also leads to 
the NLO correction ultimately becoming larger than the LO result. 
Hence, I choose to retain the whole of 
$({\cal G}^1_E(N,t))^{-1}$ in the definition of the NLO anomalous dimension,
\vadjust{\eject} thus obtaining the full 
$P_{gg}(x)=\exp(\lambda_0\xi-\Delta\lambda\xi)$ as 
$x \to 0$, even though in practice the choice makes little difference at 
the values of $x$ relevant to HERA.

So now I can use \resultinloprop\ to determine analytic expressions for 
the NLO gluon Green's function and anomalous dimension. However, the formal
definition again results in a divergent power series, and as at LO I really
truncate the series in \resultinloprop\ at $n_0=5$. This leaves the 
problem of calculating the power-suppressed corrections. 
In order to do this it is necessary to have an exact definition 
for ${\cal G}_{E}^1(N,t)$ in the form 
of an inverse Mellin transformation, as in \solrunnloii. This requires finding 
the integral expression which would lead to \resultinloprop\ if  
a power-series expansion of the integrand is performed. 
Unfortunately this is not that 
simple. The problem comes with the manner of treating the 
$-c_lN\ln(\gamma)$ term in \expXnlo. In order to have the leading
$t^{(1-c_lN)/(\bar\beta_0N)}$ factor in \resultinloprop, and hence obtain
the correct expression for the ${\cal O}(\alpha_s(Q^2))$ part of the 
anomalous dimension, it is necessary to keep $-c_lN\ln(\gamma)$ in the 
exponential in the integrand, giving a factor $\gamma^{-c_l/\bar\beta_0}$.
Expanding out $\exp(-c_l\ln(\gamma)/\bar\beta_0)$ to first order would lead to
$\ln(t)$ contributions to the anomalous dimension. However, keeping the 
full $\gamma^{-c_l/\bar\beta_0}$ factor results in the argument of the 
$\Delta_n$ being $-(1-c_lN)/(\bar\beta_0N)$ as in \resultinlo. Hence, there is
no simple way to generate only NLO corrections from this term.          
In order to obtain an expression equivalent to \resultinloprop\ I choose to 
effectively put the known factor of $t^{(-c_lN/(\bar\beta_0N))}$ in by
hand and to generate the derivatives of the $\Delta_n$ within the integral 
with respect to $\gamma$. 

In order to see how to do this I consider the LO expressions \solruniv\ 
and \result.  
It is quite simple to generate the first part of \resultinloprop. All one
needs do is insert the series expansion $1+1/(\bar\beta_0)
\sum_{n=1}^{\infty}c_n\gamma^n$ expanded to first order in $1/\bar\beta_0$
into the integral representation, i.e.
\eqn\contouri{{\cal G}^{1,I}_E(N,t)=\int_{C} 
\gamma^{-1/(\bar\beta_0N)-1}f^{\beta_0}(\gamma)\exp\biggl(\gamma t
-{1\over (\bar\beta_0N)}\sum_{n=1}^{\infty}a_n\gamma^{2n+1}\biggr)
\biggl(1+\sum_{m=0}
(1/\bar\beta_0)c_m \gamma^m\biggr)  d \gamma,}
where the integral is over the full, unspecified contour,
and generates the $t$-independent factor $\sin\biggl({-\pi\over
(\bar\beta_0N)} \biggr)\Gamma\biggl({-1\over(\bar\beta_0N)}\biggr)$, 
as well as the $t$-dependent parts explicitly in \resultinloprop.
On top of this one must also insert the $t^{-c_lN/(\bar\beta_0N)}$ factor
by hand. If one is also concerned with the $N$-dependent normalization it
is probably most consistent to also multiply by the factor
\eqn\normfact{{\sin(\pi(1-c_lN)/(\bar\beta_0N))
\Gamma(-(1-c_lN)/(\bar\beta_0N))
\exp(-(\gamma_E-c_0N)/(\bar\beta_0N))
\over \sin(\pi/(\bar\beta_0N))\Gamma(-1/(\bar\beta_0N))},}
in order to obtain the overall factor of
\eqn\nlofact{-\sin((\pi(1-c_lN)/(\bar\beta_0N))\Gamma(-(1-c_lN)/(\bar\beta_0N))
\exp\biggl({-\gamma_E+c_0N\over \bar\beta_0N}\biggr).}

Generating the second part of \resultinloprop\ is rather more complicated. 
One has to somehow modify the integral representation so that the derivatives 
of the $\Delta_n(-1/(\bar\beta_0N))$ are obtained.\vadjust{\eject} 
To see how to do this 
we let $1/(\bar\beta_0N)=z$, in which case the equivalence of \contour\ and
\resulti\ (ignoring the divergence of the series) is     
\eqn\equival{\int_{C} 
\gamma^{-z-1}\exp\biggl(\gamma t
-z\sum_{n=1}^{\infty}a_n\gamma^{2n+1}\biggr)d \gamma =
-\sin(\pi z)\Gamma(-z) t^{z}
\biggl(1+\sum_{n=3}^{\infty}A_n(z)t^{-n}
\Delta(-z+n)\biggr),}
where I have removed the trivial factor of $\exp(-\gamma_E/(\bar\beta_0N))$
from each side. Differentiating both sides with respect to $z$ we obtain
\eqn\diffequiv{\eqalign{ -\int_{C} 
\ln(\gamma)\gamma^{-z-1}\exp\biggl(\gamma t
-z\sum_{n=1}^{\infty}&a_n\gamma^{2n+1}\biggr)d \gamma\cr
& -\int_{C} 
\gamma^{-z-1} \sum_{m=1}^{\infty}a_m\gamma^{2m+1} \exp\biggl(\gamma t
-z\sum_{n=1}^{\infty}a_n\gamma^{2n+1}\biggr)d \gamma \cr
& = \Psi(-z)\sin(\pi z)\Gamma(-z) t^{z}
\biggl(1+\sum_{n=3}^{\infty}A_n(z)t^{-n}
\Delta(-z+n)\biggr) \cr
&-\pi\cot(\pi z)\sin(\pi z)\Gamma(-z) t^{z}
\biggl(1+\sum_{n=3}^{\infty}A_n(z)t^{-n}
\Delta(-z+n)\biggr)\cr
&-\ln(t)\sin(\pi z)\Gamma(-z) t^{z}
\biggl(1+\sum_{n=3}^{\infty}A_n(z)t^{-n}
\Delta(-z+n)\biggr)\cr
&-\sin(\pi z)\Gamma(-z) t^{z}
\biggl(1+\sum_{n=3}^{\infty}A_n(z)t^{-n}
{ d\Delta(-z+n) \over dz}\biggr)\cr
& -\sin(\pi z)\Gamma(-z) t^{z}
\biggl(\sum_{n=3}^{\infty}{ d A_n(z) \over d z} t^{-n}
\Delta(-z+n)\biggr).\cr}}
The last term on each side are equivalent, and rearranging the rest 
we obtain an expression for a series containing the derivatives of the 
$\Delta_n(z)$ -
\eqn\diffdel{\eqalign{\sin(\pi z)\Gamma(-z)& t^{z}
\biggl(\sum_{n=3}^{\infty}A_n(z)  t^{-n}
{d\,\Delta(-z+n)\over dz}\biggr) = \int_{C} 
\ln(\gamma)\gamma^{-z-1}\exp\biggl(\gamma t
-z\sum_{n=1}^{\infty}a_n\gamma^{2n+1}\biggr)d \gamma \cr
&+(\Psi(-z)-\pi\cot(\pi z)-\ln t)\sin(\pi z)\Gamma(-z) t^{z}
\biggl(1+\sum_{n=3}^{\infty}A_n(z)t^{-n}
\Delta(-z+n)\biggr),\cr}}  
which using \equival\ becomes
\eqn\diffdels{\eqalign{\sin(\pi z)\Gamma(-z) t^{z}
\biggl(\sum_{n=3}^{\infty} &A_n(z)  t^{-n}
{ d\Delta(-z+n)\over dz}\biggr) \cr
&= \int_{C}\biggl[\ln(\gamma)
-(\Psi(-z)-\pi\cot(\pi z)-\ln t)\biggl] \gamma^{-z-1}\exp\biggl(\gamma t
-z\sum_{n=1}^{\infty}a_n\gamma^{2n+1}\biggr)d \gamma.}}  
Therefore, the right-hand-side of \diffdels, multiplied by
$-c_l/(\bar\beta_0)t^{(-c_lN/(\bar\beta_0N))}$, gives the second term in
\resultinloprop\ with some $t$-independent normalization which should be
multiplied by \nlofact\ to be consistent \vadjust{\eject}
with the first term in the preceding paragraph. Thus, we have a  
prescription for the full calculation at NLO which is equivalent to the 
series expansion in \resultinloprop, i.e 
\eqn\intnloresult{\eqalign{{\cal G}^1_{E}&(N, t) \propto  
\,t^{-c_l/\bar\beta_0}\int_{C}\biggl[ 
\gamma^{-1/(\bar\beta_0N)-1}f^{\beta_0}(\gamma)\exp\biggl(\gamma t
-{1\over (\bar\beta_0N)}\sum_{n=1}^{\infty}a_n\gamma^{2n+1}\biggr)
\biggl(1+\sum_{m=0}
(1/\bar\beta_0) c_m \gamma^m\biggr) \cr
&\hskip -0.3in -{c_l \over \bar\beta_0}\biggl[\ln(\gamma t)
-\Psi\biggl(-{1\over \bar\beta_0N}\biggr)+\pi\cot\biggl({\pi\over
\bar\beta_0N}\biggr)\biggl] 
\gamma^{-(1/(\bar\beta_0N))-1}f^{\beta_0}(\gamma)\exp\biggl(\gamma t
-{1\over \bar\beta_0N}\sum_{n=1}^{\infty}a_n\gamma^{2n+1}\biggr)
\biggr]d \gamma,\cr}}  
and once again one should multiply by \nlofact\ to get the most suitable 
normalization. We can now insert the above expression into 
$\gamma^{LO+NLO}(N,t)= (d \ln({\cal G}^1_E(N,t))/dt)$ 
and evaluate numerically in order to get the NLO
anomalous dimension without recourse to the truncated series expansion. 

We are now in a position to solve for the anomalous dimension and 
splitting function at NLO. Unlike the case of fixed
coupling, or the simplistic results of the saddle-point evaluation, the
NLO corrections to the LO anomalous dimension are 
under control. This is simply illustrated by the positions of the leading 
pole in the anomalous dimensions which are shown in \intercepts, 
and one can see that they change from about $0.25$ for $\gamma_{gg}(N,t)$ at 
LO to $0.17$ at NLO, and that the $Q^2$-dependence reduces a little.
However, as already noted at LO, the value of the intercepts has little to do 
with physics at HERA -- the power-like behaviour only really settling down 
for lower $x$, and this is even more true at NLO.  
Being more particular one notices that the anomalous dimension 
$\gamma_{gg}(N,t)$ over a wide range
of $N$ shows only a relatively small change going from LO to NLO. 
This is shown in \anom.b where the part of the NLO anomalous dimension at 
first order in $\alpha_s(Q^2)$, i.e $-0.935\alpha_s(Q^2)$,
is not included, since this should properly be included at LO in a
combined leading order in $\alpha_s(Q^2)$ and $\alpha_s(Q^2)\ln(1/x)$
expansion scheme. Alternative definitions of NLO lead to very similar
results except at very high values of $N$, where less sophisticated 
definitions lead to blowing up at large $N$, as already mentioned. For this
case of the gluon structure function the NLO correction is negative except
for very large $N$. I should also note that the power-like 
correction to the purely analytic result is a larger proportion of the 
NLO correction than of the LO contribution, but would still be almost 
impossible to spot if shown on \anom.b. The correction to the analytic 
value for the intercept is about $7\%$ at $t=6$ however.           
 
One can also make the transformation to $x$-space and calculate the
NLO corrected splitting function. Unfortunately, due to the increase in 
size of the $c_n$ coefficients compared to the $a_n$ (particularly the
absence of zeros) and also to the factors
of $n$ invoked by differentiating the $\Delta_n$ in \resultinlo\ the 
power-series in $\alpha_s(Q^2)$ is much less convergent than at LO. In 
order to obtain an expression which is reliable down to $x=0.00001$ at
$Q^2=1\Gev^2$ it is necessary to go to $20$th order 
in $\alpha_s(Q^2)$.
Hence we can write the NLO correction to the splitting function as 
\eqn\nlosplit{xP^{NLO}_{gg}(\xi,\alpha_s(Q^2)) = \bar\alpha_s(Q^2)
\sum_{n=1}^{19}\sum_{m=0}^{m_{max}} \bar\alpha^n_s(Q^2)\biggl(K_{nm}{\xi^m 
\bar\beta_0^{n-m-1}\over m!}+K_{n\delta}\bar\beta_0^{n}\delta(1-x)\biggr),} 
where because we have truncated the series for the gluon structure function
$m_{max}$ can be greater than the naive expectation of $m_{max}=n-1$.
The coefficients for the series are shown in table 1. If one is only 
concerned with $x>0.0001$ or $Q^2> 4\Gev^2$ then the series can be truncated
at about $12th$ order. 
 
As at LO we also have to model the $N$ dependence of the power-suppressed 
correction by an analytic function. Fortunately, exactly the same type of 
function is sufficient and we obtain the power-suppressed NLO correction 
to the splitting function of the form
\eqn\modelnlogg{\eqalign{-2.86\exp(-1.02t)&\delta(1-x)+
\exp(-t)\biggl( 13.59 
\biggl({\alpha_s(t)\over \alpha_s(4.5)}\biggr)^{0.88}
-29.61 \biggl({\alpha_s(t)\over \alpha_s(4.5)}\biggr)^{1.21}\xi\cr
&+39.76 \biggl({\alpha_s(t)\over \alpha_s(4.5)}\biggr)^{1.315}{\xi^2 \over 2!}
-33.765 \biggl({\alpha_s(t)\over \alpha_s(4.5)}\biggr)^{1.48}{\xi^3 \over 3!}
+16.89 \biggl({\alpha_s(t)\over \alpha_s(4.5)}\biggr)^{1.77}{\xi^4 \over 4!}\cr
&-4.479 \biggl({\alpha_s(t)\over \alpha_s(4.5)}\biggr)^{2.16}{\xi^5 \over 5!}
+0.4839 \biggl({\alpha_s(t)\over \alpha_s(4.5)}\biggr)^{2.63}{\xi^6 \over 6}
\biggr).\cr}}

The full NLO correction $xP^{NLO}_{gg}(x)$ and its 
power-series and power-suppressed contributions are shown in 
\fig\splitfnlo{a. The splitting function $xP^{NLO}_{gg}(x)$ and its 
power-series and power-suppressed contributions
plotted as functions of $x$ for
$t=6$. b. The splitting functions $xP^{LO+NLO}_{gg}(x)$ 
plotted as a function of $x$ for
$t=6$ ($Q^2 \sim 6\Gev^2$). Also shown is the ${\cal O}(\alpha_s(Q^2))$ 
contribution $\bar\alpha_s(Q^2)$, and the LO contribution 
$xP^{LO}_{gg}(x)$.}.a, where the relatively unimportant terms 
$\propto \delta(1-x)$ are absent.
As at LO the power-suppressed correction is proportionally much larger in 
$x$-space than in moment space and certainly needs to be considered at $t=
6$ and below. Also as at LO it tends to oppose the form of the power-series 
expression, hence reducing the total NLO correction. The powers of $\alpha_s$
in \modelnlogg\ are slightly smaller than for LO, and hence the 
power-suppressed correction does not fall quite so quickly with $Q^2$. 

The total NLO  
splitting function, i.e LO plus the NLO correction, is shown for $t=6$ in
\splitfnlo.b, where the 
contributions $\propto \delta(1-x)$ both from the
${\cal O}(\alpha_s(Q^2))$ part and the running coupling corrections to this
are absent. The latter of these is a very small contribution. The NLO
corrected splitting function is clearly not qualitatively  different 
from that at LO, though it is quite a lot smaller at small $x$.
Hence it seems as though by including the infinite series of 
running coupling corrections the perturbative expansion of the BFKL 
splitting function has been stabilized. However, the real importance
of the NLO corrections as far as
physics is concerned is the effect they have on the evolution of the
gluon structure function. This is demonstrated in \fig\evolutiong{The values
of $d G(x,Q^2)/d\ln Q^2$, for $G(x,Q^2)=x^{-0.2}(1-x)^6$, due to the 
LO splitting function $P^{LO}_{gg}(x)$ and the LO+NLO splitting
function $P^{LO+NLO}_{gg}(x)$, plotted as functions of $x$ for
$t=6$ ($Q^2 \sim 6\Gev^2$). Also shown is the evolution due to the
${\cal O}(\alpha_s(Q^2))$ contribution $P(x)=\bar\alpha_s(Q^2)/x$.} 
where the evolution of a suitable model
for the structure function $G(x,Q^2)$, i.e. $(1-x)^6x^{-0.2}$, is
shown both for the LO running coupling splitting function, and for the
NLO corrected one (all $\delta(1-x)$ contributions other
that at
first order in $\alpha_s(Q^2)$ one are included). Also shown is the evolution 
due just to the double-leading-log term $P(x)=\bar\alpha_s(Q^2)/x$. 
As one sees, at this (fairly low) value of $t$, i.e. $Q^2\sim 6\Gev^2$, 
the evolution driven by the LO splitting function is very similar to that 
from the double-leading-log contribution, and is even slightly smaller
for $x$ from $0.007$ to $0.00001$, corresponding to the dip in the splitting 
function seen in \splitf. Below this the growth of the splitting function 
increases the evolution above the double-leading-log result. One also sees 
that the effect of the NLO corrections is certainly significant, and 
increases relatively with falling $x$, but it is clearly a correction rather 
than the complete change in qualitative behaviour induced by the NLO
corrections without resummation of running coupling effects. 
        
A further way often used to investigate the perturbative stability of a 
fixed order perturbative calculation is to investigate the 
renormalization-scale dependence. This is often used fallaciously, 
e.g. if one calculates
$P_{gg}(\alpha_s,x)$ to NLO in the standard perturbative expansion 
and then investigates variation of renormalization scales one will never
notice the influence of the terms at higher orders in $\alpha_s$ 
which are also of higher 
order in $\ln(1/x)$. This is symptomatic of the fact that the expansion 
purely in powers
of $\alpha_s$ is not really a correct expansion scheme for splitting 
functions (for a full discussion see \LORSC). However, once we have performed
a resummation of large logarithms, as here, renormalization-scale
variation should be more reliable.
The renormalization scheme dependence may be investigated by letting
\eqn\rsdep{\alpha_s(Q^2) \to \alpha_s(kQ^2)+\beta_0\ln(k)\alpha_s^2(kQ^2)}
and in the LO part of the splitting function expanding out to first order 
in $\ln(k)$, whilst in the NLO part using only the zeroth order, i.e.
just letting $\alpha_s(Q^2) \to \alpha_s(kQ^2)$. In this case we must also 
use a similar procedure for the power-suppressed corrections, i.e. these are 
really of the form $(\Lambda^2/\mu_R^2)$ rather than $(\Lambda^2/Q^2)$. 
The results for $k=0.5$ and $k=2$ are shown in \fig\rsvariation{a. The 
renormalization scale variation of the LO+NLO splitting function 
$P^{LO+NLO}_{gg}$.
Shown are the three choices of scale $Q^2$, $0.5Q^2$ and $2Q^2$ for $t=6$,
i.e. $Q^2 \sim 6\Gev^2$. b. The same for the LO+NLO physical splitting 
function 
$P^{LO+NLO}_{LL}$.} for $Q^2\sim 6\Gev^2$. As with the NLO corrections 
to LO the variation is significant but leads only to a correction rather 
than a qualitative change. This implies that the series expansion is stable, 
if not as rapidly converging as one might ideally hope for.    
        
Hence, the NLO corrections to the running coupling BFKL derived
splitting function are well under control, both in terms of the
asymptotic power-like behaviour of the splitting functions and in terms
of the evolution in the range currently accessible to experiments.
For deep-inelastic scattering, or indeed any process where there is 
factorization of the infrared physics into the input parton distributions,
e.g. Drell-Yan scattering in proton-proton collisions, no further resummation 
is necessary, or even useful,
beyond the running coupling corrections. 
This in distinct contrast to the case
where both ends of the gluon ladder are associated with a hard scale.
In this case the conventional BFKL expansion is fundamentally flawed due to
progressively higher  order poles at $\gamma=0$ and $\gamma=1$ 
(corresponding to large logs in
the ratios of the two scales $k_1^2$ and $k_2^2$) as shown in \gavin. These
large order poles need to be resummed, and without this resummation 
calculations are badly 
behaved over the whole range of $N$ (in fact 
explicit calculation shows that this is particularly the case at large $N$). 
In the case of deep inelastic 
scattering the collinear factorization procedure automatically orders
the poles at $\gamma=0$ correctly, and the above problem shows up in
high order poles at $\gamma=1$ only. The anomalous dimension is
totally dominated by the region very close to $\gamma=0$, as this paper
shows, and is very insensitive to effects at $\gamma=1$. Including the
type of resummation in \gavin\ciafpcol\ alters results from the NLO
corrected case by only a very small amount, and is likely to be no more
influential than the remaining NNLO effects for which it does not account. 
Resummation of poles near $\gamma=1$ would be essential if one attempted to
obtain information about the input form of the gluon,
i.e. ${\cal G}_I(Q_0^2,N)$. 
However, as well as the fact that $Q_0^2$ is an essentially nonperturbative 
scale, this type of calculation, along with the whole subject of
single-scale processes, is also plagued by the infrared ambiguity problem
caused by behaviour of the coupling at low scales. 
A discussion of such issues can be found in \ciafsalam\ and \salamlo.

I close this section by noting that although the above results all 
look promising it is important to realize that 
they are all in a sense ambiguous because they deal with a
particular way of defining the gluon parton distribution, which is a 
factorization scheme-dependent quantity. In this paper it is defined in 
a manner which is natural from the point of view of the solution of the 
BFKL equation, and which one may think of as perhaps a good ``physical''
definition of the gluon. However, it is very different from, for example, 
the gluon defined in the ${\overline{\rm MS}}$ scheme. In order to 
investigate the real success of the approach in this paper it is necessary to
look at the results for the real physical quantities -- the structure 
functions.   

\newsec{Small $x$ Structure Functions.}

One may define a real structure function by a simple extension of the above 
methods, i.e. by including a hard scattering cross section at the top of the 
gluon ladder. This modifies \gluondef\ to 
\eqn\strcfun{{\cal F}_{i}(Q^2,N)=\alpha_s\int_{0}^{\infty}{dk^2\over k^2} 
\sigma_{i,g}(k^2/Q^2) f(N,k^2,Q_0^2)
g_B(N,Q_0^2),}
where $\sigma_{i,g}(k^2/Q^2)$ is the cross-section for scattering of a 
virtual photon from a gluon with transverse momentum $k^2$. For the case 
of the longitudinal structure function this cross-section is well defined 
even in the limit $k^2 \to 0$, but for ${\cal F}_2(N,Q^2)$ the cross-section 
diverges like $\ln(Q^2/k^2)$ as $k^2 \to 0$ (for details see 
\ref\cathaut{S. Catani and F. Hautmann, \PL \vyp{B315}{1993}{157}; \NP
\vyp{B427}{1994}{475}.}). This demonstrates
that for ${\cal F}_L(x,Q^2)$ the solution in the leading $1/N$ limit 
factorizes neatly into the gluon distribution and a multiplicative coefficient
function, while for ${\cal F}_2(N, Q^2)$ there is interference at this order 
between the coefficient function and the result of solving the evolution 
equation including the anomalous dimension $\alpha_s\gamma^0_{qg}
(\alpha_s,N)$. In this latter case it is simplest instead to differentiate 
with respect to $\ln(Q^2)$ obtaining
\eqn\difstrcfun{{d{\cal F}_{2}(Q^2,N)\over d \ln Q^2}
=\alpha_s\int_{0}^{\infty}{dk^2\over k^2} 
{d\sigma_{i,g}(k^2/Q^2)\over d\ln Q^2} f(N,k^2,Q_0^2)
g_B(N,Q_0^2),}
where ${d\sigma_{i,g}(k^2/Q^2)\over d\ln Q^2}$ is finite as $k^2 \to 0$. 
In this case, if we work in an DIS-type scheme, i.e. one in which the 
quark-gluon coefficient function vanishes beyond zeroth order, there is 
a simple factorization between the anomalous dimension 
$\alpha_s\gamma^0_{qg}(\alpha_s, N)$
and the gluon distribution.\foot{Note that in this article I ignore the 
mixing with the quark input distribution in general for simplicity. 
However, it does implicitly appear in the NLO correction to the kernel, 
i.e. it is the NLO correction to the anomalous dimension eigenvalue rather 
than to $\gamma_{gg}$ which I use since this is the quantity directly
calculated in \NLOBFKLlf\NLOBFKLcc. The contribution to this due
to the quark mixing is very small in practice.}

In order to progress it is first necessary to consider the overall factor of
$\alpha_s$ in the above expressions, and particularly its scale. One might 
think that it should be $\alpha_s(k^2)$, and thus appear within the integrals
with respect to $k^2$. However, this could only come about due to double
counting of diagrams, since the resummation of bubble diagrams required to 
make this equal to $\alpha_s(k^2)$ has already been performed in defining
the coupling in the BFKL equation as $\alpha_s(k^2)$. $Q^2$ is the only 
remaining scale, so it must be the scale of this coupling. One can also 
justify this by considering the fact that there is a NLO correction to the 
input of the BFKL equation of the form $-\beta_0\alpha_s \ln(Q_0^2/\mu_R^2)
\delta(k^2-Q_0^2)$ (coming from bubbles in a gluon propagator). Introducing 
this into calculations leads to multiplying each result by a factor 
$(1-\beta_0\alpha_s \ln(Q_0^2/\mu_R^2))$. This splits into 
$-\beta_0\alpha_s \ln(Q^2/\mu_R^2)+\beta_0\alpha_s\ln(Q^2/Q_0^2)$, and the 
latter term is an infrared divergence which contributes to the one-loop
gluon-gluon splitting function while the second goes into making the overall
factor of $\alpha_s$ have renormalization scale $Q^2$.     
    
Now removing the overall factor of $\alpha_s(Q^2)$ (or in fact
the normalization factor $\alpha_s(Q^2)N_f/(3\pi)$) from \strcfun,
and taking the Mellin transformation with respect to 
$(Q^2/\Lambda^2)$ leads to the simple expression
\eqn\strcfunmell{\tilde {\cal F}_{i}(\gamma,N)= 
h_{i,g}(\gamma) \tilde {\cal G}(\gamma,N).}
Thus we may solve for ${\cal F}_{i}(N,t)$ in exactly the same way as for 
${\cal G}(N,t)$, obtaining exactly the same divergent $Q^2$-independent part 
and a $Q^2$-dependent part given by solving 
\eqn\solrunf{{\cal F}_{E,i}(N,t)={1\over 2\pi i}
\int_{\half -i\infty}^{\half+i\infty}
{h_{i,g}(\gamma)\over \gamma}f^{\beta_0}(\gamma)
\exp(\gamma t-X_0(\gamma)/(\bar\beta_0 N))
d\gamma.}
This may be evaluated numerically, using the same contour as for the gluon,
or in order to find the power-series solution
we may proceed as with the gluon structure function
by expanding the $h_{i,g}(\gamma)$ (which were calculated in
\cathaut) as a power series about
$\gamma=0$. For the two cases  we discussed above we have
\eqn\serfl{h_{L,g}(\gamma)f^{\beta_0}(\gamma)=1-0.33\gamma+2.13\gamma^2
+0.67\gamma^3+2.58\gamma^4+2.99\gamma^5+1.92\gamma^6+\cdots,}
and
\eqn\serftwo{h_{2,g}(\gamma)f^{\beta_0}(\gamma)=1+2.17\gamma+2.30\gamma^2
+6.67\gamma^3+7.05\gamma^4+12.92\gamma^5+15.47\gamma^6+\cdots.}
It seems natural to absorb the (in some sense) NLO corrections from 
$f^{\beta_0}(\gamma)$ into the contributions from the $h_{i,g}(\gamma)$
since they are of exactly the same form, whereas the other NLO corrections
have inverse powers of $\beta_0$. 
Following the same steps as in section 2.2 then  results in 
an expression
\eqn\resultif{{\cal F}_{E,i}(N,t)=t^{1/(\bar\beta_0N)}\biggl(1+
\sum_{n=1}^{n_0}B_{i,n}(1/(\bar\beta_0N))t^{-n}\Delta_n(-1/(\bar\beta_0N))
\biggr),}
where the $B_{i,n}(1/(\bar\beta_0N))$ are now determined not only by
the power series in $\gamma$ obtained from the expansion 
of $X_0(\gamma)$, but also from the expansion of
$h_{i,g}(\gamma)$. In particular they now contain parts at zeroth
order in $1/(\bar\beta_0N)$. 

Using these results it is now a simple matter to derive the
longitudinal gluon coefficient function at leading powers of $\ln(1/x)$
plus running coupling corrections and similarly for the quark-gluon anomalous 
dimension, i.e.
\eqn\coeffdef{C_{L,g}(\alpha_s(Q^2),N) = {\alpha_s(Q^2)N_f  \over3\pi}
{ {\cal F}_{E,L}(N,t)
\over {\cal G}_E(N,t)},}
with obvious generalization to $\gamma_{qg}(\alpha_s(Q^2),N)$.
These moment space expressions may easily be converted to $x$-space. 
Truncating the series for the structure functions and the gluon at $n_0=5$
results in the perturbative series for $xC_{L,g}(\alpha_s(Q^2),x)$, 
\eqn\expclg{\eqalign{x&C_{L,g}(\alpha_s(Q^2),x)={\alpha_s(Q^2)N_f  \over3\pi}
\biggl[\delta(1-x)-0.33\alpha_s(Q^2)+2.13\alpha_s^2(Q^2)
\bigl(\xi-\bar\beta_0\bigr)\cr
&+\alpha_s^3(Q^2)\bigl(-0.933{\xi^2 \over 2!}+2.79\bar\beta_0\xi-1.86\bar
\beta_0^2\bigr)+\alpha_s^4(Q^2)\bigl(2.32{\xi^3\over 3!}-14.69\bar\beta_0
{\xi^2 \over 2!}+27.85\bar\beta^2_0\xi-15.48\bar\beta^3_0\bigr)\cr
&+\alpha_s^5(Q^2)\bigl(8.41{\xi^4\over 4!}-54.45\bar\beta_0
{\xi^3 \over 3!}+125.2\bar\beta^2_0{\xi^2\over 2!}-121.2\bar\beta^3_0\xi
+42.0\bar\beta_0^4\bigr)\cr
&+\alpha_s^6(Q^2)\bigl({-0.89\over \bar\beta_0}{\xi^6 \over 6!}
+7.76{\xi^5\over 5!}
-27.53\bar\beta_0{\xi^4 \over 4!}+49.48\bar\beta^2_0{\xi^3\over 3!}
-44.59\bar\beta^3_0{\xi^2 \over 2!}+15.77\bar\beta_0^4\xi\bigr)\cr
&+\alpha_s^7(Q^2)\bigl({2.74 \over \bar\beta_0}{\xi^7 \over 7!}-33.41
{\xi^6 \over 6!}+164.8\bar\beta_0{\xi^5\over 5!}
-419.3\bar\beta^2_0{\xi^4 \over 4!}+577.2\bar\beta^3_0{\xi^3\over 3!}
-404.9\bar\beta^4_0{\xi^2 \over 2!}+112.9\bar\beta_0^5\xi\bigr)\cr
&+\alpha_s^8(Q^2)\bigl({6.48 \over \bar\beta_0}{\xi^8 \over 8!}
-72.27{\xi^7 \over 7!}+335.7\beta_0{\xi^6 \over 6!}
-838.2\bar\beta^2_0{\xi^5\over 5!}
+1210\bar\beta^3_0{\xi^4 \over 4!}-1004\bar\beta^4_0{\xi^3\over 3!}\cr
&+441.7\bar\beta^5_0{\xi^2 \over 2!}-79.05\bar\beta_0^6\xi\bigr)\biggl].\cr}}
However, as for the gluon splitting function we have to calculate the 
power-suppressed correction 
by evaluating the inverse Mellin transformations 
numerically. This is done in precisely the same way as for the gluon, and
results in the correction to $xC_{L,g}(\alpha_s(Q^2),x)$ of the form
\eqn\modelclg{\eqalign{{\alpha_s(Q^2)N_f  \over3\pi}\biggl[&
(-1.168-0.482t+0.1106)\exp(-t)\delta(1-x)+
\exp(-t)\biggl( -4.685 
\biggl({\alpha_s(t)\over \alpha_s(4.5)}\biggr)^{-3.026}\cr
&+34.25 \biggl({\alpha_s(t)\over \alpha_s(4.5)}\biggr)^{-0.875}\xi
-59.47 \biggl({\alpha_s(t)\over \alpha_s(4.5)}\biggr)^{0.074}{\xi^2 \over 2!}
+45.81 \biggl({\alpha_s(t)\over \alpha_s(4.5)}\biggr)^{0.78}{\xi^3 \over 3!}\cr
&-17.94 \biggl({\alpha_s(t)\over \alpha_s(4.5)}\biggr)^{1.37}{\xi^4 \over 4!}
+3.365 \biggl({\alpha_s(t)\over \alpha_s(4.5)}\biggr)^{1.77}{\xi^5 \over 5!}
-0.2942 \biggl({\alpha_s(t)\over \alpha_s(4.5)}\biggr)^{1.78}{\xi^6 \over 6}
\biggr)\biggr],\cr}}
where in this case it was necessary to model the $N\to \infty$,
i.e. the $\delta(1-x)$ part with a 
slightly more complicated form than previously. Both expressions have been 
shown in a form which is sufficient for $Q^2 > 1\Gev^2$ and $x>0.00001$.
The full $xC_{L,g}(x,t)$ is shown in \fig\coefffl{a. The full 
leading $\ln(1/x)$ plus running coupling corrections coefficient function
$xC_{L,g}(x,t)$ plotted as a function of $x$ for $t=6$ and $N_f=4$. Also shown 
are the contributions from the power-series and the power-suppressed part.
Note that the term $\propto \delta(1-x)$ in the power-series is 
replaced by the full ${\cal O}(\alpha_s(Q^2))$ contribution $6x^2(1-x)$,
and the terms $\propto \delta(1-x)$ in the power-suppressed part are absent.
b. $xC^{LO}_{L,g}(x,t)$ plotted as a function of $x$ for $t=6$ and $N_f=4$.
Also shown is the coefficient function obtained from the naive 
LO BFKL calculation, and the contribution at ${\cal O}(\alpha_s(Q^2))$ alone.}
a. along with the two contributions above. Note that the $\delta(1-x)$ term 
at ${\cal O}(\alpha_s(Q^2))$ in the power-series is obtained from the inverse 
Mellin transformation of the limit as $N \to 0$ of the full 
${\cal O}(\alpha_s(Q^2))$ coefficient function and in the figure
we replace it by the full ${\cal O}(\alpha_s(Q^2))$ contribution, 
$6x^2(1-x)$, for ease of presentation
(it not being easy to represent the normalization of the $\delta(1-x)$ term). 
The $\delta(1-x)$ term is simply
missing from the power-suppressed part, though this is insignificant.
We see that the power-suppressed contribution is now a much larger fraction of 
the total than for the gluon, though it does not increase as quickly with 
falling $Q^2$. 
In \coefffl\ b. we show $xC_{L,g}(x,t)$ along with the 
${\cal O}(\alpha_s(Q^2))$ contribution and with the naive LO BFKL result
in this factorization scheme, which grows far more quickly than the resummed 
result.
     
Similarly we can calculate the perturbative series  
$xP_{qg}(\alpha_s(Q^2),x)$, 
\eqn\exppqg{\eqalign{x&P_{qg}(\alpha_s(Q^2),x)={\alpha_s(Q^2)N_f  \over3\pi}
\biggl[\delta(1-x)+2.17\alpha_s(Q^2)+2.30\alpha_s^2(Q^2)
\bigl(\xi-\bar\beta_0\bigr)\cr
&+\alpha_s^3(Q^2)\bigl(5.07{\xi^2 \over 2!}-15.21\bar\beta_0\xi+10.14\bar
\beta_0^2\bigr)+\alpha_s^4(Q^2)\bigl(8.80{\xi^3\over 3!}-47.50\bar\beta_0
{\xi^2 \over 2!}+81.02\bar\beta^2_0\xi-42.30\bar\beta^3_0\bigr)\cr
&+\alpha_s^5(Q^2)\bigl(18.88{\xi^4\over 4!}-156.7\bar\beta_0
{\xi^3 \over 3!}+478.0\bar\beta^2_0{\xi^2\over 2!}-620.4\bar\beta^3_0\xi
+280.3\bar\beta_0^4\bigr)\cr
&+\alpha_s^6(Q^2)\bigl({4.95 \over \bar\beta_0}{\xi^6 \over 6!}
-44.15\bar{\xi^5\over 5!}
+159.9\bar\beta^1_0{\xi^4 \over 4!}-293.4\bar\beta^2_0{\xi^3\over 3!}
+269.7\bar\beta^3_0{\xi^2 \over 2!}-97.03\bar\beta_0^4\xi\bigr)\cr
&+\alpha_s^7(Q^2)\bigl({7.98 \over \bar\beta_0}{\xi^7 \over 7!}
-86.53{\xi^6 \over 6!}
+385.6\bar\beta_0{\xi^5\over 5!}
-899.9\bar\beta^2_0{\xi^4 \over 4!}+1153\bar\beta^3_0{\xi^3\over 3!}
-764.0\bar\beta^4_0{\xi^2 \over 2!}+203.8\bar\beta_0^5\xi\bigr)\cr
&+\alpha_s^8(Q^2)\bigl({17.15\over \bar\beta_0}{\xi^8 \over 8!}
-234.6\bar{\xi^7 \over 7!}+1354\beta_0{\xi^6 \over 6!}
-4263\bar\beta^2_0{\xi^5\over 5!}
+7882.9\bar\beta^3_0{\xi^4 \over 4!}-8519\bar\beta^4_0{\xi^3\over 3!}\cr
&+4962\bar\beta^5_0{\xi^2 \over 2!}-1199\bar\beta_0^6\xi\bigr)\cr
&+\alpha_s^9(Q^2)\bigl({3.97\over \bar\beta_0^2}{\xi^{10} \over 10!}
-{51.57 \over \bar\beta_0}
{\xi^9 \over 9!}+269.5{\xi^8 \over 8!}-647.5\bar\beta_0
{\xi^7 \over 7!}+258.8\beta^2_0{\xi^6 \over 6!}
+2451\bar\beta^3_0{\xi^5\over 5!}\cr
&-6962\bar\beta^4_0{\xi^4 \over 4!}+8473\bar\beta^5_0{\xi^3\over 3!}
-5145\bar\beta^6_0{\xi^2 \over 2!}+1259\bar\beta_0^7\xi\bigr)\biggl].\cr}}
and we have a power-suppressed contribution to $xP_{qg}(\alpha_s(Q^2),x)$ 
of the form
\eqn\modelpqg{\eqalign{{\alpha_s(Q^2)N_f  \over3\pi}\biggl[&
12.86\exp(-1.521t)\delta(1-x)+\exp(-t)\biggl( -14.31 
\biggl({\alpha_s(t)\over \alpha_s(4.5)}\biggr)^{2.695}
+36.297 \biggl({\alpha_s(t)\over \alpha_s(4.5)}\biggr)^{2.93}\xi\cr
&-41.14 \biggl({\alpha_s(t)\over \alpha_s(4.5)}\biggr)^{3.03}{\xi^2 \over 2!}
+25.34 \biggl({\alpha_s(t)\over \alpha_s(4.5)}\biggr)^{3.20}{\xi^3 \over 3!}
-9.096 \biggl({\alpha_s(t)\over \alpha_s(4.5)}\biggr)^{3.44}{\xi^4 \over 4!}\cr
&+1.85 \biggl({\alpha_s(t)\over \alpha_s(4.5)}\biggr)^{3.695}{\xi^5 \over 5!}
-0.1693 \biggl({\alpha_s(t)\over \alpha_s(4.5)}\biggr)^{3.80}{\xi^6 \over 6}
\biggr)\biggr].\cr}}

\eject

The full $xP_{qg}(\alpha_s(Q^2),x)$ is shown in \fig\quarkanom{a. The full 
leading $\ln(1/x)$ plus running coupling corrections coefficient function
$xP_{qg}(x,t)$ plotted as a function of $x$ for $t=6$ and $N_f=4$. Also shown 
are the contributions from the power-series and the power-suppressed part.
Note that the term $\propto \delta(1-x)$ in the power-series is 
replaced by the full ${\cal O}(\alpha_s(Q^2))$ contribution 
$1.5x(x^2+(1-x)^2)$,
and the terms $\propto \delta(1-x)$ in the power-suppressed part are absent.
b. $xP^{LO}_{qg}(x,t)$ plotted as a function of $x$ for $t=6$ and $N_f=4$.
Also shown is the coefficient function obtained from the naive 
LO BFKL calculation, and the contribution at ${\cal O}(\alpha_s(Q^2))$ 
alone.} a.
along with the two contributions above.
As with $xC_{L,g}(x,t)$ the $\delta(1-x)$ term 
at ${\cal O}(\alpha_s(Q^2))$ in the power-series is replaced by the 
full ${\cal O}(\alpha_s(Q^2))$ contribution 
which is $1.5x(x^2+(1-x)^2)$. Again the $\delta(1-x)$ term is 
missing from the power-suppressed part, and again this is insignificant.
In this case the power-suppressed part is tiny at $t=6$, though from the
large powers of $\alpha_s(Q^2)$ in \modelpqg\ we see that it grows very 
quickly at lower $Q^2$.  
In \quarkanom\ b. we show $xP_{qg}(x,t)$ along with the 
${\cal O}(\alpha_s(Q^2))$ contribution and with the naive LO BFKL result
in this factorization scheme, which again grows far more quickly than the 
resummed result.

These above results, along with the LO gluon splitting function, allow for a 
LO in $\ln(1/x)$ (with running coupling corrections) calculation and 
analysis of structure functions. In previous papers \LORSC\ I have strongly 
warned against the use of factorization-scheme dependent splitting functions 
and coefficient functions within the $\ln(1/x)$ expansion. It is still true
that it is always possible to make huge redefinitions of the unphysical 
parton distributions by factorization scheme changes at a given order (or even
at all orders) but the changes invoked by transfer between the commonly used 
schemes are diminished somewhat by the reduction of the size of the 
splitting functions and coefficient functions by the inclusion of the running 
coupling effects. It is also true that many of the changes invoked by 
factorization scheme changes are themselves due to running coupling effects,
and the resummation of these stabilizes the whole procedure a great deal. 
Hence, it is now possible to work in terms of these unphysical quantities if 
one wishes, without potential disasters, as long as the ordering of the 
expressions is done with particular care. Nevertheless, it is still very 
convenient in some ways to eliminate the partons completely and work directly
in terms of the structure functions $F_L(x,Q^2)$ and $F_2(x,Q^2)$ and the 
physical anomalous dimensions \physanom. In fact we can easily argue a case 
for improved stability. At LO the longitudinal coefficient function is 
positive and quite large at small $x$, and hence $F_L(x,Q^2)$ will be 
enhanced compared to the gluon at small $x$. At NLO the gluon evolution 
is smaller than at LO. Hence, evolving down from a given gluon
at very high $Q^2$ (where everything is simpler and more reliable) the 
NLO gluon will be larger 
at small $Q^2$ than the LO gluon. However, we expect the 
NLO corrections to $C_{L,g}(x,Q^2)$ to be negative, and thus counteract this 
increase in the NLO gluon in the calculation of $F_L(x,Q^2)$. Hence 
$F_L(x,Q^2)$ is (probably) a more stable perturbative quantity at small 
$x$ than $G(x,Q^2)$.   
       
\eject

The physical anomalous dimension which is 
most closely related to the gluon anomalous dimension is 
\eqn\physlong{\Gamma_{LL}(N,t)={d \ln({\cal F}_L(N,t))\over d t}.}  
Ignoring the mixing with the quark sector this is given in terms of the parton
related quantities by
\eqn\physfldef{\Gamma_{LL}(N,t)=\gamma_{gg}(N,t) +{d \ln(C_{L,g}(N,t)) 
\over d t},}
where I will use the convention of ignoring the overall power of 
$\alpha_s(Q^2)$ in the coefficient function which would just result
in a single contribution of $-\beta_0\alpha_s^2(Q^2)$ to \physfldef.
Using the LO $\gamma_{gg}(N,t)$ plus running coupling corrections, and 
similarly for $C_{L,g}(N,t)$ we see that the latter gives entirely running 
coupling corrections, and the total is the LO $\gamma_{gg}(N,t)$ with an
extended set of running coupling corrections. This total expression could
be calculated from the $\gamma_{gg}(N,t)$  and $C_{L,g}(N,t)$ already 
calculated, but part of the advantage in using physical anomalous dimensions 
is that it reduces the number of perturbative quantities governing the 
structure function evolution, i.e. the four splitting functions and four 
coefficient functions used to define $F_2(x,Q^2)$ and $F_L(x,Q^2)$ are 
reduced to four truly independent physical splitting functions. Hence, we 
notice that using \solrunf\ for the longitudinal structure function we can
calculate $\Gamma_{LL}(N,t)$ and $P_{LL}(x,t)$ directly, rather than from
\physfldef. Of course, the two definitions are equivalent, but the latter 
allows a single power-suppressed correction to be calculated rather than
having to combine those for $\gamma_{gg}(N,t)$ and $C_{L,g}(N,t)$
and thus the potential error is minimized. 
The asymptotic power-like behaviour for $P_{LL}^{LO}(x,t)$ is not identical
to that of $P^{LO}_{gg}(x,t)$ and is shown in \intercepts. The difference is
only relatively minor, but one sees that the power-like growth for 
$F_L(x,Q^2)$ is slightly smaller than for the gluon, and is also slightly less
$Q^2$-dependent. The result for the 
LO in $\ln(1/x)$ power-series solution $xP^{LO}_{LL}(\alpha_s(Q^2),x)$ 
is unfortunately a little less convergent
than the previous LO quantities, due to large coefficients generated in 
taking the derivative with respect to $t$ of the expression for 
${\cal F}_L(N,t)$ (or of $C_{L,g}(N,t)$). Hence, in order to obtain an 
expression which is sufficiently accurate for $Q^2 > 1\Gev^2$ and 
$x>0.00001$ we need to go to about $12$th order. This results in the explicit 
expression 
\eqn\losplitfl{\eqalign{x&P^{LO}_{LL}(\alpha_s(Q^2),x)=
\bar\alpha_s(Q^2)+0.333\alpha_s^2(Q^2)\bar\beta_0+\alpha_s^3(Q^2)
\bigl(-4.157\bar\beta_0\xi+4.266\bar\beta^2_0\bigr)\cr
&+\alpha_s^4(Q^2)\bigl(2.4{\xi^3 \over 3!} - 11.29\bar\beta_0{\xi^2 \over 2!}
+12.94\bar\beta^2_0\xi-4.02\bar
\beta_0^3\bigr)\cr
&+\alpha_s^5(Q^2)\bigl(0.121\bar\beta_0{\xi^3\over 3!}
+37.85\bar\beta^2_0
{\xi^2 \over 2!}-99.88\bar\beta^3_0\xi+61.92\bar\beta^4_0\bigr)\cr
&+\alpha_s^6(Q^2)\bigl(2{\xi^5 \over 5!} -75.14\bar\beta_0{\xi^4\over 4!}
+454.7\bar\beta^2_0
{\xi^3 \over 3!}-1034\bar\beta^3_0{\xi^2\over 2!}+1011\bar\beta^4_0\xi
-358.8\bar\beta_0^5\bigr)\cr
&+\alpha_s^7(Q^2)\bigl({1.92 \over \bar\beta_0}{\xi^7 \over 7!}-13.94
{\xi^6 \over 6!}+23.68\bar\beta_0{\xi^5\over 5!}
-39.48\bar\beta^2_0{\xi^4 \over 4!}+121.9\bar\beta^3_0{\xi^3\over 3!}
-155.2\bar\beta^4_0{\xi^2 \over 2!}+61.14\bar\beta_0^5\xi\bigr)\cr
&+\alpha_s^8(Q^2)\bigl(-16.91{\xi^7 \over 7!}+348.8\beta_0{\xi^6 \over 6!}
-2087\bar\beta^2_0{\xi^5\over 5!}
+5522\bar\beta^3_0{\xi^4 \over 4!}-7305\bar\beta^4_0{\xi^3\over 3!}
+4754\bar\beta^5_0{\xi^2 \over 2!}-1215\bar\beta_0^6\xi\bigr)\cr
&+\alpha_s^9(Q^2)\bigl({2.56 \over \bar\beta_0}{\xi^9 \over 9!}
-119.5{\xi^8 \over 8!}
+1173\bar\beta_0{\xi^7 \over 7!}-5052\beta^2_0{\xi^6 \over 6!}
+12044\bar\beta^3_0{\xi^5\over 5!}
-17444\bar\beta^4_0{\xi^4 \over 4!}\cr
&+15528\bar\beta^5_0{\xi^3\over 3!}
-7859\bar\beta^6_0{\xi^2 \over 2!}+1728\bar\beta_0^7\xi\bigr)\cr
&+\alpha_s^{10}(Q^2)\bigl({1.536 \over \bar\beta_0^2}{\xi^{11} \over 11!}-
{16.73 \over \bar\beta_0}{\xi^{10} \over 10!}+83.37
{\xi^9 \over 9!}-492.2\bar\beta_0{\xi^8 \over 8!}+1559\bar\beta_0^2
{\xi^7 \over 7!}+2043\beta^3_0{\xi^6 \over 6!}
-24427\bar\beta^4_0{\xi^5\over 5!}\cr
&+61280\bar\beta^5_0{\xi^4 \over 4!}-72753\bar\beta^6_0{\xi^3\over 3!}
+42720\bar\beta^7_0{\xi^2 \over 2!}-9998\bar\beta_0^8\xi\bigr)\cr
&+\alpha_s^{11}(Q^2)\bigl(-{18.53 \over \bar\beta_0}{\xi^{11} \over 11!}+
444.0 {\xi^{10} \over 10!}-2988\bar\beta_0
{\xi^9 \over 9!}+5290\bar\beta^2_0{\xi^8 \over 8!}-22253\bar\beta_0^3
{\xi^7 \over 7!}-135896\beta^4_0{\xi^6 \over 6!}\cr
&+321404\bar\beta^5_0{\xi^5\over 5!}
-425485\bar\beta^6_0{\xi^4 \over 4!}+330620\bar\beta^7_0{\xi^3\over 3!}
-141370\bar\beta^8_0{\xi^2 \over 2!}+25747\bar\beta_0^9\xi\bigr)\cr
&+\alpha_s^{12}(Q^2)\bigl({2.82 \over \bar\beta_0^2}{\xi^{13} \over 13!}
-{141.7 \over \bar\beta_0}{\xi^{12} \over 12!}+1757{\xi^{11} \over 11!}-
10347\bar\beta_0{\xi^{10} \over 10!}+39345\bar\beta^2_0
{\xi^9 \over 9!}-119096\bar\beta^3_0{\xi^8 \over 8!}\cr
&+295058\bar\beta_0^4{\xi^7 \over 7!}-538834\beta^5_0{\xi^6 \over 6!}
+658339\bar\beta^6_0{\xi^5\over 5!}
-499685\bar\beta^7_0{\xi^4 \over 4!}+211914\bar\beta^8_0{\xi^3\over 3!}
-38311\bar\beta^9_0{\xi^2 \over 2!}\bigr).\cr}}
The power-suppressed correction is calculated in the usual manner 
and is of the form  
\eqn\modellofl{\eqalign{36.57&\exp(-1.75t)\delta(1-x)+
\exp(-t)\biggl( 4.626 
\biggl({\alpha_s(t)\over \alpha_s(4.5)}\biggr)^{-2.78}
-37.84 \biggl({\alpha_s(t)\over \alpha_s(4.5)}\biggr)^{-0.58}\xi\cr
&+67.22 \biggl({\alpha_s(t)\over \alpha_s(4.5)}\biggr)^{0}{\xi^2 \over 2!}
-51.30 \biggl({\alpha_s(t)\over \alpha_s(4.5)}\biggr)^{0.17}{\xi^3 \over 3!}
+18.82 \biggl({\alpha_s(t)\over \alpha_s(4.5)}\biggr)^{-0.01}
{\xi^4 \over 4!}\cr
&-3.136 \biggl({\alpha_s(t)\over \alpha_s(4.5)}\biggr)^{-0.69}{\xi^5 \over 5!}
+0.1706 \biggl({\alpha_s(t)\over \alpha_s(4.5)}\biggr)^{-2.27}{\xi^6 \over 6}
\biggr).\cr}}
The anomalous dimension $\Gamma^{LO}_{LL}(N,t)$ is plotted in  
\fig\fphysanom{a. The anomalous dimensions for the gluon structure function 
at LO and for ${\cal F}_{L}(N,t)$ at LO plotted as functions of $N$ for $t=6$. 
Also shown is the ${\cal O}(\alpha_s(Q^2))$ contribution common to each. 
b. The anomalous dimensions for ${\cal F}_L(N,t)$ at LO and ``NLO'' plotted as 
functions of $N$ for $t=6$.}.a. Until $N$ is very small it is similar to 
$\gamma^{LO}_{gg}(N,t)$ and both are close to the common $\alpha_s(Q^2)/N$ 
contribution, though $\Gamma^{LO}_{LL}(N,t)$ is a little larger at large $N$. 
However, at lower $N$ $\Gamma^{LO}_{LL}(N,t)$ dips below the 
others before \vadjust{\eject} eventually rising above $\alpha_s(Q^2)/N$ but 
staying below $\gamma^{LO}_{gg}(N,t)$.    
Clearly the effect of the additional coefficient function, and
hence additional running coupling corrections, is to make
$\Gamma_{LL}(N,t)$ dip significantly below the ${\cal O}(\alpha_s(Q^2))$
contribution $\bar\alpha_s(Q^2)/N$
for a region and to reduce the value of the intercept compared to the
gluon structure function. The effective
splitting function $xP^{LO}_{LL}(x,t)$ is shown in 
\fig\fphysspli{a. The splitting functions $xP^{LO}_{LL}(x)$ and its 
power series and power-suppressed contributions
plotted as a function of $x$ for
$t=6$. b. The splitting function $xP^{LO}_{LL}(x)$ 
plotted as a function of $x$ for
$t=6$ ($Q^2 \sim 6\Gev^2$). Also shown is the ${\cal O}(\alpha_s(Q^2))$ 
contribution $\bar\alpha_s(Q^2)$, the gluon splitting function
$P^{LO}_{gg}(x)$ and the naive LO BFKL splitting function with coupling
$\alpha_s(Q^2)$.}. In \fphysspli.a we see that the power-suppressed 
contribution is 
larger for $xP^{LO}_{LL}(x,t)$ than it was for $xP^{LO}_{gg}(x,t)$.
In \fphysspli.b we see the outcome of the comparison of the anomalous 
dimensions for 
$F_L$ and the gluon. $xP^{LO}_{LL}(x,t)$ starts a little higher at $x=0$ and 
the dip below the ${\cal O}(\alpha_s(Q^2))$ part is considerably more
pronounced than for $xP^{LO}_{gg}(x,t)$. Also, going to $x \sim 10^{-5}$, 
we see that the splitting function dips again, showing that 
the subleading poles in the
anomalous dimension may have large residues compared to the leading
pole, and that the increase in $xP^{LO}_{LL}(x)$ with decreasing $x$ is not
monotonic. This corresponds to the significant fall of $\Gamma_{LL}(N,t)$
below $\bar\alpha_s(Q^2)/N$ at $N \sim 0.6$. The eventual rise of 
$\Gamma_{LL}(N,t)$ guarantees that the splitting function will eventually 
rise again with the calculated intercept, i.e. like $x^{-0.23}$, at even 
smaller $x$.
However, for $t=6$ this asymptotic power behaviour does not set in until 
$x< 10^{-10}$ and in the region of $x\sim 10^{-7}$ $xP^{LO}_{LL}(x)$ even 
becomes slightly negative.
For higher $t$ even smaller $x$ is required, e.g. $t=8$ 
($Q^2 \approx 30 \Gev^2$) needs $x$ to become as low as $10^{-13}$
before the power-like behaviour sets in, though the size of the dip 
before this is smaller than for $t=6$. 
This illustrates very clearly that as far as phenomenology 
at HERA, or any foreseeable collider, is concerned the value of the 
intercept for the anomalous dimension is simply not relevant to the evolution 
of structure functions. Indeed, it is very possible that before the 
power-like behaviour has set it unitarization effects have already become
important. For collider phenomenology it is the splitting functions over the 
relevant $x$ and $Q^2$ range which one needs, and this requires the sort of 
detailed calculation in this paper.   

One can follow exactly the same procedure for the other important physical
anomalous dimension defined by
\eqn\physanomtwo{{\partial {\cal F}_2(N,Q^2)\over \partial \ln Q^2} = 
\Gamma_{2L}(Q^2,N) {\cal F}_L(N,Q^2),}
simply by using the LO expressions for ${d {\cal F}_2(N,Q^2)\over d \ln Q^2}$
and ${\cal F}_L(N,t)$.
The power-like behaviour as $x \to 0$ is governed by the
poles in ${\cal F}(N,t)$ as in the previous case, so the position of the 
intercepts is identical.  
The power-series expression requires the first 10 powers in order to be valid 
over the required range of $x$ and $Q^2$, so I write it as
\eqn\losplitftwo{\eqalign{x&P^{LO}_{2L}(\alpha_s(Q^2),x)= 
\biggl[\delta(1-x)+2.5\alpha_s(Q^2)+\alpha_s^2(Q^2)
\bigl(\xi-0.167\bar\beta_0\bigr)\cr
&+\alpha_s^3(Q^2)\bigl({\xi^2 \over 2!}-12.72\bar\beta_0\xi+12.0\bar
\beta_0^2\bigr)+\alpha_s^4(Q^2)\bigl(7.007{\xi^3\over 3!}-41.41\bar\beta_0
{\xi^2 \over 2!}+61.42\bar\beta^2_0\xi-26.82\bar\beta^3_0\bigr)\cr
&+\alpha_s^5(Q^2)\bigl(5.78{\xi^4\over 4!}-52.95\bar\beta_0
{\xi^3 \over 3!}+253.0\bar\beta^2_0{\xi^2\over 2!}-444.1\bar\beta^3_0\xi
+238.32\bar\beta_0^4\bigr)\cr
&+\alpha_s^6(Q^2)\bigl({5.80 \over \bar\beta_0}{\xi^6 \over 6!}
-87.30{\xi^5\over 5!}
+409.7\bar\beta_0{\xi^4 \over 4!}-773.3\bar\beta^2_0{\xi^3\over 3!}
+621.7\bar\beta^3_0{\xi^2 \over 2!}-176.6\bar\beta_0^4\xi\bigr)\cr
&+\alpha_s^7(Q^2)\bigl({9.348 \over \bar\beta_0}{\xi^7 \over 7!}
-117.8{\xi^6 \over 6!}+591.4\bar\beta_0{\xi^5\over 5!}
-1701\bar\beta^2_0{\xi^4 \over 4!}+2792\bar\beta^3_0{\xi^3\over 3!}
-2315\bar\beta^4_0{\xi^2 \over 2!}+741.5\bar\beta_0^5\xi\bigr)\cr
&+\alpha_s^8(Q^2)\bigl({-4.380 \over \bar\beta_0}{\xi^8 \over 8!}
+52.41{\xi^7 \over 7!}+70.68\beta_0{\xi^6 \over 6!}
-1954\bar\beta^2_0{\xi^5\over 5!}
+6623\bar\beta^3_0{\xi^4 \over 4!}-9500\bar\beta^4_0{\xi^3\over 3!}\cr
&+6307\bar\beta^5_0{\xi^2 \over 2!}-1596\bar\beta_0^6\xi\bigr)\cr
&+\alpha_s^9(Q^2)\bigl({4.64 \over \bar\beta_0^2}{\xi^{10} \over 10!}
-{95.79\over \bar\beta_0}{\xi^9 \over 9!}+657.3{\xi^8 \over 8!}
-1775.7\bar\beta_0{\xi^7 \over 7!}+450.0\beta^2_0{\xi^6 \over 6!}
+9410\bar\beta^3_0{\xi^5\over 5!}\cr
&-26327\bar\beta^4_0{\xi^4 \over 4!}+33805\bar\beta^5_0{\xi^3\over 3!}
-21743\bar\beta^6_0{\xi^2 \over 2!}+5614\bar\beta_0^8\xi\bigr)\cr
&+\alpha_s^{10}(Q^2)\bigl({7.478 \over \bar\beta_0^2}{\xi^{11} \over 11!}
-{115.7\over \bar\beta_0}{\xi^{10} \over 10!}+765.7{\xi^9 \over 9!}
-3293\bar\beta_0{\xi^8 \over 8!}+8687\beta^2_0{\xi^7 \over 7!}
-5511\bar\beta^3_0{\xi^6\over 6!}\cr
&-35038\bar\beta^4_0{\xi^5 \over 5!}+104536\bar\beta^5_0{\xi^4\over 4!}
-127198\bar\beta^6_0{\xi^3 \over 3!}+74260\bar\beta_0^8{\xi^2\over 2!}
-17101\bar\beta_0^9\xi\bigr)\biggl].\cr}}
The power-suppressed correction is   
\eqn\modellofl{\eqalign{(3.558&+0.4216t-0.1542t^2)\exp(-t)\delta(1-x)+
\exp(-t)\biggl(72.17 
\biggl({\alpha_s(t)\over \alpha_s(4.5)}\biggr)^{0.93}x\cr
&-78.03 \biggl({\alpha_s(t)\over \alpha_s(4.5)}\biggr)^{1.66}
+56.85 \biggl({\alpha_s(t)\over \alpha_s(4.5)}\biggr)^{2.66}\xi 
-24.16 \biggl({\alpha_s(t)\over \alpha_s(4.5)}\biggr)^{0.58}{\xi^2 \over 2!}\cr
&+13.50 \biggl({\alpha_s(t)\over \alpha_s(4.5)}\biggr)^{2.50}{\xi^3 \over 3!}
-10.32 \biggl({\alpha_s(t)\over \alpha_s(4.5)}\biggr)^{-2.27}{\xi^4 \over 4!}
+3.918 \biggl({\alpha_s(t)\over \alpha_s(4.5)}\biggr)^{-1.584}
{\xi^5 \over 5!}\cr
&-0.5141 \biggl({\alpha_s(t)\over \alpha_s(4.5)}\biggr)^{-1.05}
{\xi^6 \over 6!}
\biggr),\cr}}
where it is necessary to introduce a term $\propto x$ in order to get a good 
description at high $N$. 
The full $xP_{2L}(x,t)$ is shown in \fig\ftwosplit{a. The full 
leading $\ln(1/x)$ plus running coupling corrections physical splitting 
function $xP_{2L}(x,t)$ plotted as a function of $x$ for $t=6$. Also shown 
are the contributions from the power-series and the power-suppressed part.
In the power-series the part $\propto \delta(1-x)$ is replaced by
$1.5x(x^2+(1-x)^2)$ while in the power-suppressed part this contribution is 
simply absent. b. The physical splitting function
$xP_{2L}(x,t)$ plotted as a function of $x$ for $t=6$ along with 
the physical splitting function obtained from the naive LO BFKL calculation
with coupling $\alpha_s(Q^2)$ and the zeroth order contribution.}.
a along with the two contributions above. The $\delta(1-x)$ term 
is replaced in the power series by the $x$-dependence in the 
${\cal O}(\alpha_s(Q^2))$ quark-gluon splitting function,
i.e. $x(x^2+(1-x)^2)$, normalized by 1.5 to give 
the correct $N \to 0$ limit. This corresponds to a slight modification of the 
usual physical anomalous dimension in terms of the ${\cal O}(\alpha_s(Q^2))$
longitudinal gluon coefficient function, but may be viewed as an analytic 
function with the correct $N \to 0$ limit which aids presentation 
here.\foot{This modification to the physical splitting function will be 
discussed in a future paper.} The $\delta(1-x)$ terms in the 
power-suppressed contribution are very small, and are simply left out. 
In \ftwosplit.b we see $xP_{2L}(x,t)$ plotted as a function of $x$ along 
with the
naive LO BFKL calculation with coupling $\alpha_S(Q^2)$, and in order to 
illustrate the contribution of the higher-order terms, 
also the zeroth order contribution $1.5x(x^2+(1-x)^2)$.
As with $P_{LL}(x,t)$ one can see that $P_{2L}(x,t)$ has a dip at 
small $x$ before the eventual power-like growth sets 
in, again only for $x < 10^{10}$, and as with all calculated quantities 
the running coupling corrections
severely diminish the strength of the small $x$ growth. 

\bigskip

We can also try to investigate the effect of NLO corrections on physical 
quantities. In terms of partons the only known NLO correction is that to
the gluon splitting function -- there is simply no information on the NLO
corrections to coefficient functions or the quark splitting functions. In terms
of the physical anomalous dimensions, similarly there is no real information
for $\Gamma_{2L}(N,t)$, but the situation is better for $\Gamma_{LL}(N,t)$.
Let us look at the expression in terms of the partonic quantities \physfldef,
for the moment in the leading $\ln(1/x)$ expansion without resummed 
running coupling corrections.
At LO in $1/N$, $\Gamma_{LL}^{LO}(N,t)$ is equal to $\gamma^{LO}_{gg}(N,t)$ 
since the differentiation of the $\log$ of the coefficient function with 
respect to $t$ automatically introduces an extra factor of 
$\beta_0\alpha_s(Q^2)$. At NLO in $1/N$ $\Gamma_{LL}^{NLO}(N,t)$ picks up
a contribution from $\gamma^{NLO}_{gg}(N,t)$ which is (largely) independent
of the running coupling, and the contribution from the derivative of the LO
coefficient function, which is entirely running coupling dependent. 
Hence, by knowing $\gamma^{NLO}_{gg}(N,t)$ we know the whole of 
$\Gamma_{LL}^{NLO}(N,t)$ before resuming running coupling corrections. 
Hence, we might hope that using an expression of the form \solrunf, but 
corrected in the way described in the previous section for the NLO 
corrections to the kernel, we might calculate the full NLO, running coupling
corrected BFKL expression for $\Gamma_{LL}(N,t)$. Unfortunately, this 
is not quite the case. This can be appreciated by again using \physfldef. 
When solving this NLO corrected expression for ${\cal F}_{E,L}(N,t)$ one
includes all the running coupling corrections to $\gamma^{NLO}_{gg}(N,t)$
just by the manner of solving the equation. But without knowing the NLO
correction to the coefficient function one misses a whole series of terms of
the form $\alpha_s(Q^2)(\bar\beta_0\alpha_s(Q^2))^nf(\bar\alpha_s(Q^2)/N)$
which would come form the ${d \ln(C_{L,g}(N,t)) \over d t}$ term.\foot{Some
of these are automatically generated by using the NLO kernel in our solution,
but the full set requires also the NLO correction to the hard scattering
cross-section which will lead to NLO corrections to $h_{L,g}(\gamma)$.}  
Thus, we do not yet know the full running coupling corrections to the 
NLO contribution to $\Gamma_{LL}(N,t)$.   

I will proceed to calculate the ``NLO'' corrected $\Gamma_{LL}(N,t)$ on the 
assumption that since the resummation of the running coupling corrections
stabilizes the perturbative expansion the missing running coupling 
corrections will not lead to anything other than minor corrections. 
It is straightforward to generalize the results of section 4 to the case 
of the physical quantity. Essentially we just replace \intnloresult\ by
\eqn\intnloresultfl{\eqalign{{\cal F}^1_{E,L}&(N, t) \propto  
\,t^{-c_l/\bar\beta_0}\int_{C}\!\biggl[ 
\gamma^{-1/(\bar\beta_0N)-1}h_{L,g}(\gamma)f^{\beta_0}(\gamma)
\exp\biggl(\gamma t
-{1\over (\bar\beta_0N)}\sum_{n=1}^{\infty}a_n\gamma^{2n+1}\biggr)
\biggl(1+\sum_{m=0}
(1/\bar\beta_0) c_m \gamma^m\biggr) \cr
&\hskip -0.35in -{c_l \over \bar\beta_0}\biggl[\ln(\gamma t)
-\Psi\biggl(-{1\over \bar\beta_0N}\biggr)+\pi\cot\biggl({\pi\over
\bar\beta_0N}\biggr)\biggl] 
\gamma^{-(1/(\bar\beta_0N))-1}h_{L,g}(\gamma)f^{\beta_0}(\gamma)
\exp\biggl(\gamma t
-{1\over \bar\beta_0N}\sum_{n=1}^{\infty}a_n\gamma^{2n+1}\biggr)
\biggr]d \gamma,\cr}}  
where we are currently missing a further term of the form
\eqn\miss{-Nt^{-c_l/\bar\beta_0}\int_{C} 
\gamma^{-1/(\bar\beta_0N)-1}\delta h_{L,g}(\gamma,\bar\beta_0N)
f^{\beta_0}(\gamma)\exp\biggl(\gamma t
-{1\over (\bar\beta_0N)}\sum_{n=1}^{\infty}a_n\gamma^{2n+1}\biggr).}
Using \intnloresultfl\ we can calculate both the power-series and 
power-suppressed NLO contributions to $\Gamma_{LL}(N,t)$ and hence 
$P_{LL}(x,t)$. The LO +``NLO'' values of the intercept for the asymptotic 
power-like behaviour are shown in \intercepts. These lie very slightly 
below the LO+NLO intercepts for the gluon, and hint at perhaps a more
rapid convergence for the physical $F_L$ than for the gluon. However, we 
would expect the missing contributions to lower the intercept a little more.
The ``NLO'' corrected anomalous dimension $\Gamma^{LO+NLO}_{LL}(N,t)$ 
is shown as a 
function of $N$ for $t=6$ in \fphysanom.b. It is very similar to that at LO
until very low $N$ where the difference in the leading intercept starts 
to become apparent.   

As for the NLO correction to $xP_{gg}(x,t)$ the power-series
is not very convergent an to work all the way down to $Q^2=1\Gev^2$ and
$x=0.00001$ we again need the first 20 or so terms. Hence the 
power-series contribution is 
\eqn\nlosplitfl{xP^{NLO}_{LL}(\alpha_s(Q^2),x) = \bar\alpha_s(Q^2)
\sum_{n=1}^{19}\sum_{m=0}^{m_{max}} \bar\alpha^n_s(Q^2)\biggl(K_{nm}{\xi^m 
\bar\beta_0^{n-m-1}\over m!}+K_{n\delta}\bar\beta_0^{n}\delta(1-x)\biggr),} 
where the coefficients are listed in table 2. The power-suppressed 
contribution is 
\eqn\modelnlofl{\eqalign{(-0.183&\exp(-0.51t)\delta(1-x)+
\exp(-t)\biggl(31.90 \biggl({\alpha_s(t)\over \alpha_s(4.5)}\biggr)^{-0.274}
-80.22 \biggl({\alpha_s(t)\over \alpha_s(4.5)}\biggr)^{0.346}\xi\cr
&+56.67 \biggl({\alpha_s(t)\over \alpha_s(4.5)}\biggr)^{0.60}{\xi^2 \over 2!}
+9.017 \biggl({\alpha_s(t)\over \alpha_s(4.5)}\biggr)^{3.15}{\xi^3 \over 3!}
-25.925 \biggl({\alpha_s(t)\over \alpha_s(4.5)}\biggr)^{1.715}
{\xi^4 \over 4!}\cr
&+10.28 \biggl({\alpha_s(t)\over \alpha_s(4.5)}\biggr)^{1.875}{\xi^5 \over 5!}
-1.298 \biggl({\alpha_s(t)\over \alpha_s(4.5)}\biggr)^{2.09}{\xi^6 \over 6}
\biggr).\cr}}
The NLO correction to the splitting function $xP^{NLO}_{LL}(x,t)$ is shown, 
minus the contributions $\propto \delta(1-x)$, in 
\fig\splitflnlo{a. The splitting functions $xP^{NLO}_{LL}(x)$ and its 
power series and power-suppressed contributions
plotted as functions of $x$ for
$t=6$. b. The splitting function $xP^{LO+NLO}_{LL}(x)$ 
plotted as a function of $x$ for
$t=6$ ($Q^2 \sim 6\Gev^2$). Also shown is the ${\cal O}(\alpha_s(Q^2))$ 
contribution $\bar\alpha_s(Q^2)$, and the LO contribution 
$xP^{LO}_{LL}(x)$.}.a. Clearly there is a very large cancellation between
the power-series and power-suppressed contributions resulting in a relatively
small total NLO correction. We can see that unlike for the the gluon this 
NLO correction 
is actually positive in some regions of $x$, rather than everywhere negative. 
We also see from \splitflnlo.b that the NLO splitting function is 
quite similar to the LO splitting function over the whole $x$ range. 

However, as with the gluon, 
the real test of perturbative stability is the evolution of the structure 
function itself. This is shown in \fig\evolutionfl{The values
of $d F_L(x,Q^2)/d\ln Q^2$, for $F_L(x,Q^2)=x^{-0.2}(1-x)^6$, due to the 
LO splitting functions $P^{LO}_{LL}(x)$ and the LO+NLO splitting
function $P^{LO+NLO}_{LL}(x)$, 
plotted as a function of $x$ for
$t=6$ ($Q^2 \sim 6\Gev^2$). Also shown is the evolution due to the
${\cal O}(\alpha_s(Q^2))$ contribution $P(x)=\bar\alpha_s(Q^2)/x$.} 
where the evolution of a model
for the structure function $F_L(x,Q^2)$, i.e. $(1-x)^6x^{-0.2}$, is
shown both for the LO running coupling splitting function, and for the
``NLO'' corrected one (all $\delta(1-x)$ contributions other that at
first order in $\alpha_s(Q^2)$ are included). Also shown is the evolution 
due just to the double-leading-log term $P(x)=\bar\alpha_s(Q^2)/x$.
Compared to the evolution of the gluon shown in the previous section we see 
that the additional running coupling contributions due to the
$t$-derivative of the coefficient function have slowed the LO evolution
below that of the double-leading-log result over the whole range of $x$
(except very high $x$), and this will only cease to be true at very small 
$x$ indeed, when the power-like growth of the physical splitting function 
finally sets in. In this case, however, the difference between LO and
LO+``NLO'' is much smaller than for the gluon, and the perturbative expansion
seems very stable indeed. 
As with the NLO corrections to the intercepts
this might be a sign that the expansion 
converges more quickly for the physical structure functions than for the 
unphysical gluon structure function. However, as a note of caution, the 
missing contributions at NLO are likely to be negative in general, and this
difference between LO and NLO evolution will probably be increased a little.
In fact it is desirable for these missing contributions to be non-negligible.
Whilst if we decrease $t$ to $4.5$, i.e. $Q^2 \sim 1\Gev^2$, at
NLO everything remains relatively stable for the gluon, the physical 
splitting function 
starts to develop extreme behaviour at this low scale -- the minimum at
$x\sim 0.01$ becomes much lower and the peak at $x\sim 0.0001$ becomes very 
much higher. This trend is illustrated in \rsvariation\ b. which shows the 
renormalization scale dependence of $P^{LO+NLO}_{LL}(x,t)$
for $t=6$. Clearly there is 
very good stability for an increase in scale, but it is not so good 
for a decrease in scale (though since the splitting function oscillates,
the variation washes out to a large extent when evolution is calculated). 
There is very good stability in both directions if one examines the 
variation for a slightly higher $t$, say $t=8$ ($Q^2 \sim 30\Gev^2$).  
This instability in the physical splitting function  
results in instabilities in the evolution at 
$t=4.5$, even though it appeared to be very stable at $t=6$. Hopefully, the
inclusion of the missing terms will help stabilize this evolution, though 
it may simply be a sign that at this low $Q^2$ some non-perturbative 
contribution is becoming essential.

\newsec{Conclusions.}

In this paper I have shown that it is possible to obtain analytic solutions 
to the LO running coupling BFKL equation for the 
$Q^2$-dependent parts of the gluon structure 
function and for the real physical structure functions $F_2(x,Q^2)$ and 
$F_L(x,Q^2)$. This results in a resummation of the leading $\ln(1/x)$ terms at
each power in $\alpha_s(Q^2)$ and also of the leading powers in $\beta_0$ at
each power of $\alpha_s(Q^2)$ and $\ln(1/x)$.
However, the $Q_0^2$-dependent gluon input is plagued by contamination from
infrared nonperturbative physics, and has an inherent ambiguity of 
${\cal O}(\Lambda^2/Q_0^2)$. The analytic expressions may be expressed 
in form of a power-series in $\alpha_s(Q^2)$. 
In practice the main features of 
the solution are almost completely determined by only the first
handful ($\sim 5$) of terms in the expansion, in
complete contrast with the case of fixed coupling, where an all orders
summation is needed. In fact the perturbative series for the structure 
functions is not convergent, and the analytic expression is most accurately
obtained by this truncation. The small remainder, which roughly speaking
is suppressed by powers of $(\Lambda^2/Q^2)$, may be calculated from the 
difference between a numerical solution with the analytic solution,
and then modelled by an analytic expression of $Q^2$ and $N$, which may 
easily be transformed to $x$-space. 
There are two points to note here. Firstly, this power-suppressed condition 
is both well-defined and is nothing to do with higher twist operators. 
Even though there are infrared  (and ultraviolet) renormalons  in the 
untruncated perturbative expansion, they only appear due to the impossibility
of expressing the $Q^2$-dependent part of the the structure functions as 
a power series in $\alpha_s(Q^2)$, not because of  some inherent ambiguity at 
leading twist as is often the case with renormalons. Hence, they are 
circumvented completely by this manner of calculation. Secondly, this 
procedure of an analytic calculation as a truncated power series 
plus a numerical calculation of the 
power-suppressed part, which is then modelled,
seems to allow for the most accurate determination of $x$-space
quantities. Transformation of numerical moment space expressions to $x$-space
are subject to errors, and the magnification of the power-suppressed 
contributions in $x$-space, compared to moment space, seen in this paper 
highlights the potential effect of small errors in moment space  when 
ultimately working in $x$-space. Hence, obtaining as accurate an analytic 
moment space expression as possible is vital in ultimately obtaining good
accuracy for splitting functions and the evolution of structure functions.

It is also demonstrated that there are well-defined, calculable higher-twist 
contributions due to the transverse degrees of freedom of the 
two-gluon operator. However, both the normalization and 
splitting functions of these genuinely higher twist operators decrease 
quickly as $x\to 0$ (roughly like $x^{0.5}\cos(0.5\ln(1/x))$) when the small 
$x$ resummation is performed. Unlike leading twist, this is largely 
insensitive 
to the running coupling corrections. This result is only apparent from 
resummation, and a fixed (small) order in $\alpha_s(Q^2)$, particularly 
first order
only, gives very misleading results. Hence, this one form of higher 
twist does not lead to any sizable correction at all at 
small $x$ and $Q^2$. It is possible that this unambiguous, small-$x$ vanishing
higher-twist contribution to the two-gluon operator is responsible for the 
absence of a genuine ambiguity in the leading twist anomalous dimensions.  
However, I note that this paper has nothing to say about the 
size of shadowing corrections coming from four gluon operators,
except to point out that the double-leading-log type calculations often
performed are likely to lead to huge overestimations. Neither does it 
consider the power-suppressed corrections due to nonperturbative effects
which mix with higher twist, leading to mixing with leading twist,
and may well be important at extremely small $x$ \ciafsalam\salamlo.

The calculated expressions for leading twist structure functions may be 
used to produce LO expressions for the splitting functions and coefficient 
functions for 
physical processes, and also the physical splitting functions which allow 
one to work directly in terms of physical quantities. 
My results prove that the effect of the running of the coupling is to
weaken the asymptotic power-like growth of the splitting
functions severely compared to the naive BFKL results, and even to lower the
splitting function below the $\alpha_s(Q^2)/x$ contribution for
$0.001\gsim x \gsim 0.2$. 
It is also noted that the asymptotic behaviour of the form $x^{-\lambda}$
is often not approached even approximately until $x<<0.00001$, with the 
required $x$ decreasing with increasing $Q^2$, and is 
therefore by no means a good indicator of physics at present or future 
colliders. In fact it is very likely that unitarization will stop this 
true power-like behaviour ever being seen.   
Rather than the intercept, the detailed expressions for the splitting 
functions and coefficient functions are needed in order to really calculate
the evolution at realistic values of $x$.

The procedure  can also be extended to NLO without any real modification, 
though there is some ambiguity in precisely what the best definition of NLO
is.\foot{The power-series expressions also become very complicated at NLO. It 
will probably ultimately be more convenient to model them accurately with some 
simpler function of $x$ and $t$ similar to the manner in which the 
power-suppressed contributions are treated at present.} 
The choice is made so that the expressions for the structure functions
are genuinely only a single power of $\alpha_s(Q^2)$ down on LO, up to
$\beta_0\alpha_s(Q^2)$ corrections, but in $\gamma(N,t) =d\ln(G(N,t))/dt$
the full NLO expression for $(G(N,t))^{-1}$ is used, rather than truncating 
its expansion at NLO, and hence the full NLO correction to the 
intercept is obtained. This has little effect until extremely small $x$.   
Unlike leading
$\ln(1/x)$ calculations without resummation of running coupling effects the 
NLO correction to the gluon splitting function here is moderate, 
both for the value of the intercept and for the exact size of the splitting 
function and the evolution of gluon structure  
function for $x>10^{-5}$. Hence, this running coupling resummation does 
a great deal to 
stabilize the perturbative series. Unfortunately it is not yet possible to 
calculate the complete NLO correction to any real physical quantity, 
though one may 
come close for $P_{LL}(x,t)$, the splitting function governing the 
evolution of the longitudinal structure function in terms of itself,
which is very similar to $P_{gg}(x,t)$. 
In this case only a subset of the running coupling corrections to the 
NLO in $\ln(1/x)$ part are still unknown. For $F_L$ the stability of 
the perturbative series looks even better than for the gluon as long as 
$Q^2\gsim 4\Gev^2$, but begins to deteriorate below this, perhaps due to
the missing corrections. 

Let me also comment briefly on other methods which attempt to incorporate 
the NLO corrections (and beyond) to the BFKL equation.  
Firstly I note that my previous conjecture that the effect of 
the running coupling in the BFKL equation could be accounted for using an
$x$-dependent scale for the coupling \xscale, resulting in falling coupling
for decreasing $x$, turns out to be essentially correct so long as 
the change in the scale of the coupling is moderate compared to the 
scale itself, though it fails if this condition is not satisfied. 
In practice this condition is identical to that specifying that
diffusion in the fixed coupling BFKL equation is not too large, and 
therefore that the virtualities sampled in the running coupling
equation are not too far away
from $Q^2$. This results in the requirement that $t^3\gsim 20
\ln(1/x)$ \ref\mueller{A.H. Mueller, \PL \vyp{B396}{1997}{251}.}. This is 
true for all but the lowest $x$ and $Q^2$ at HERA. I also note that my 
approach is completely consistent with that in \ciafpcol\ciafsalam, 
with both being 
built upon the running coupling BFKL equation essentially introduced long
ago \glr\liprun\jan\janpcoll\ and generalized beyond LO in \ciafpcol. 
The differences to this approach 
are that I ignore the collinear resummation which is a central theme in this
work, since as I stress it is an unnecessary complication in the 
calculation of splitting functions  - the running coupling effects being the 
most important and stabilizing the calculation themselves; that I concentrate 
on solving very accurately and precisely for the $Q^2$-dependent part of 
the gluon and structure functions, obtaining splitting functions over 
the range of $x$ and $Q^2$ relevant for a phenomenological treatment; 
and that I also ignore the complication of a real regularization 
of the coupling in the 
infrared region (this latter point is also considered in \ref\jeff{J.R. 
Forshaw, D.A. Ross and A. Sabio-Vera, \PL \vyp{498}{2001}{149}.}). 
Hence, I obtain detailed accurate results for all splitting functions and 
coefficient functions in 
closed form, but ignore contributions considered in these papers which  
are necessary if investigating single-scale processes and/or 
potential nonperturbative effects (which may be important for splitting 
functions at low $Q^2$ and very small $x$ \salamlo).
There is less similarity with other approaches. Even though that in 
\ref\brolip{S.J. Brodsky, V.S. Fadin, V.T. Kim, L.N. Lipatov, G.B. Pivovarov,
{\it JETP Lett.} \vyp{70}{1999}{155}.} claims to in some sense be dealing
with the scale appropriate for the coupling in this problem, it has no 
overlap with the approach in this paper, and comments on this approach can be 
found in \xscale. Also there is no connection with the approach in 
\ref\altball{G. Altarelli, R.D. Ball and S. Forte, \NP \vyp{B575}{313}{2000}
\semi G. Altarelli, R.D. Ball and S. Forte, \NP \vyp{B599}{2001}{383}.} which
adopts a phenomenological approach to resummation beyond fixed orders 
in $\ln(1/x)$ in terms of the asymptotic power-like behaviour, which is a free
parameter, and which consequently 
loses true predictive power for the evolution at small $x$. Finally, there 
also seems to be no overlap 
with the approach in the first part of 
\ref\kms{J. Kwiecinski, A.D. Martin and P.J. Sutton,
\ZP \vyp{C71}{1996}{585}.} which incorporates subleading effects via a 
kinematic constraint while solving the BFKL equation - resulting in an
anomalous dimension which includes a 
resummation of some subset of higher order contributions, none of which is 
concerned with the running of the coupling, but which stabilizes the 
calculation. (The latter part of \kms\ also includes a running coupling and
infrared regularization, but concentrates on the normalization rather than 
the evolution.) In this
sense it has some similarities to the resummation of collinear logs in 
\gavin, which also stabilizes results even with fixed coupling (and which 
is essential in single scale processes). Hence, there appear to be a 
number of ways in which the apparent poor convergence of the perturbative 
series at small $x$ can be improved.   
However, since one must ultimately deal with the contribution of the running 
coupling in all perturbative QCD calculations I prefer to concentrate on 
this feature and 
consider just the resulting $\beta_0$ resummation combined with the $\ln(1/x)$ 
resummation, which results in explicit results in
terms of an ordered power-series in the well-defined quantities 
$\alpha_s(Q^2)$, $\ln(1/x)$ and $\beta_0$. This
stabilizes the small $x$ expansion without consideration of these other 
effects, indeed it leads to the most divergent terms as $x \to 0$ \xscale\
and alters the complete singularity structure, 
and moreover is easy to directly incorporate 
into the usual calculation of partons and structure functions in terms of the 
coefficient functions and splitting functions.         

It will, of course, be interesting to examine the effect of incorporating 
my resummed corrections to splitting function in a global fit to structure 
function and related data. Such an analysis will also need to 
include a precise explanation of how the small $x$ relevant expansions 
derived in this paper must be combined with the normal order-by-order in 
$\alpha_s(Q^2)$ expansion, and potentially large $\ln(1-x)$ expansions.
Full details of such a fit, and the complete procedure used, will appear in a 
future paper which awaits the release of new data from a number of experimental
collaborations. From the analysis of presently published data it is clear that
the quality of such a fit is improved by inclusion of these small $x$ 
resummed corrections\foot{A prediction for $d F_2(x,Q^2)/d\ln Q^2$ for 
preliminary ZEUS data is shown in fig. 35 of \ref\foster{B. Foster, 
{\it Phil. Trans. Roy. Soc. Lond} \vyp{A359}{2001}{325} 
[{\tt hep-ex/0008669}].}
which shows the clear improvement compared to a convention 
NLO-in-$\alpha_s(Q^2)$ treatment.}, and that the predicted $F_L(x,Q^2)$ is 
smaller than that form a NLO-in-$\alpha_s(Q^2)$ fit, but much more regular 
in shape at low $Q^2$ than that seen in 
\mrstb.\foot{Figures showing this can be 
seen in \ref\mandy{A. Cooper-Sarkar, {\tt hep-ph/0102151}, figures borrowed 
from talk by R.S. Thorne at DIS 2000, Liverpool, 2000.}.} 
This can be seen as a solution to 
the lack of convergence of $F_L(x,Q^2)$ apparent as one goes from LO to NLO
to NLO in the conventional expansion scheme which is seen in \mrstb.
 
Hence, I conclude by claiming that this paper outlines a method for 
including the most 
complete resummation of splitting functions (and coefficient functions) 
which is needed at small $x$, and satisfies the theoretical requirements of 
stability of the perturbative expansion and the minimum of model dependence 
as well as the more practical considerations of being in a closed form which
is easy to implement. It will prove useful in an analysis of structure function
data, and in a prediction of related quantities relevant for the Tevatron 
and the LHC. However, at present it only really exists at LO (and not even 
that for many quantities), and for full implementation the calculation of the 
NLO impact factors within the BFKL framework is urgently needed. Once this 
is done, a truly full NLO analysis of structure functions, which will be
equally valid over the full perturbative range will be possible.  
       
\bigskip

\noindent{\bf Acknowledgements.}

\medskip

I would like to thank P.V. Landshoff, A.D. Martin, R.G. Roberts, 
W.J. Stirling, A. Sabio-Vera and B. Webber for useful discussions. 
I would also like to thank the Royal Society for awarding me a 
University Research Fellowship enabling me to produce this work.  
   
\vfill 
\eject

\noindent {\bf Table 1.} \hfil

\medskip

The coefficients $K_{nm}$ in  
\eqn\eqtabone{xP^{NLO}_{gg}(\xi,\alpha_s(Q^2)) 
= \bar\alpha_s(Q^2)
\sum_{n=1}^{19}\sum_{m=1}^{m_{max}} \bar\alpha^n_s(Q^2)\biggl(K_{nm}{\xi^m 
\bar\beta_0^{n-m}\over m!}+K_{n\delta}\bar\beta_0^{n}\delta(1-x)\biggr).}

\medskip

\hfil\vtop{{\offinterlineskip
\halign{ \strut\tabskip=0.6pc
\vrule#& \hfil#&  \vrule#& \hfil#&  
\vrule#& \hfil#& \vrule#&
\hfil#& \vrule#& \hfil#& \vrule#& \hfil#& \vrule#& \hfil#& 
\vrule#\tabskip=0pt\cr
\noalign{\hrule}
& $n$ && $m=5$ && $m=4$ && $m=3$ && $m=2$ &&
$m=1$ && $m=0$ &\cr
& && $m=11$ && $m=10$ && $m=9$ && $m=8$ &&
$m=7$ && $m=6$ &\cr
& && $m=17$ && $m=16$ && $m=15$ && $m=14$ &&
$m=13$ && $m=12$ &\cr
& && $m=23$ && $m=22$ && $m=21$ && $m=20$ &&
$m=19$ && $m=18$ &\cr
\noalign{\hrule}
& 1 && && && && && && -0.4236 &\cr
\noalign{\hrule}
& 2 && && && && && -1.354 && 1.611 &\cr
\noalign{\hrule}
& 3 && && && && -7.000 && 30.22 && -34.63 &\cr
\noalign{\hrule}
& 4 && && && -5.686 && 46.92 && -103.2 && 63.85 &\cr
\noalign{\hrule}
& 5 && && -16.14 && 193.5 && -797.2 && 1373 && -918.0 &\cr
\noalign{\hrule}
& 6 && 186.0 && -971.0 && 2518 && -3323 && 2045 && -458.9 &\cr
&   && && && && && && -14.35 &\cr
\noalign{\hrule}
& 7 &&  -1386 && 5051 && -9865 && 10113 && -4281 && 709.2 &\cr
&   && && && && && -10.60 && 192.0 &\cr
\noalign{\hrule}
& 8 && 21431 && -59800 && 99225 && -95325 && 49058 && -11483 &\cr
&   && && && && -24.48 && 511.5 && -4497 &\cr
\noalign{\hrule}
& 9 && 70532 && -46099 && -25896 && 59631 && -29684 && 2798 &\cr
&   && && -17.21 && 349.4 && -3100 && 15284 && -44034 &\cr
\noalign{\hrule}
& 10 && -126387 && -261087 && 735693 && -761882 && 373984 && -77690 &\cr
&   && -12.01 && 326.1 && -3758 && 23801 && -88010 && 179647 &\cr
\noalign{\hrule}
& 11 && 8688676 && -9665206 && 6981022 && -3087487 && 771318 && -102010 &\cr
&   && 1117.6 && -15044 && 119789 && -620744 && 2179220 && -5256680 &\cr
&   && && && && && && -37.57 &\cr
\noalign{\hrule}
& 12 && -1.621$\times 10^7$ && 1.864$\times 10^7$ && -1.288$\times 10^7$ 
&& 5044618 && -962638 && 64963 &\cr
&  && 46536 && -211318 && 563130 && -547416 && -1766356 && 8225690 &\cr
&   && && && && -18.36 && 506.4 && -6353 &\cr
\noalign{\hrule}
& 13 &&  1.139$\times 10^8$ && - 1.236$\times 10^8$ && 8.576$\times 10^7$ 
&& -3.677$\times 10^7$ && 9106782 && -1186015 &\cr
&  && -400076 && 1467162 && -2855626 && -645255 && 2.136$\times 10^7$ &&
-6.623$\times10^7$ &\cr
&   && && && -12.58 && 453.6 && -7298 && 68149 &\cr
\noalign{\hrule}
& 14 && 4.425 $\times 10^8$ && -2.089$\times 10^8$ && 6.225$\times 10^7$ 
&& -1.129$\times 10^7$ && 2061380 && -386008 &\cr
&  && 1.425$\times 10^7$ && -5.541$\times 10^7$ && +1.589$\times 10^8$ 
&& -3.345$\times 10^8$ && 5.135$\times 10^8$ && -5.625$\times10^8$ &\cr
&   && && -46.85 && 180.1 && -32842 && 359398 && -2678626 &\cr
\noalign{\hrule}
& 15 && -7.723$\times 10^8$ && 1.340$\times 10^8$ && 1.391$\times 10^8$ 
&& -1.075$\times 10^8$ && 3.170$\times 10^7$ && -4331143 &\cr
&  && -1.146$\times 10^7$ && 1.108$\times 10^8$ && -4.028$\times 10^8$ 
&& 9.069$\times 10^8$ && -1.361$\times 10^9$ && 1.339$\times 10^9$ &\cr
&   && 639.2&& -103283 && 101509 && -640342 && 2523533 && -4527424 &\cr
&   && &&  &&  &&  &&  && -18.33 &\cr
\noalign{\hrule}}}}\hfil

\vfil\eject

\hfil\vtop{{\offinterlineskip
\halign{ \strut\tabskip=0.6pc
\vrule#& \hfil#&  \vrule#& \hfil#&  
\vrule#& \hfil#& \vrule#&
\hfil#& \vrule#& \hfil#& \vrule#& \hfil#& \vrule#& \hfil#& 
\vrule#\tabskip=0pt\cr
\noalign{\hrule}
& $n$ && $m=5$ && $m=4$ && $m=3$ && $m=2$ &&
$m=1$ && $m=0$ &\cr
& && $m=11$ && $m=10$ && $m=9$ && $m=8$ &&
$m=7$ && $m=6$ &\cr
& && $m=17$ && $m=16$ && $m=15$ && $m=14$ &&
$m=13$ && $m=12$ &\cr
& && $m=23$ && $m=22$ && $m=21$ && $m=20$ &&
$m=19$ && $m=18$ &\cr
\noalign{\hrule}
& 16 && 1.403$\times 10^{10}$ && -8.066$\times 10^9$ && 3.172$\times 10^9$ 
&& -8.188$\times 10^8$ && 1.324$\times 10^8$ && -1.227$\times 10^7$ &\cr
&  && 2.987$\times 10^9$ && -1.490$\times 10^9$ && 4.598$\times 10^9$ 
&& -9.865$\times 10^9$ && 1.529$\times 10^{10}$ && -1.724$\times 10^{10}$ &\cr
&   && -11408&& 137101 && -1076349 && 5431540 && -1.481$\times 10^7$ 
&& -9014694 &\cr
&   && &&  &&  &&  &&  -12.37 && 552.7 &\cr
\noalign{\hrule}
& 17 && -1.108$\times 10^{10}$ && 7.809$\times 10^9$ && -3.096$\times 10^9$ 
&& 6.807$\times 10^8$ && -7.184$\times 10^7$ && 1672556 &\cr
&  && 9.306$\times 10^9$ && -1.676$\times 10^{10}$ && 2.172$\times 10^{10}$ 
&& -1.824$\times 10^{10}$ && 6.057$\times 10^{9}$ && 6.571$\times 10^{9}$ &\cr
&   && 803905&& -7783462 && 5.544$\times 10^7$ && -2.982$\times 10^8$ 
&& 1.227$\times 10^{9}$ && -3.869$\times 10^{9}$ &\cr
&   && &&  &&  && -57.64 &&  2645 && -58085 &\cr
\noalign{\hrule}
& 18 && 7.129$\times 10^{10}$ && -4.890$\times 10^{10}$ 
&& 2.107$\times 10^{10}$ 
&& -5.733$\times 10^9$ && 9.531$\times 10^8$ && -8.884$\times 10^7$ &\cr
&  && -3.690$\times 10^{10}$ && 6.574$\times 10^{10}$ 
&& -8.024$\times 10^{11}$ 
&& 5.604$\times 10^{10}$ && 3.195$\times 10^{9}$ && -5.773$\times 10^{10}$ &\cr
&   && 7633219&& -2.006$\times 10^7$ && -4.467$\times 10^7$ && 
7.555$\times 10^8$ && -4.185$\times 10^{9}$ && 1.480$\times 10^{10}$ &\cr
&   && && -17.59 &&  741.8 && -14788 &&  180754 && -1460081 &\cr
\noalign{\hrule}
& 19 && 4.467$\times 10^{11}$ && -1.573$\times 10^{11}$ 
&& 3.986$\times 10^{10}$ 
&& -7.340$\times 10^9$ && 9.737$\times 10^8$ && -8.398$\times 10^7$ &\cr
&  && 4.962$\times 10^{11}$ && -9.477$\times 10^{11}$ 
&& 1.402$\times 10^{12}$ && -1.602$\times 10^{12}$ && 1.399$\times 10^{12}$ 
&& -9.198$\times 10^{11}$ &\cr
&   && -4.778$\times 10^7$ && -6.248$\times 10^7$ && 1.948$\times 10^9$ && 
-1.372$\times 10^{10}$ && 6.153$\times 10^{10}$ && -2.005$\times 10^{11}$ &\cr
&   && -11.75&& 620.5 && -15217 && 225792 && -2200274 
&& 1.398$\times 10^7$ &\cr
\noalign{\hrule}}}}\hfil

\bigskip

The series for the part proportional to $\delta(1-x)$ is more convergent in
$\alpha_s(Q^2)$ and for all $Q^2 \gsim 1\Gev^2$ is given accurately by
\eqn\delserg{\eqalign{\bar\alpha_s(Q^2)\delta(1-x)\bigl[&
9.0(\bar\beta_0\bar\alpha_s(Q^2))^3 + 139.5(\bar\beta_0\bar\alpha_s(Q^2))^5
+38.88(\bar\beta_0\bar\alpha_s(Q^2))^6+964.2(\bar\beta_0\bar\alpha_s(Q^2))^8\cr
&+167.0(\bar\beta_0\bar\alpha_s(Q^2))^9 
+5605(\bar\beta_0\bar\alpha_s(Q^2))^{10}
\bigr].\cr}}

\vfil
\eject

\noindent {\bf Table 2.} \hfil

\medskip

The coefficients $K_{nm}$ in  
\eqn\eqtabone{xP^{NLO}_{LL}(\xi,\alpha_s(Q^2)) 
= \bar\alpha_s(Q^2)
\sum_{n=1}^{19}\sum_{m=1}^{m_{max}} \bar\alpha^n_s(Q^2)\biggl(K_{nm}{\xi^m 
\bar\beta_0^{n-m}\over m!}+K_{n\delta}\bar\beta_0^{n}\delta(1-x)\biggr).}

\medskip

\hfil\vtop{{\offinterlineskip
\halign{ \strut\tabskip=0.6pc
\vrule#& \hfil#&  \vrule#& \hfil#&  
\vrule#& \hfil#& \vrule#&
\hfil#& \vrule#& \hfil#& \vrule#& \hfil#& \vrule#& \hfil#& 
\vrule#\tabskip=0pt\cr
\noalign{\hrule}
& $n$ && $m=5$ && $m=4$ && $m=3$ && $m=2$ &&
$m=1$ && $m=0$ &\cr
& && $m=11$ && $m=10$ && $m=9$ && $m=8$ &&
$m=7$ && $m=6$ &\cr
& && $m=17$ && $m=16$ && $m=15$ && $m=14$ &&
$m=13$ && $m=12$ &\cr
& && $m=23$ && $m=22$ && $m=21$ && $m=20$ &&
$m=19$ && $m=18$ &\cr
\noalign{\hrule}
& 1 && && && && && && -0.4236 &\cr
\noalign{\hrule}
& 2 && && && && && -1.354 && 9.494 &\cr
\noalign{\hrule}
& 3 && && && && -7.040 && 25.89 && -29.49 &\cr
\noalign{\hrule}
& 4 && && && -5.672 && 63.20 && -222.22 && 251.90 &\cr
\noalign{\hrule}
& 5 && && -15.84 && 310.8 && -1504 && 2766 && -1964 &\cr
\noalign{\hrule}
& 6 && 243.4 && -1444 && 4540 && -7293 && 5206 && -1100 &\cr
&   && && && && && && -17.45 &\cr
\noalign{\hrule}
& 7 &&  -5265 && 24975 && -61945 && 82368 && -55633 && 16210 &\cr
&   && && && && && -19.57 && 521.2 &\cr
\noalign{\hrule}
& 8 && 27358 && -84630 && 162654 && -187932 && 116668 && -31108 &\cr
&   && && && && -6.545 && 448.6 && -5158 &\cr
\noalign{\hrule}
& 9 && 215634 && -122925 && -266550 && 550451 && -383797 && 100196 &\cr
&   && && -20.94 && 468.9 && -5027 && 31574 && -114142 &\cr
\noalign{\hrule}
& 10 && -1552522 && 1019004 && 567195 && -1582395 && 1103037 && -286332 &\cr
&   && -22.36 && 814.7 && -12094 && 91607 && -396924 && 1031187 &\cr
\noalign{\hrule}
& 11 && 3.965$\times 10^7$ && -5.343$\times 10^7$ && 4.627$\times 10^7$ 
&& -2.425$\times 10^7$ && 7013331 && -1016798 &\cr
&   && 1291 && -23513 && 232273 && -1492972 && 6537249 
&& -1.956$\times 10^7$ &\cr
&   && && && && && && -20.00 &\cr
\noalign{\hrule}
& 12 && -2.424$\times 10^8$ && 2.970$\times 10^8$ && -1.702$\times 10^8$ 
&& 1.214$\times 10^8$ && -3.553$\times 10^7$ && 5390954 &\cr
&  && 60402 && -153923 && -952951 && 1.057$\times 10^7$ 
&& -4.805$\times 10^7$ && 1.328$\times 10^8$ &\cr
&   && && && && -22.33 && 641 && -8519 &\cr
\noalign{\hrule}
& 13 &&  -1.514$\times 10^8$ && - 7.7974$\times 10^7$ && 1.993$\times 10^8$ 
&& -1.426$\times 10^8$ && 4.814$\times 10^7$ && -7741384 &\cr
&  && -2256260 && 1.199$\times 10^7$ && -4.586$\times 10^7$ 
&& 1.252$\times 10^8$ && -2.356$\times 10^8$ && 2.796$\times10^8$ &\cr
&   && && && -23.25 && 1171 && -25002 && 296133 &\cr
\noalign{\hrule}
& 14 && 4.934 $\times 10^9$ && -2.902$\times 10^9$ && 9.784$\times 10^8$ 
&& -1.227$\times 10^8$ && -1.883$\times 10^7$ && 6764209 &\cr
&  && 5.595$\times 10^7$ && -2.675$\times 10^8$ && 9.236$\times 10^8$ 
&& -2.317$\times 10^9$ && 4.213$\times 10^9$ && -5.475$\times10^9$ &\cr
&   && && -33.40 && 2462 && -61331 && 876031 && -8361477 &\cr
\noalign{\hrule}
& 15 && -2.576$\times 10^{10}$ && 1.702$\times 10^{10}$ && -7.397$\times 10^9$ 
&& 1.968$\times 10^9$ && -2.961$\times 10^8$ && 2.201$\times 10^7$ &\cr
&  && -3.263$\times 10^8$ && 1.473$\times 10^9$ && -4.775$\times 10^9$ 
&& 1.147$\times 10^{10}$ && -2.052$\times 10^{10}$ 
&& 2.708$\times 10^{10}$ &\cr
&   && 781.7&& -12818 && 114646 && -395405 && -2857628 
&& 4.706$\times 10^7$ &\cr
&   && &&  &&  &&  &&  && -22.34 &\cr
\noalign{\hrule}}}}\hfil

\vfil\eject

\hfil\vtop{{\offinterlineskip
\halign{ \strut\tabskip=0.6pc
\vrule#& \hfil#&  \vrule#& \hfil#&  
\vrule#& \hfil#& \vrule#&
\hfil#& \vrule#& \hfil#& \vrule#& \hfil#& \vrule#& \hfil#& 
\vrule#\tabskip=0pt\cr
\noalign{\hrule}
& $n$ && $m=5$ && $m=4$ && $m=3$ && $m=2$ &&
$m=1$ && $m=0$ &\cr
& && $m=11$ && $m=10$ && $m=9$ && $m=8$ &&
$m=7$ && $m=6$ &\cr
& && $m=17$ && $m=16$ && $m=15$ && $m=14$ &&
$m=13$ && $m=12$ &\cr
& && $m=23$ && $m=22$ && $m=21$ && $m=20$ &&
$m=19$ && $m=18$ &\cr
\noalign{\hrule}
& 16 && 1.725$\times 10^{11}$ && -1.209$\times 10^{11}$ 
&& 5.738$\times 10^{10}$ 
&& -1.764$\times 10^{10}$ && 3.314$\times 10^9$ && -3.434$\times 10^8$ &\cr
&  && -1.550$\times 10^9$ && -7.835$\times 10^8$ && 1.583$\times 10^{10}$ 
&& -5.601$\times 10^9$ && 1.187$\times 10^{11}$ && -1.710$\times 10^{11}$ &\cr
&   && -40001&& 623658 && -6411756 && 4.638$\times 10^7$ 
&& -2.376$\times 10^8$ && 8.205$\times 10^8$ &\cr
&   && &&  &&  &&  &&  -2.90 && 1464 &\cr
\noalign{\hrule}
& 17 && -3.968$\times 10^{11}$ && 3.104$\times 10^{11}$ 
&& -1.572$\times 10^{11}$ && 5.059$\times 10^{10}$ && -9.869$\times 10^9$ 
&& 1.057$\times 10^9$ &\cr
&  && 5.446$\times 10^{10}$ && -1.036$\times 10^{11}$ && 1.304$\times 10^{11}$ 
&& -6.795$\times 10^{10}$ && 1.099$\times 10^{11}$ 
&& 3.145$\times 10^{11}$ &\cr
&   && 1959861&& -2.351$\times 10^7$ && 2.013$\times 10^8$ 
&& -1.264$\times 10^9$ && 5.914$\times 10^{9}$ && -2.077$\times 10^{10}$ &\cr
&   && &&  &&  && -45.09 &&  3674 && -110727 &\cr
\noalign{\hrule}
& 18 && -2.261$\times 10^{12}$ && 8.240$\times 10^{11}$ 
&& -1.747$\times 10^{11}$ 
&& 1.319$\times 10^{10}$ && 2.293$\times 10^9$ && -5.601$\times 10^8$ &\cr
&  && -8.652$\times 10^{11}$ && 2.018$\times 10^{12}$ 
&& -3.648$\times 10^{12}$ 
&& 5.066$\times 10^{12}$ && -5.319$\times 10^{12}$ 
&& 4.119$\times 10^{12}$ &\cr
&   && -1.287$\times 10^7$&& 2.343$\times 10^8$ && -1.611$\times 10^9$ && 
1.497$\times 10^{10}$ && -7.492$\times 10^{10}$ && 2.888$\times 10^{11}$ &\cr
&   && && -21.44 &&  885.6 && -171708 &&  177961 && -514627 &\cr
\noalign{\hrule}
& 19 && 1.562$\times 10^{13}$ && -6.662$\times 10^{12}$ 
&& 1.980$\times 10^{12}$ 
&& -3.942$\times 10^{11}$ && 4.967$\times 10^{10}$ && -3.565$\times 10^9$ &\cr
&  && 5.886$\times 10^{12}$ && -1.334$\times 10^{13}$ 
&& 2.338$\times 10^{13}$ && -3.179$\times 10^{13}$ && 3.328$\times 10^{13}$ 
&& -2.645$\times 10^{13}$ &\cr
&   && -6.036$\times 10^8$ && 1.710$\times 10^9$ && 4.335$\times 10^9$ && 
-7.920$\times 10^{10}$ && 4.895$\times 10^{11}$ && -1.982$\times 10^{12}$ &\cr
&   && -21.75&& 1680 && -54959 && 1031287 && -1.271$\times 10^7$ 
&& 1.077$\times 10^8$ &\cr
\noalign{\hrule}}}}\hfil

\bigskip

The series for the part proportional to $\delta(1-x)$ is more convergent in
$\alpha_s(Q^2)$ and for all $Q^2 \gsim 1\Gev^2$ is given accurately by
\eqn\delserfl{\eqalign{\bar\alpha_s&(Q^2)\delta(1-x)\bigl[-0.3094
(\bar\beta_0\bar\alpha_s(Q^2))-3.856(\bar\beta_0\bar\alpha_s(Q^2))^2 +
6.376(\bar\beta_0\bar\alpha_s(Q^2))^3 
-50.36(\bar\beta_0\bar\alpha_s(Q^2))^4\cr
&+ 340.0(\bar\beta_0\bar\alpha_s(Q^2))^5
+55.51(\bar\beta_0\bar\alpha_s(Q^2))^6-1600(\bar\beta_0\bar\alpha_s(Q^2))^7
+2838(\bar\beta_0\bar\alpha_s(Q^2))^8-8457(\bar\beta_0\bar\alpha_s(Q^2))^9\cr 
&+24526(\bar\beta_0\bar\alpha_s(Q^2))^{10}
+57602(\bar\beta_0\bar\alpha_s(Q^2))^{11}
-325984(\bar\beta_0\bar\alpha_s(Q^2))^{12}
+477536(\bar\beta_0\bar\alpha_s(Q^2))^{13}\bigr].\cr}}

\vfil
\eject

\vfill\eject\immediate\closeout\ffile{\parindent40pt
\baselineskip14pt\centerline{{\bf Figure Captions}}\nobreak\medskip
\escapechar=` \input figs.tmp\vfill\eject}

\footatend\vfill\supereject\immediate\closeout\rfile\writestoppt
\baselineskip=14pt\centerline{{\bf References}}\bigskip{\frenchspacing%
\parindent=20pt\escapechar=` \input refs.tmp\vfill\eject}\nonfrenchspacing

\end

\ref\bluvogt{J. Bl\"umlein, V. Ravindran, W.L.
van Neerven and  A. Vogt, Proceedings of DIS 98, 
Brussels, April 1998, p. 211, {\tt hep-ph/9806368}.}%
\nref\ballforte{R.D. Ball 
and S. Forte Proceedings of DIS 98, Brussels, April 1998, p. 770,
{\tt hep-ph/9805315}.}%
\nref\ross{D.A. Ross, \PL \vyp{B431}{1998}{161}.}%
\nref\mplusk{Yu. V. Kovchegov 
and A.H. Mueller, \PL \vyp{B439}{1998}{423}.}%
\nref\levin{E.M. Levin, 
Tel Aviv University, Report No. TAUP 2501-98, 
{\tt hep-ph/9806228}.}%
--
\nref\hancock{R.E. Hancock and 
D.A. Ross, \NP \vyp{B383}{1992}{575}; \NP \vyp{B394}{1993}{200}.}%
\nref\nik{N.N. Nikolaev and B.G. 
Zakharov, \PL \vyp{B327}{1994}{157}.}%
\nref\levrun{E.M. Levin, \NP \vyp{B453}{1995}{303}.}%
\nref\Andersson{B. Andersson, G. Gustafson and 
H. Kharraziha, \PR \vyp{D57}{1998}{5543}.}%
\nref\haakman{L.P.A. Haakman, O.V. 
Kancheli, and J.H. Koch, \PL \vyp{B391}{1997}{157}; \NP 
\vyp{B518}{1998}{275}.}%
-- 
\ref\diff{J.Bartels and
H. Lotter, \PL \vyp{B309}{1993}{400}\semi
J. Bartels, H. Lotter and M. Vogt
\PL \vyp{B373}{1996}{215}.} 

\ref\BLM{S.J. Brodsky, G.P. Lepage and 
P.B. Mackenzie, \PR \vyp{D28}{1983}{228}.}  
\ref\kti{S. Catani, M. Ciafaloni 
and F. Hautmann, \PL  \vyp{B242}{1990}{97}; \NP \vyp{B366}{1991}{135};
\PL\vyp{B307}{1993}{147}.}\ref\ktii{J.C. Collins and R.K. Ellis, \NP 
\vyp{B360}{1991}{3}.}

\eqn\melltranspd{f(N,Q^2)= \int_0^1\,x^{N}{\rm f}(x,Q^2)dx.}

\liprun\diff\ref\harriman{J.R. Forshaw, P.N. Harriman and
P.N. Sutton, \NP \vyp{B416}{1994}{739}.}\haakman\mueller. 

\eqn\nblm{{d {\cal G}(N,Q^2) \over d\ln Q^2} \approx 
\Gamma(N,\bar\alpha_s(s(N)Q^2)) G(N,Q^2),}
\eqn\xblm{{d G(x,Q^2) \over d\ln(Q^2)} \approx \int_x^1
P(z,\bar\alpha_s(s(z),Q^2))G(x/z,Q^2)dz.}
\eqn\nlosplit{(x/\bar\alpha_s(Q^2))P(x,Q^2)=p^0(\bar\alpha_s(Q^2)\xi)
-\beta_0\alpha_s(Q^2)\hat
p^1(\bar\alpha_s(Q^2)\xi)+{\cal O}((\beta_0\alpha_s(Q^2))^2) 
r(\bar\alpha_s(Q^2)
\xi),}
this is the same as 
\eqn\absorb{(x/\bar\alpha_s(Q^2s(\xi
\bar \alpha_s(Q^2))))P(x,Q^2)=p^0(\bar\alpha_s(Q^2s(\xi
\bar \alpha_s(Q^2)))+{\cal O} 
((\beta_0\alpha_s(Q^2))^2) \hat r(\bar\alpha_s(Q^2)\xi),}
As $x \to 0$
\eqn\losplit{p^0(\bar\alpha_s(Q^2)\xi) \to {1\over 
(56\pi\zeta(3))^{\half}} \exp(\lambda(Q^2)\xi)(\bar\alpha_s(Q^2)\xi)^{-3/2},}
So far I have simply assumed that an accurate way to account for the running
of the coupling in the LO BFKL equation is to use \bfklruni. This is an 
assumption which involves the resummation of an infinite number of terms, i.e.
it assumes that at all orders in $\alpha_s(\mu^2)$ the dominant contribution
to the BFKL equation due to the running coupling is 
\eqn\runassump{{\bar \alpha_s \over  N}(-1)^n
(\beta_0\alpha_s(\mu^2)\ln(k^2/\mu^2))^n 
\int_0^{\infty}{dq^2\over q^2}K^0(q^2,k^2)f(q^2),}
Formally the NLO BFKL equation may be
written as   
\eqn\NLOBFKL{\eqalign{f(k^2,Q_0^2,\bar\alpha_s(\mu^2)/N)= f^0(k^2,Q_0^2)
&+\biggl({\bar\alpha_s(\mu^2) \over N}\biggr)
\int_0^{\infty}{dq^2\over q^2}(K^0(q^2,k^2)\cr
&-\beta_0\alpha_s(\mu^2)
\ln(k^2/\mu^2)K^0(q^2,k^2)-\alpha_s(\mu^2)K^1(q^2,k^2))f(q^2),\cr}}
where $K^1(q^2,k^2)$ can be found in \NLOBFKLlf. 
\eqn\NLOBFKLalt{f(k^2,Q_0^2,\mu^2)= f^0(k^2,Q_0^2)
+\biggl({\bar\alpha_s(k^2) \over N}\biggr)
\int_0^{\infty}{dq^2\over q^2}(K^0(q^2,k^2)
-\alpha_s(\mu^2)K^1(q^2,k^2))f(q^2).}
This is identical to \NLOBFKL\ up to NNLO in $\alpha_s(\mu^2)$ and is
a common way for the NLO BFKL equation to be written.
\eqn\bfklrunnlo{{d \tilde f(\gamma,N)\over d \gamma}={d\tilde 
f^0(\gamma, Q_0^2) \over
d\gamma}-{1\over \bar\beta_0 N} (\chi_0(\gamma)-\alpha_s(\mu^2)\chi_1(\gamma))
\tilde f(\gamma, N),}
\eqn\wrongnloan{\alpha_s{\chi_1(\gamma^0(\bar\alpha_s/N))\over 
-\chi'_0(\gamma^0(\bar\alpha_s/N))} \equiv \alpha_s \gamma^1(\bar\alpha_s/N),}
is often called the NLO-BFKL anomalous dimension.

\eqn\nloanom{\Gamma(N,Q^2/\mu^2)=\gamma^0-
\beta_0\alpha_s\biggl({\partial \gamma^0\over
\partial \ln(\alpha_s)}\ln(Q^2/\mu^2)+  
{\partial \gamma^0 \over \partial\ln(\alpha_s)}
\biggl({-\chi''(\gamma^0) \over
2\chi'(\gamma^0)}-{1\over \gamma^0}\biggr)\biggr)-\alpha_s\gamma^1.}
\eqn\nlospli{xP(x,Q^2) =\bar\alpha_s\exp(\lambda\xi)\biggl(
{0.068 \over (\bar\alpha_s\xi)^{3/2}}-\beta_0\alpha_s\biggl(\biggl({0.188 \over
(\bar\alpha_s\xi)^{\half}}\biggr)\ln(Q^2/\mu^2)+0.69\biggr)-
\alpha_s\biggl({1.18 \over
(\bar\alpha_s\xi)^{\half}}\biggr)\biggr).}
\eqn\nloconfan{-\alpha_s(Q^2)\gamma^1(\alpha_s(Q^2)/N)\equiv  
- \alpha_s(Q^2){\chi_1(\gamma^0(\bar\alpha_s(Q^2)/N))
\over -\chi'_0(\gamma^0(\bar\alpha_s(Q^2)/N))}}

\noindent {\bf Table 1.} \hfil\break

The coefficients in the power series $p^i_{LL}(\bar\alpha_s(Q^2)\xi)=
\sum_{0}^{\infty}a_n
(\bar\alpha_s(Q^2)\xi)^{n}/ n!$ for the
various LO and NLO contributions to the physical splitting function 
$P_{LL}(x,Q^2)$. 

\hfil\vtop{{\offinterlineskip
\halign{ \strut\tabskip=0.6pc
\vrule#&  #\hfil&  \vrule#&  \hfil#& \vrule#& \hfil#& \vrule#& \hfil#&
\vrule#& \hfil#& \vrule#\tabskip=0pt\cr
\noalign{\hrule}
& $n$ && $p^0_{LL}$ && $p^{1,tot}_{LL}$ && $p^{1,\beta}_{LL}$ &&
$p^{1,conf}_{LL}$ &\cr
\noalign{\hrule}
& 0 && 1.00 && 0.23 && -2.00 && 1.57 &\cr
& 1 && 0.00 && 4.38 && 4.15 && 1.60 &\cr
& 2 && 0.00 && 15.87 && 11.32 && 8.29 &\cr
& 3 && 2.40 && 13.41 && -16.18 && 24.25 &\cr
& 4 && 0.00 && 86.26 && 76.03 && 35.31 &\cr
& 5 && 2.07 && 252.92 && 167.34 && 140.81 &\cr
& 6 && 17.34 && 323.08 && -81.51 && 377.69 &\cr
& 7 && 2.01 && 1699.65 && 1472.42 && 713.25 &\cr
& 8 && 39.89 && 4338.69 && 2665.07 && 2553.16 &\cr
& 9 && 168.75 && 7592.65 && 1674.16 && 6470.97 &\cr
& 10 && 69.99 && 33409.13 && 28319.16 && 14435.29 &\cr
& 11 && 661.25 && 79427.26 && 47284.56 && 47746.61 &\cr
& 12 && 1945.31 && 173361.43 && 81792.97 && 118560.14 &\cr
& 13 && 1717.68 && 657395.79 && 543255.72 && 293414.46 &\cr
& 14 && 10643.26 && 1527235.16 && 927749.64 && 905642.90 &\cr
& 15 && 25266.78 && 3833618.50  && 23539999.61 && 2256438.84 &\cr
\noalign{\hrule}}}}\hfil

\vfil
\eject

\noindent {\bf Table 2.} \hfil\break

The coefficients in the power series $p^i_{2L}(\bar\alpha_s(Q^2)\xi)=
\sum_{0}^{\infty}a_n
(\bar\alpha_s(Q^2)\xi)^{n}/ n!$ for the
LO and $\beta_0$-dependent NLO contributions to the physical splitting
function $P_{2L}(x,Q^2)$. 

\hfil\vtop{{\offinterlineskip
\halign{ \strut\tabskip=0.6pc
\vrule#&  #\hfil&  \vrule#& \hfil#&
\vrule#& \hfil#& \vrule#\tabskip=0pt\cr
\noalign{\hrule}
& $n$ && $p^0_{2L}$ && $p^{1,\beta}_{2L}$ &\cr
\noalign{\hrule}
& 0 && 2.50 && -4.00 &\cr
& 1 && 1.00 && 9.39 &\cr
& 2 && 1.00 && 36.60 &\cr
& 3 && 7.01 && 6.27 &\cr
& 4 && 5.81 && 239.73 &\cr
& 5 && 13.40 && 687.03 &\cr
& 6 && 58.11 && 771.35 &\cr
& 7 && 64.74 && 5281.50 &\cr
& 8 && 196.83 && 13213.51 &\cr
& 9 && 649.89 && 24043.80 &\cr
& 10 && 930.65 && 111578.92 &\cr
& 11 && 3034.70 && 265509.09 &\cr
& 12 && 8527.87 && 613964.05 &\cr
& 13 && 15046.02 && 2311855.03 &\cr
& 14 && 48434.53 && 5521425.31 &\cr
& 15 && 124600.51 && 14458201.96 &\cr
\noalign{\hrule}}}}\hfil

\vfil
\eject

\noindent Table 3\hfil\break
\noindent Comparison of quality of fits using full leading order (including 
$\ln (1/x)$ terms) renormalization scheme consistent expression, with BLM 
scale setting and the NLO in $\alpha_s(Q^2)$ fit \mrst. The references to the 
data can be found in \mrst.   
\medskip

\hfil\vtop{{\offinterlineskip
\halign{ \strut\tabskip=0.6pc
\vrule#&  #\hfil&  \vrule#&  \hfil#& \vrule#& \hfil#& \vrule#& \hfil#&
\vrule#\tabskip=0pt\cr
\noalign{\hrule}
& Experiment && data &&$\chi^2$&\omit& \omit &\cr
&\omit&& points && LO(x) &\omit& MRST &\cr
\noalign{\hrule}
& H1 $F^{ep}_2$ && 221 && 149 && 164 &\cr
& ZEUS $F^{ep}_2$ && 204 && 246 && 270 &\cr
\noalign{\hrule}
& BCDMS $F^{\mu p}_2$ && 174 && 241 && 249 &\cr
& NMC $F^{\mu p}_2$ && 130 && 118 && 141 &\cr
& NMC $F^{\mu d}_2$ && 130 && 81 && 101 &\cr
& NMC $F^{\mu n}_2/F^{\mu p}_2$ && 163 && 176 && 187 &\cr
& SLAC $F^{\mu p}_2$ && 70 && 87 && 119 &\cr
& E665 $F^{\mu p}_2$ && 53 && 59 && 58 &\cr
& E665 $F^{\mu d}_2$ && 53 && 61 && 61 &\cr
\noalign{\hrule}
& CCFR $F^{\nu N}_2$ && 66 && 57 && 93 &\cr
& CCFR $F^{\nu N}_3$ && 66 && 65 && 68 &\cr
\noalign{\hrule}
& total && 1330 && 1339 && 1511 &\cr
\noalign{\hrule}}}}\hfil
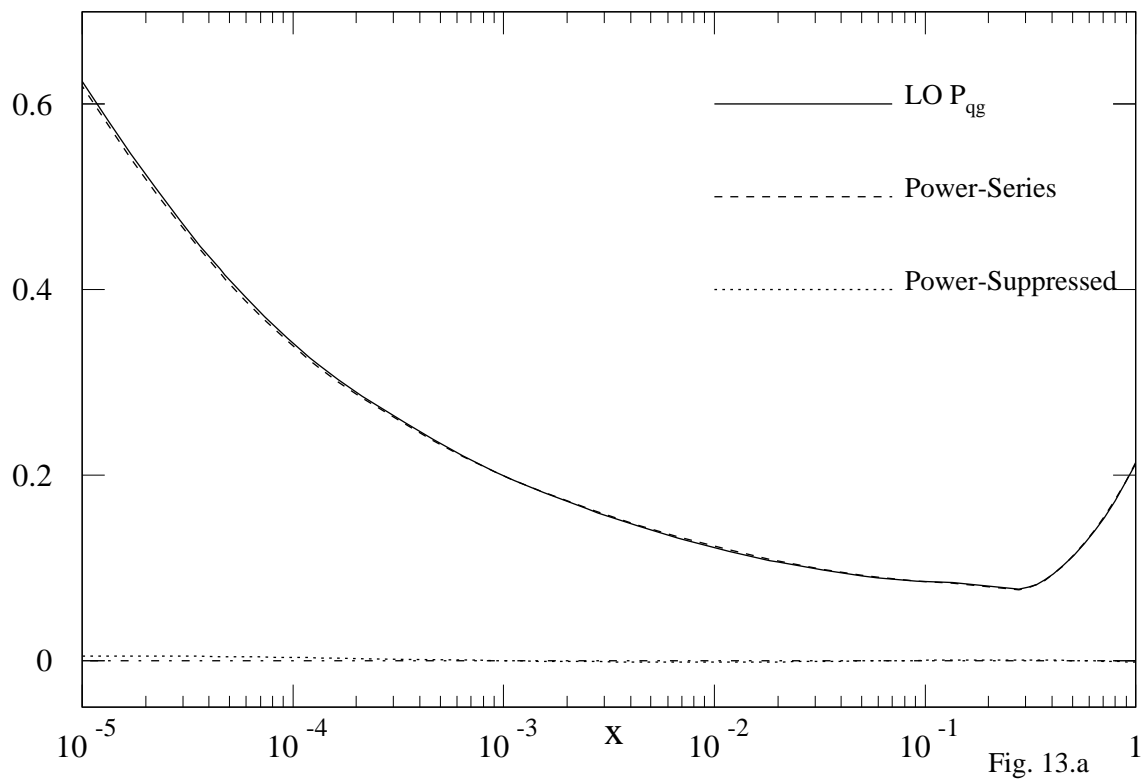
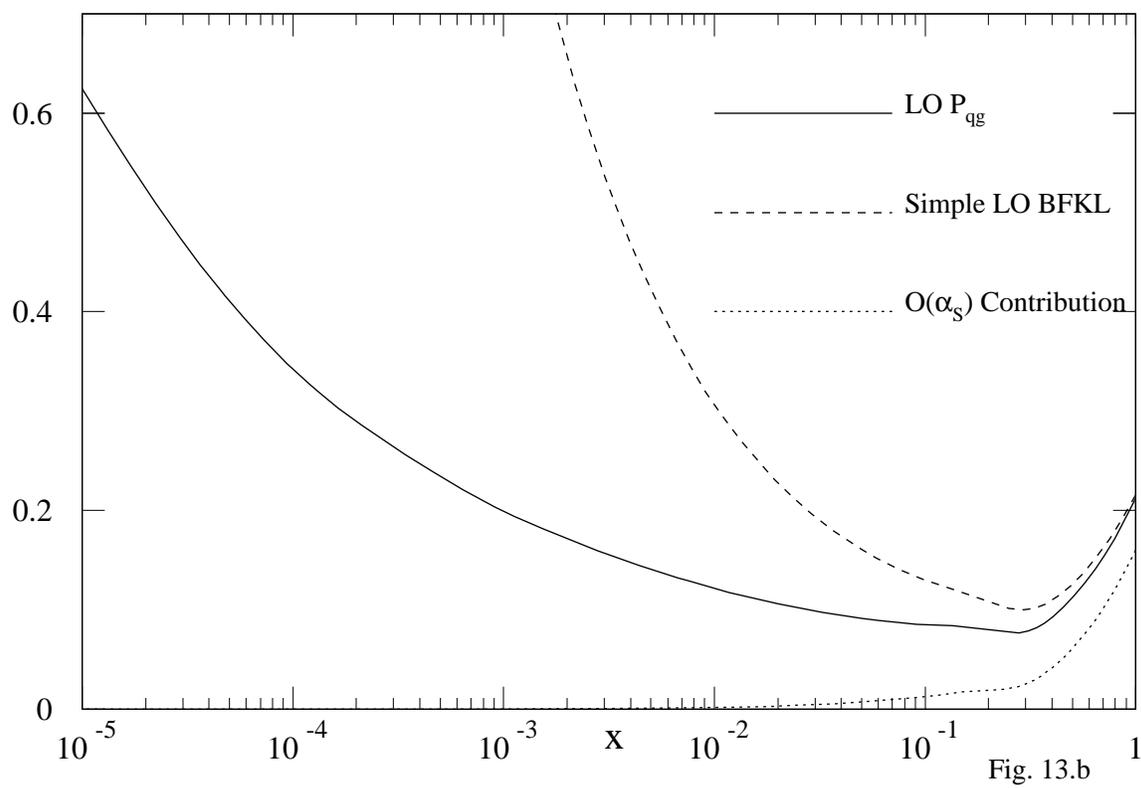
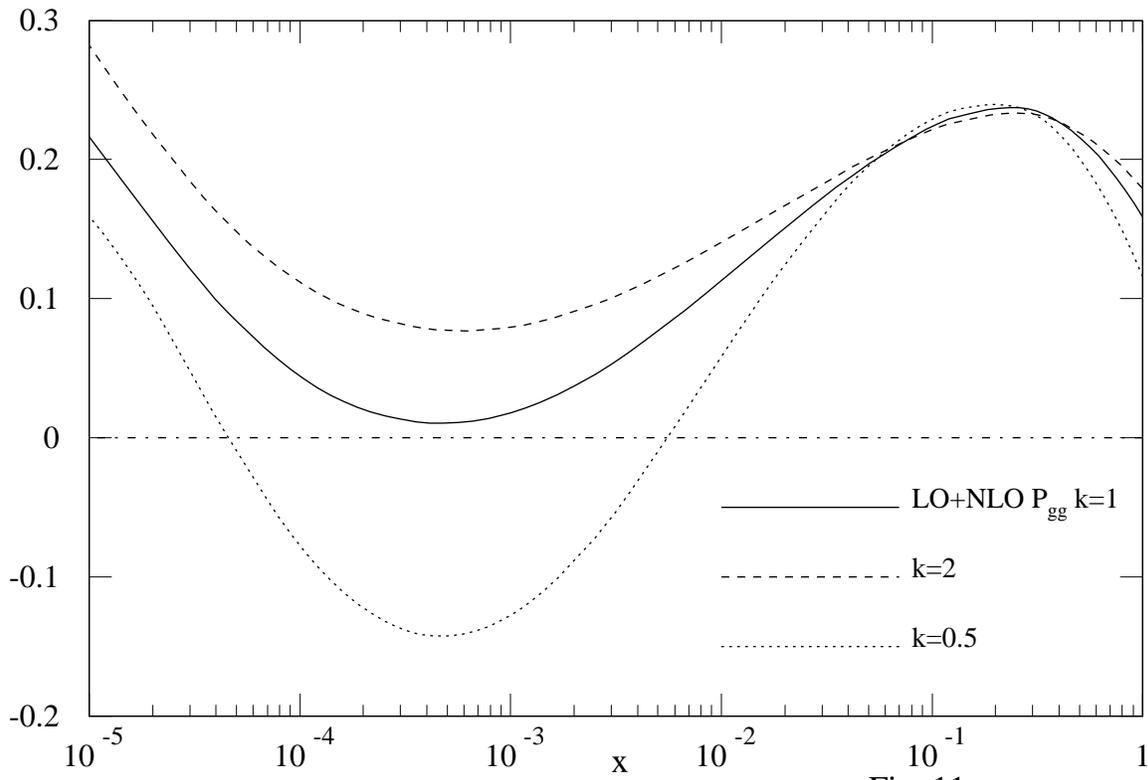
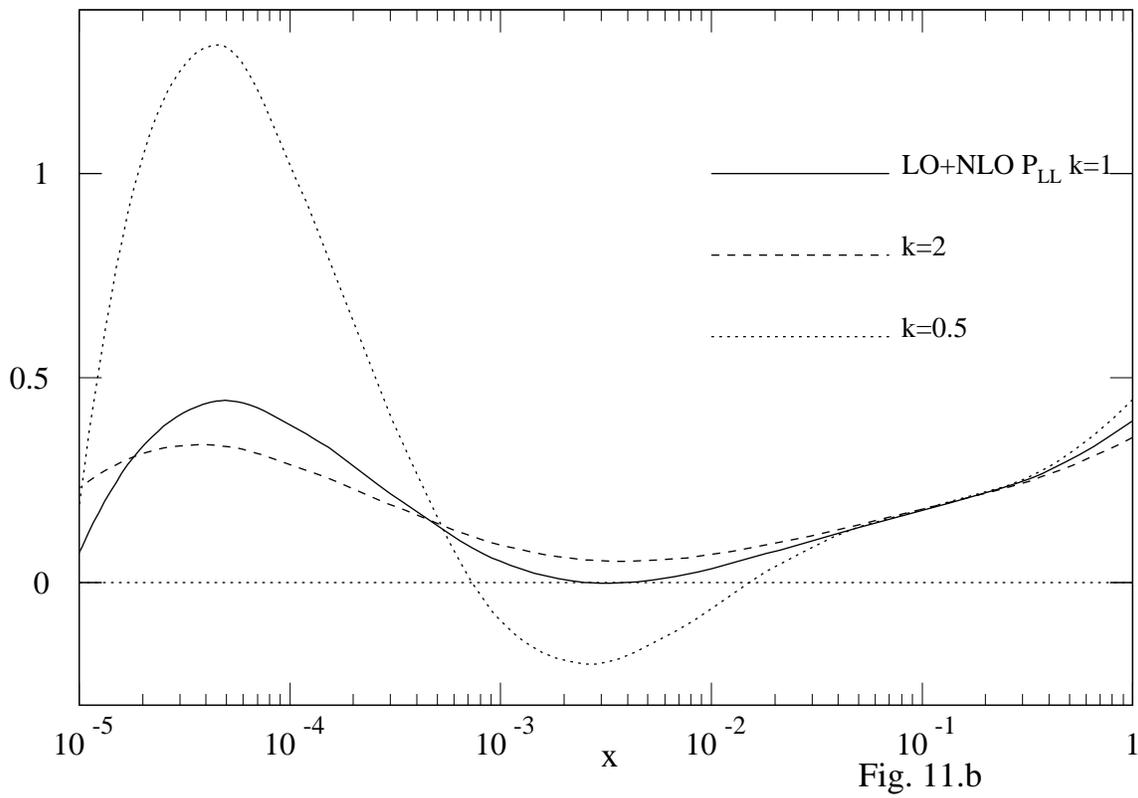
